\documentclass[10pt,superscriptaddress,tightenlines,twocolumn,amsmath,amssymb,aps, prx]{revtex4-2}

\usepackage[utf8]{inputenc}
\usepackage{CJKutf8}

\UseRawInputEncoding
\usepackage{amsmath}
\usepackage{amssymb}
\usepackage{amsthm}
\usepackage{graphicx}
\usepackage{dcolumn}
\usepackage{bm}
\usepackage{bbm}
\usepackage{color}

\usepackage{subfigure}
\usepackage{indentfirst}
\usepackage{mathrsfs}
\usepackage{enumerate}
\usepackage{enumitem}
\usepackage{algorithm}
\usepackage{algpseudocode}
\floatname{algorithm}{Protocol}
\usepackage{amsfonts}
\usepackage[normalem]{ulem} %package for delete

\usepackage{makecell}
\usepackage{lmodern}
\usepackage{lineno}
\usepackage{multirow}

\usepackage{overpic}
\usepackage{braket}
\usepackage{physics}
\usepackage[dvipsnames]{xcolor}
\usepackage{hyperref}

\newcommand{\be}{\begin{equation}}
\newcommand{\ee}{\end{equation}}

\newcommand{\RNum}[1]{\uppercase\expandafter{\romannumeral #1\relax}}

\def \proj#1{|#1\rangle\! \langle #1|}
\def \ket#1{|#1\rangle}
\def \ketbra#1#2{|#1\rangle\! \langle #2|}

\def \id {\mathbbm{I}}

\def \E {\mathcal{E}}

\def \H {\mathcal{H}}

\def \Enc {\normalfont{\texttt{Enc}}}
\def \Dec {\normalfont{\texttt{Dec}}}
\def \KeyGen {\normalfont{\texttt{KeyGen}}}

\newcommand{\naturals}{\ensuremath{\mathbb{N}}}

\newcommand{\N}{\naturals}

\makeatletter

\newcommand{\Rmnum}[1]{\expandafter\@slowromancap\romannumeral #1@}

\newcommand {\br} [1] {\ensuremath{ \left( #1 \right) }}

\makeatother

\colorlet{RED}{red}
\colorlet{BLACK}{black}

\providecommand{\U}[1]{\protect\rule{.1in}{.1in}}

\hypersetup{
    colorlinks=true, 
    urlcolor=ForestGreen,   
    linkcolor=red,   
    citecolor=blue
}

\algnewcommand{\Input}{\item[\textbf{Input:}]} 
\algnewcommand{\Output}{\item[\textbf{Output:}]}

\makeatletter
\newcommand{\newreptheorem}[2]{\newtheorem*{rep@#1}{\rep@title}\newenvironment{rep#1}[1]{\def\rep@title{#2 \ref{##1}}\begin{rep@#1}}{\end{rep@#1}}}
\makeatother

\newtheorem{theorem}{Theorem}
\newreptheorem{theorem}{Theorem}
\newtheorem{lemma}[theorem]{Lemma}
\newreptheorem{lemma}{Lemma}
\newtheorem{proposition}[theorem]{Proposition}
\newreptheorem{proposition}{Proposition}
\newtheorem{definition}{Definition}
\newtheorem{remark}{Remark}
\newtheorem{corollary}{Corollary}

\newcommand{\restoj}[1]{{#1|_{\H_j}}}

\def \Accept {\rm Accept}
\def \Abort {\rm Abort}

\begin{document}
\begin{CJK*}{UTF8}{gbsn}

\title{Consecutive Measurement Tradeoffs in Quantum Cryptography}

\author{Chen-Xun Weng}%\thanks{These authors contributed equally to this work.}
\email{chenxun.weng@outlook.com}
\affiliation{Centre for Quantum Technologies, National University of Singapore, Singapore 117543, Singapore}
\affiliation{Department of Physics, National University of Singapore, Singapore 117551, Singapore}
\affiliation{National Laboratory of Solid State Microstructures and School of Physics, Collaborative Innovation Center of Advanced Microstructures, Nanjing University, Nanjing 210093, China}

\author{Minglong Qin}%\thanks{These authors contributed equally to this work.}
\email{mlqin@nus.edu.sg}
\affiliation{Centre for Quantum Technologies, National University of Singapore, Singapore 117543, Singapore}

\author{Yanglin Hu(胡杨林)}%\thanks{These authors contributed equally to this work.}
\email{yanglin.hu@u.nus.edu}
\affiliation{Centre for Quantum Technologies, National University of Singapore, Singapore 117543, Singapore}
\affiliation{QICI Quantum Information and Computation Initiative, School of Computing and Data Science, The University of Hong Kong, Pokfulam Road, Hong Kong, China}

\author{Marco Tomamichel}
%\email{marco.tomamichel@nus.edu.sg}
\affiliation{Centre for Quantum Technologies, National University of Singapore, Singapore 117543, Singapore}
\affiliation{Department of Electrical and Computer Engineering, National University of Singapore, Singapore 117583, Singapore}

\date{\today}
\begin{abstract}

Measuring a quantum system can reveal information while disturbing its state. Consequently, the probabilities of obtaining desired outcomes in individual measurements constrain the probability of obtaining them in sequence. This interplay arises naturally in mistrustful quantum cryptography, where a dishonest participant may perform sequential measurements to extract more information than the protocol allows. We develop consecutive measurement theorems (CMTs) to quantify this constraint, relating the average success probability of a single measurement to that of an ordered pair of distinct measurements. For measurements on a common state, we derive the optimal lower bound on consecutive-measurement success as a function of single-measurement success and provide matching constructions over the entire parameter range. We also establish the first robust CMTs for measurements on different but nearby states, with closeness quantified by fidelity or trace distance. We use these results to strengthen sum-binding guarantees for relativistic bit commitment and soundness bounds for relativistic zero-knowledge proofs, and to obtain improved no-go results for quantum oblivious transfer and quantum private query. Our results establish CMTs as a versatile tool for translating consecutive-measurement constraints into concrete cryptographic bounds and suggest applications in other quantum-information settings.

\end{abstract}

\maketitle
\end{CJK*}

\section{Introduction}

Fundamental principles of quantum mechanics, most notably measurement incompatibility and the uncertainty principle, lie at the heart of quantum cryptography~\cite{pirandola2020advances}. Unlike quantum key distribution~\cite{Bennett_2014,Scarani2009security,2020_Xu}, where honest Alice and Bob cooperate against an external eavesdropper, mistrustful quantum cryptography (MQC) considers protocols in which the legitimate participants do not fully trust one another and their security objectives may compete, and either participant may be the dishonest party in the corresponding security analysis~\cite{broadbent2016quantum,bozzio2025quantum}. A dishonest party is therefore typically an internal participant, rather than an external interceptor, and may deviate from the prescribed protocol to improve its own outcome. Prominent MQC tasks include quantum coin flipping~\cite{Bennett_2014,Lo_1997,Spekkens_2002,Kitaev_2003,Ambainis_2004,Mochon_2007,Chailloux_2009,Aharonov_2016,Arora_2019,Bozzio_2020,Arora_2021,Arora_2025}, quantum bit commitment~\cite{brassard1993quantum,Buhrman2006security,Silman_2011,Kent_2012,Ng_2012,hardy2004cheat}, quantum oblivious transfer~\cite{Crepeau_1994,Chailloux_2013,Chailloux_2016,Amiri_2021,Hu_2023,Stroh_2023}, blind quantum computing~\cite{Broadbent_2009,Barz_2012,Fitzsimons_2017}, and quantum or relativistic zero-knowledge proofs~\cite{watrous2006zeroknowledge,chailloux2021relativistic,alikhani2021experimental,shi2024relativistic,weng2026experimental}, among many others. As a result, the attempt by a dishonest party to access multiple pieces of information in such protocols faces a fundamental limitation: measurements that reveal one observable necessarily disturb the others.

Many of the security analyses share a common structure: a participant attempts to measure two distinct properties of the state in sequence. The participant first performs the measurement associated with one property and then pursues the other on the resulting post-measurement state. These attempts cannot be analyzed independently because the first measurement changes the state available to the second. This raises a basic quantitative question: how does the average probability of succeeding in an individual measurement constrain the average probability of succeeding in two distinct measurements in sequence?

This tradeoff has appeared---often implicitly---in a range of prior works. In relativistic bit commitment (RBC), security proofs bound the probability that a dishonest party can successfully unveil both bit values, which can be interpreted through consecutive measurements~\cite{kent1999unconditionally,Unruh2012quantum}. Related reasoning also appears in relativistic zero-knowledge proofs (RZKPs), where coupled-game analyses constrain entangled provers attempting to answer two different challenges~\cite{chailloux2017relativistic,crepeau_2023}. These arguments, however, employ protocol-specific consecutive-measurement bounds rather than developing consecutive measurement theorems (CMT) as a broadly applicable analytical tool. Similarly, in quantum oblivious transfer (QOT) and quantum private query (QPQ), security hinges on an adversary's ability to extract multiple pieces of information encoded in incompatible observables. Existing CMT bounds, however, are generally not tight and do not remain applicable under perturbations away from the idealized setting.

Despite their recurring role, consecutive measurement tradeoffs have not been systematically studied as a standalone object. In particular, prior results do not provide a tight tradeoff relation, nor do they offer robust formulations that remain valid under perturbations of the measurement strategy.

In this work, we develop a sharp and robust analytical toolkit based on consecutive measurement theorems, with three main contributions. First, for binary projective measurements on a common state, under uniform sampling of ordered pairs of distinct measurements, we determine the tight tradeoff bound on the average probability $E$ of succeeding in two distinct measurements at every value of the average probability $V$ of succeeding in an individual measurement and give matching constructions throughout the parameter range. Second, we extend the tradeoff to quantitatively close states, obtaining robustness bounds in fidelity and trace distance. Third, we combine CMTs with protocol-specific reductions in two complementary directions, i.e., strengthening upper bounds in security analyses and constructing explicit sequential attacks that lead to no-go results.

To make the scope of these applications precise, we distinguish two routes by which a cryptographic security experiment can be reduced to a consecutive-measurement tradeoff. The first is the coupled-game route. It applies when the cheating probability relevant to the protocol's security can first be upper-bounded by the quantum value of a projective nonlocal game with uniform inputs, and when the associated coupled game admits an independent nontrivial upper bound. Our tight CMT (Theorem~\ref{thm:consec_mmt}) converts the resulting upper bound on the consecutive-measurement success probability $E$ into an upper bound on the original single-measurement success probability $V$. For the $\mathrm{CHSH}_{q}(p)$ games and their parallel repetitions, no-signaling supplies the required coupled-game bound, yielding Propositions \ref{thm:chsh_upper} and \ref{thm:n_upper}, and consequently, improved sum-binding guarantees for the corresponding RBC protocols (Theorems \ref{thm:sumbinding_RBC} and \ref{thm:sum_binding_parallel_RBC}). The same route also applies to  relativistic zero-knowledge protocols, leading to the better soundness bounds in Propositions \ref{prop:rzko-hamiltonian}, \ref{prop:rzko-subset-sum} and \ref{prop:rzko-3sat}.

The second is the explicit-attack route, for which the new robust CMTs are essential in the imperfect settings considered below. This route uses CMTs to analyze explicit sequential attacks and derive impossibility results. In QOT and QPQ, the correctness condition guarantees that each of several alternative recovery measurements succeeds with high probability, while the corresponding privacy requirement ensures that the residual states associated with these alternatives are sufficiently close for a robust CMT to apply. A dishonest participant can then perform two recovery measurements consecutively, and the robust CMT converts their high individual success probabilities into a lower bound on the probability of succeeding in both. Using this approach, our fidelity-based robust CMT (Theorem \ref{thm:RCMT_F}) gives an improved no-go theorem for QOT (Theorem \ref{thm:bound_RCMT}), while the trace-distance-based robust CMT (Theorem \ref{thm:RCMT_TD}) yields an improved impossibility bound for QPQ (Theorem \ref{thm:QPQ}). The QOT bound can further be transferred to many other MQC primitives through existing cryptographic reductions, as illustrated by QHE.

\section{Tight CMT}

\begin{figure}[t]
   \centering
    \begin{overpic}[width=0.85\columnwidth]{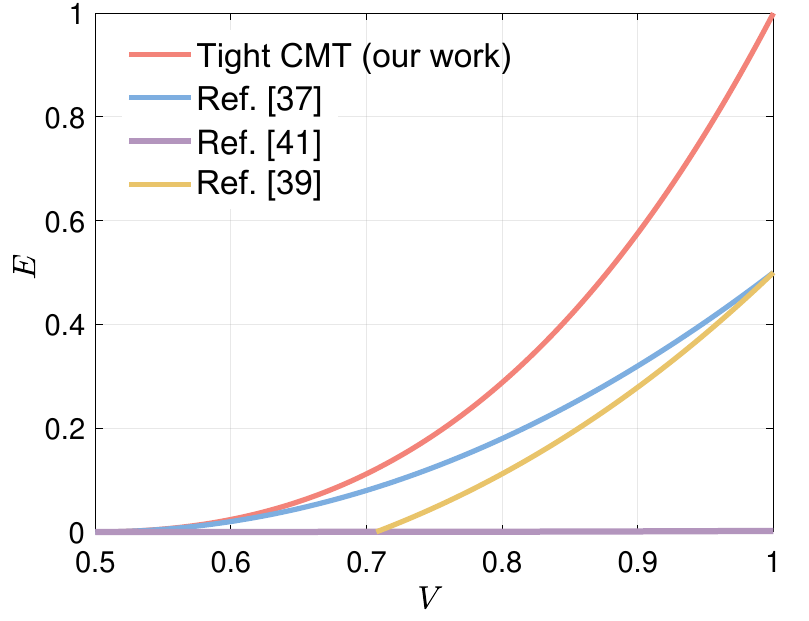}
    \put(24.25,67){\color{white}\rule{85pt}{12pt}} % 白块遮住原文字
    \put(24.5,69){\small Tight CMT (our work)}
    \put(24.25,50){\color{white}\rule{40pt}{36pt}} % 白块遮住原文字
    \put(24.5,63.25){\small Ref.~\cite{shi2024relativistic}}
    \put(24.5,57.75){\small Ref.~\cite{chailloux2017relativistic}}
    \put(24.5,52){\small Ref.~\cite{Unruh2012quantum}}
    \end{overpic}
   \caption{\textbf{Comparison of different lower bounds with $n=2$.} Our tight CMT provides a significantly tighter bound than all previous works for all $V \in [\frac{1}{n}, 1]$. Note that Ref.~\cite{shi2024relativistic} only applies to the special case $n=2$, while Refs.~\cite{Unruh2012quantum, chailloux2017relativistic} and our work are applicable for all $n \geq 2$. Ref.~\cite{Unruh2012quantum} imposes an extra condition that $V \geq \frac{1}{\sqrt{n}}$. Ref.~\cite{chailloux2017relativistic} provides a positive but small bound of the order $10^{-3}$, which is visually indistinguishable from the horizontal axis on a plot with vertical range $[0,1]$. }
   \label{fig:consec_mmt}
\end{figure}

We consider a setting where a party can perform one of $ n $ possible binary projective measurements $\{P_1,\id-P_1\},\dots,\{P_n,\id-P_n\}$ on a quantum state. We consider the case where the party first applies the $ i $-th measurement to a state $ \sigma $, and then, on the resulting post-measurement state, applies the $ j $-th measurement, where $ i \neq j $ is chosen uniformly at random. We study the relationship between the average probabilities of obtaining preferred outcomes ($P_i$ without loss of generality) in the first measurement and in the two consecutive measurements. Our analysis leads to a tight CMT, which improves significantly on previous work~\cite{Unruh2012quantum,chailloux2017relativistic,shi2024relativistic}, as shown in Table~\ref{tab:consec_mmt} and Fig.~\ref{fig:consec_mmt}.

\begin{theorem}[Tight CMT]
\label{thm:consec_mmt}
Let $n\geq 2$ be an integer, let $P_1,\ldots,P_n$ be $n$ projectors in a finite-dimensional Hilbert space $\H$, let $\sigma$ be any quantum state in $\H$, and let 
\begin{equation}
\begin{aligned}
    V & = \frac{1}{n} \sum_{i=1}^n \Tr(P_i\sigma),  \\
    E & = \frac{1}{n(n-1)} \sum_{i\neq j} \Tr(P_j P_i \sigma P_i P_j) .
\end{aligned}
\end{equation}
It holds that
\begin{align}\label{eqn:CMT_inequality}
    E\geq  \frac{n^2}{(n-1)^2 }V \left(\max\left\{0,V-\frac{1}{n}\right\}\right)^2.
\end{align}
Moreover, for any $n\geq 2$ and $v\in[0,1]$, there exists a state $\sigma$ and projectors $P_1,\ldots,P_n$ in some finite-dimensional Hilbert space $\H$, such that $V=v$ and Eq.~\eqref{eqn:CMT_inequality} holds with equality. 
\end{theorem}

It is worth noting that the density operator $\sigma$ need not represent a single physical carrier. In a cryptographic application, it may represent the joint residual state of multiple quantum systems held by the relevant protocol participant when the alternative measurements are applied, and each $P_i$ may be a joint projector on this composite system. Applying the CMT to a particular protocol nevertheless requires identifying these measurements with the corresponding security events. The proof of Theorem~\ref{thm:consec_mmt} can be found in Appendix~\ref{apd:TCMT}.

\section{Applications of Tight CMT: Relativistic Bit Commitment and Zero-Knowledge Proofs}

Quantum bit commitment (QBC)~\cite{brassard1993quantum,hardy2004cheat} is a foundational cryptographic primitive that enables a party to commit to a chosen bit while keeping it concealed, with the binding guarantee that the committed value cannot subsequently be changed. QBC also serves as a building block for more complex secure multiparty primitives, including quantum zero-knowledge proofs~\cite{watrous2006zeroknowledge,yan2015quantum}. However, unconditionally secure QBC is impossible when security relies solely on quantum mechanics~\cite{lo1997isquantum,mayers1997unconditionally}. This impossibility motivated relativistic bit commitment (RBC), in which spatially separated agents are prevented from communicating during designated protocol stages and security is enforced by the no-signaling principle~\cite{kent1999unconditionally,Lunghi2013Experimental,liu2014Experimental,lunghi2015practical,chakraborty2015arbitrarily,Verbanis201624-Hour}.

\begin{table}[t]
    \centering
    \setlength{\tabcolsep}{4pt} 
    \renewcommand{\arraystretch}{1.3} 
    \caption{\textbf{Different CMTs.} Our CMT is tighter than those in previous works~\cite{Unruh2012quantum,chailloux2017relativistic,shi2024relativistic}. The indicated ranges are those for which the lower bounds are non-trivial. }

    \begin{tabular}{cc}
        \hline\hline
         & Statement\\
        \hline
         Ref.~\cite{Unruh2012quantum} &  $E\geq V(V^2-\frac{1}{n})$, $\forall n\geq 2$, $\frac{1}{\sqrt{n}}\leq V \leq 1$ \\
        Ref.~\cite{chailloux2017relativistic} & $E\geq \frac{1}{64}\br{V-\frac{1}{n}}^3$, $\forall n\geq 2$, $\frac{1}{n}\leq V\leq 1$\\
        Ref.~\cite{shi2024relativistic} &  $E\geq 2\br{V-\frac{1}{2}}^2$, $n=2$,   $\frac{1}{2}\leq V\leq 1$ \\
        Tight CMT & $E \geq  \frac{n^2}{(n-1)^2 }V \br{V-\frac{1}{n}}^2$, $\forall n\geq 2$,  $\frac{1}{n}\leq V\leq 1$ \ \\
        \hline\hline
    \end{tabular}
    \label{tab:consec_mmt}
\end{table}

Relativistic zero-knowledge proofs (RZKPs) exploit the same physical principle to achieve information-theoretic security without relying on computational assumptions~\cite{chailloux2017relativistic,alikhani2021experimental,crepeau_2023,weng2026experimental}. In a typical two-prover one-round RZKP, two spatially separated prover agents receive classical questions from the verifier, return classical answers, and are accepted according to a prescribed verification predicate. Even if dishonest provers share arbitrary entanglement and perform general local operations, their separated responses obey no-signaling, and the protocol's soundness on false statements is characterized by the quantum value of the associated nonlocal game. 

The security analyses of both RBC and RZKP can therefore be related to projective nonlocal games and their coupled games. In the coupled-game method~\cite{chailloux2017relativistic,crepeau_2023,shi2024relativistic}, one prover is required to produce valid answers to two distinct classical challenges. A CMT lower-bounds this consecutive-measurement success probability in terms of the quantum value of the original game, while an independent no-signaling argument upper-bounds the coupled-game value. Combining these two bounds yields an upper bound on the original quantum value. 
Using the coupled-game method, our tight CMT yields improved upper bounds for CHSH games and their parallel repetitions, thereby strengthening the security analysis and reducing the communication overhead for RBC protocols. For RZKP, it improves the quantum soundness analyses of relativistic protocols for Hamiltonian Cycle, Subset Sum, and 3SAT.

\subsection{Nonlocal game and coupled game}

To apply our tight CMT to RBC and RZKP, we frame the adversary's actions as a nonlocal game, and their sequential actions as a coupled game. 

We provide informal introductions to all relevant concepts of the coupled-game method and defer the formal definitions to Appendix \ref{subsec:formaldef}. A nonlocal game, which serves as a powerful witness of quantum nonlocality~\cite{Clauser1969proposed,Cirel1980quantum,Almeida2010Guess,Tomamichel2013monogamy,Brunner2014bell,li2023device,cleve2004consequences,ji2021mip,kahanamoku2022classically,zhu2023interactive,dong2025computational}, involves a referee and two cooperating but non-communicating players, Alice and Bob. The game is specified by an input distribution $ \pi $ over $ \mathcal{I}_A \times \mathcal{I}_B $ and a winning condition $ V: \mathcal{I}_A \times \mathcal{I}_B \times \mathcal{O}_A \times \mathcal{O}_B \to \{0,1\} $. The referee samples a pair of questions $ (x, y) \in \mathcal{I}_A \times \mathcal{I}_B $ according to $ \pi $, and sends $ x $ to Alice and $ y $ to Bob. Without communicating, Alice responds with an output $ a \in \mathcal{O}_A $, and Bob with $ b \in \mathcal{O}_B $. The players win the round if $ V(x, y, a, b) = 1 $. A nonlocal game has a uniform input distribution if $\pi(x,y) = \frac{1}{|\mathcal{I}_A||\mathcal{I}_B|}$ for all $(x,y) \in \mathcal{I}_A \times \mathcal{I}_B$.  
Additionally, we call a game projective if for any $(x, a, y)$, there is a unique $b$ such that $V(x,y,a,b) = 1$.

The quantum value $\omega^*(G)$ is the highest probability for the players to win the game $G$ using a quantum strategy.
Determining the exact value, or even a tight upper bound, on the quantum value is difficult. The coupled-game approach was proposed in~\cite{chailloux2017relativistic} to upper bound the quantum value based on constructing an associated coupled game without explicitly constructing quantum strategies. 

For a nonlocal game $G$ where Alice and Bob receive inputs $ x \in \mathcal{I}_A $ and $ y \in \mathcal{I}_B $, and respond with outputs $ a \in \mathcal{O}_A $ and $ b \in \mathcal{O}_B $, respectively, with the winning condition $ V(x, y, a, b) = 1 $, we can construct a coupled game $G_{\text{cp}}$ as follows. In the coupled game, Alice receives an input $ x\in\mathcal{I}_A $, while Bob receives a pair of inputs $ y, y' \in \mathcal{I}_B $ with $ y \neq y' $, and returns outputs $ b $ and $ b' $ corresponding to $ y $ and $ y' $, respectively. The players win if $ V(x, y, a, b) = V(x, y', a, b') = 1 $ holds.

As shown in~\cite[Proposition 1]{chailloux2017relativistic}, any strategy $\mathscr{S}$ for the original game $G$ induces a corresponding strategy $\mathscr{S}_{\text{cp}}$ for the coupled game $G_{\text{cp}}$. The key difference lies in Bob's action: in $\mathscr{S}$, Bob performs a single measurement $\mathcal{P}_y$, while in $\mathscr{S}_{\text{cp}}$, he sequentially performs two consecutive measurements $\mathcal{P}_y$ and $\mathcal{P}_{y'}$ for $y\neq y'$. As a POVM measurement can always be dilated into a projective measurement in a higher dimensional space, we assume that $\mathscr{S}$ uses projective measurements.

Our tight CMT can be used to quantify the relation between $ \omega^*(G_{\text{cp}})$ and $ \omega^*(G) $: 

Theorem~\ref{thm:consec_mmt} relates two key probabilities: $ V $, the probability of obtaining a preferred outcome in the first measurement, and $ E $, the probability of obtaining preferred outcomes in both consecutive measurements. To apply this theorem to our setting, we rely on the observation originally made in~\cite{chailloux2017relativistic}, that the winning probability conditioned on a specific pair $(x,a)$ for the original game corresponds to $ V $, while the winning probability conditioned on the same $(x,a)$ for the coupled game corresponds to $ E $. The proof of Proposition~\ref{thm:coupledgamevalue} follows the same strategy as that of \cite[Proposition 1]{chailloux2017relativistic}, except that we use a tighter consecutive measurement theorem.

\begin{proposition}\label{thm:coupledgamevalue}
Let $ G $ be a projective nonlocal game with uniform input distribution, and let $ n = |\mathcal{I}_B| $. Then the quantum values of $ G $ and its coupled game $G_{\text{cp}}$ satisfy:
\begin{equation}
    \label{eqn:omega_relation}
    \omega^*(G_{\text{cp}}) \geq \frac{n^2}{(n-1)^2} \omega^*(G)\br{\max\left\{0,\omega^*(G)-\frac{1}{n}\right\}}^2.
\end{equation}   
\end{proposition}

The proof of Proposition~\ref{thm:coupledgamevalue} is deferred to Appendix~\ref{apd:LBwG}. 

If a nontrivial upper bound on $ \omega^*(G_{\text{cp}}) $ can be obtained, Proposition~\ref{thm:coupledgamevalue} allows us to derive a corresponding upper bound on $ \omega^*(G) $. As an example, we apply Proposition~\ref{thm:coupledgamevalue} to the so-called $ \mathrm{CHSH}_q(p) $ game and its parallel game.

The ${\rm CHSH}_q(p)$ game generalizes the standard game ${\rm CHSH}_2(2)$ to inputs from larger finite fields. Let $p=r^t$ and $q=r^s$ be powers of the same prime $r$, with $t$ dividing $s$, and fix an embedding of $\mathbb F_p\hookrightarrow\mathbb F_q$.  Alice's input and both outputs are elements of $\mathbb F_q$, whereas Bob's input is an element of this embedded copy of $\mathbb F_p$. 
They win if $ a + b = x \cdot y $, where all arithmetic is in the finite field $ \mathbb{F}_q $. 
The ${\rm CHSH}_q(p)$ game is projective, i.e., for a fixed $(x,a,y)$, there is a unique $b$ that wins the ${\rm CHSH}_q(p)$ game. 

While the exact classical and quantum values of $ {\rm CHSH}_q(p)$ remain unknown and even upper bounds are difficult to determine, constructing its coupled game enables one to derive an upper bound on the quantum value~\cite{chailloux2017relativistic}. In this case, the winning probability of the ${\rm CHSH}_q(p)_{\text{cp}}$ game is easy to analyze due to the algebraic structure of a field. Specifically, the winning conditions, $ a + b = x \cdot y $ and $ a + b^{\prime} = x \cdot y^{\prime} $, imply that they win only if Bob can guess $x = \frac{b-b'}{y-y'}$ correctly. Therefore, the players can win with a probability of at most $\frac{1}{|\mathcal{I}_A|}$ due to the no-signaling principle. Although the bound derived in this way is not necessarily tight, it is sufficient for certain cryptographic applications, such as relativistic bit commitment and relativistic zero-knowledge proof~\cite{chailloux2017relativistic,chailloux2021relativistic,weng2026experimental}. Here we present a tighter bound for the quantum value of $ {\rm CHSH}_q(p) $, derived from Proposition \ref{thm:coupledgamevalue}. The proof is deferred to Appendix~\ref{apd:chsh_upper}. 

\begin{proposition}[Upper bound of $\omega^{*}$(${\rm CHSH}_q(p)$)]
    \label{thm:chsh_upper}
    Let $p=r^t$ and $q=r^s$ be powers of the same prime $r$, with $t$ dividing $s$. Then the upper bound on the quantum value of the ${\rm CHSH}_q(p)$ game is given by:
\begin{equation}
\label{eqn:chshqp_analytical}
\begin{aligned}
    \omega^*({\rm CHSH}_q(p)) \leq \frac{1}{3p} \br{2 + \mathcal{C}(\Delta) },
\end{aligned}
\end{equation}
where $\Delta = \frac{27p(p-1)^2}{q} - 2$ and $\mathcal{C}(\cdot)$ is a function defined by
\begin{equation}\label{eqn:defofCdelta}
    \mathcal{C}(\Delta) =
    \begin{cases}
        2\cosh\!\left(\dfrac{1}{3}\operatorname{arccosh}  \dfrac{\Delta}{2}\right), & \Delta\in [2,\infty),\\ 
        2\cos\!\left(\dfrac{1}{3}\arccos \dfrac{\Delta}{2}\right), & \Delta\in(-2,2). 
    \end{cases}
\end{equation}
\end{proposition}

\begin{table}[h]
\centering
\small 
\setlength{\tabcolsep}{7.5pt} 
\renewcommand{\arraystretch}{1.1} 
\caption{\textbf{Different explicit formula upper bounds on quantum values of ${\rm CHSH}_{2^l}(2)$ game}. The bounds in Ref.~\cite{Julia2010unique} are numerical.}
\begin{tabular}{lccccc}
\hline\hline
Value & $l=1$ & $l=2$ & $l=3$ & $l=4$ & $l=5$ \\
\hline
Ref.~\cite{chailloux2017relativistic} & 3.67  & 3.020  & 2.500  & 2.087  & 1.760 \\
Ref.~\cite{fillinger2019two} & 1.207 & 1 & 0.854  & 0.750  & 0.677 \\
Ref.~\cite{shi2024relativistic,sikora2014strong} & 1 & 0.854 & 0.750  & 0.677  & 0.625 \\
SDP~\cite{Julia2010unique} & 0.853 & 0.780 & 0.743 & 0.725 & 0.716 \\
Our work & 0.877 & 0.783 & 0.710 & 0.654 & 0.613 \\
\hline\hline
\end{tabular}
\label{tab:l}
\end{table}

To illustrate the performance of our bound, we consider a specific family of CHSH games that are widely used in relativistic bit commitment protocols~\cite{lunghi2015practical,chakraborty2015arbitrarily,Verbanis201624-Hour}, where $q=2^l$ and $p=2$. A best-known upper bound for this setting states that $\omega^{\ast}({\rm CHSH}_{2^l}(2)) \leq \frac{1}{2}+\frac{1}{\sqrt{2^{l+1}}} $~\cite{sikora2014strong}, which is also obtained in Ref.~\cite{shi2024relativistic}. However, it is not tight for small $l$. For example, when $l=1$, which corresponds to the standard CHSH game, the quantum value is $\frac{1}{2}+\frac{1}{2\sqrt{2}}\approx0.853 $, whereas the bound of Ref.~\cite{shi2024relativistic} yields only $ \omega^*({\rm CHSH}_2(2))\leq 1$. In contrast, our bound gives $ \omega^*({\rm CHSH}_2(2))\leq 0.877 $, which is significantly closer to the optimal value~\footnote{One may wonder why our bound does not achieve the exact optimal value. We believe that the reason is that the consecutive-measurement strategy used in the proof is not optimal for the coupled game.}. Our bound is strictly tighter than the bounds in~\cite{sikora2014strong,shi2024relativistic,chailloux2017relativistic} for all $l$ due to the tightness of our improved CMT. An alternative approach to upper bounding the value of this game numerically is via semidefinite programming (SDP)~\cite{Julia2010unique}. Our result outperforms the SDP-based bound starting from $l=3$, which shows the practicality of our work. A numerical comparison of these bounds for various values of $l$ in $ {\rm CHSH}_{2^l}(2) $ is presented in Table~\ref{tab:l}.

Consider the $m$-fold parallel repetition of the ${\rm CHSH}_q(p)$ game, denoted ${\rm CHSH}_{q}(p)^{\otimes m}$.  Alice and Bob receive $m$ independent and uniformly random input strings, $X=(x_1, \dots, x_m)\in\mathbb{F}_q^{m}$ and $Y = (y_1, \dots, y_m) \in \mathbb{F}_p^{m}$, respectively. They respond with output strings $A= (a_1, \dots, a_m) \in \mathbb{F}_q^{m}$ and $B=(b_1, \dots, b_m)\in \mathbb{F}_q^{m}$. The players win if and only if $a_i+b_i=x_i\cdot y_i$ for all $i\in[m]$. 

Applying Proposition~\ref{thm:coupledgamevalue}, we can also obtain the following asymptotic upper bound on the quantum value of ${\rm CHSH}_{q}(p)^{\otimes m}$:

\begin{proposition}[Upper bound of $\omega^{*}$(${\rm CHSH}_{q}(p)^{\otimes m}$)]\label{thm:n_upper}
Let $p=r^t$ and $q=r^s$ be powers of the same prime $r$, with $t$ dividing $s$. Then the quantum value of $ {\rm CHSH}_{q}(p)^{\otimes m} $ is upper bounded by:
\begin{equation}
    \begin{aligned}
        \label{eqn:chshqpm_analytical}
        \omega^*({\rm CHSH}_q(p)^{\otimes m}) \leq  \frac{1}{3p^m} \br{2 +  \mathcal{C}(\Delta_m) }, \\
    \end{aligned}
\end{equation}
where $\mathcal{C}(\cdot)$ is defined in Eq.~\eqref{eqn:defofCdelta} and 
\begin{equation}
     \Delta_m = 27p^m \br{p^m-1} \br{\br{1+\frac{p-1}{q}}^m-1} - 2. 
\end{equation}
If $m\gg \frac{q}{p-1}$, the quantum value of $ {\rm CHSH}_{q}(p)^{\otimes m} $ is asymptotically upper bounded by:
\begin{equation}
    \omega^{*}({\rm CHSH}_{q}(p)^{\otimes m})\leq \frac{1}{p^m} + \frac{1}{p^{\frac{m}{3}}}\br{1+\frac{p-1}{q}}^{\frac{m}{3}}.
\end{equation}    
\end{proposition}

The derivations of the analytical and asymptotic bounds are deferred to Appendix~\ref{apd:chsh_m_upper}. Our bound on $ \omega^{*}({\rm CHSH}_{q}(p)^{\otimes m}) $ is tighter than the one given in~\cite{chailloux2017relativistic}. In particular, our bound for $ \omega^{*}({\rm CHSH}_2(2)^{\otimes m}) $ is significantly closer to the perfect parallel repetition~\cite{cleve2008perfect}. Asymptotically, our bound for $ \omega^{*}({\rm CHSH}_2(2)^{\otimes m}) $ approaches $(\frac{3}{4})^{\frac{m}{3}}\approx 0.909^m$. Although it remains an open question whether perfect parallel repetition holds for ${\rm CHSH}_q(p)^{\otimes m}$ in general, we conjecture that our bound closely approximates the true quantum value $\omega^{*}({\rm CHSH}_{q}(p)^{\otimes m}) $.

\subsection{Sum-binding of relativistic bit commitment}
Bit commitment involves two parties, Alice and Bob. Bob (the prover) commits a value $y$ to Alice (the verifier) such that Alice cannot learn the value before the reveal phase (hiding), while Bob cannot change the committed value after the commit phase (binding). In two-prover relativistic bit commitment (RBC), both parties are split into two spatially separated agents: $A_1, A_2$ for Alice and $B_1, B_2$ for Bob. The pairs $A_1$-$B_1$ and $A_2$-$B_2$ are co-located, while the two locations are separated by a sufficient distance to enforce the no-signaling principle. RBC protocols achieve perfect hiding. However, analyzing the binding property is more complex and determines the required computational resources.

Against quantum adversaries, the binding property is captured by the sum-binding condition, which quantifies the ability of the provers $B_1$ and $B_2$ to cheat using shared quantum entanglement. Intuitively, it bounds their capability to successfully unveil both possible committed bits after the commit phase, even when employing quantum strategies. The formal definition of the sum-binding property is given below.

\begin{definition}[Sum-binding]
A quantum bit commitment scheme is sum-binding if, for all possible malicious quantum strategies, (Com$^*$, Open$^*$), employed by Bob the Prover(s) during the commitment and reveal phases, the following inequality holds:
\begin{multline}
    \sum_y \max_{\text{Open}^*}\Pr[\text{Bob successfully reveals \textit{y}}| (\text{Com}^*, \text{Open}^*)]\\ \le 1+\varepsilon_b.
\end{multline}
where $\varepsilon_b$ represents the binding error and is a positive, negligible quantity.

\end{definition}

\subsubsection{\texorpdfstring{$\mathbb{F}_p$}{Fp} relativistic bit commitment}\label{P_BC}
Standard relativistic bit commitment protocols typically involve a binary commitment ($y \in \{0, 1\}$) \cite{chakraborty2015arbitrarily}. However, more general schemes allow for commitments to values from a larger finite field. Here we take $\mathbb{F}_p$ and $\mathbb{F}_q$ to be fields of orders $p=r^t$ and $q=r^s$, respectively, where $t$ divides $s$~\cite{chailloux2021relativistic}. Note that in $\mathbb{F}_p$-string bit commitment, the ``bit'' is not binary but has $p$ possible values where $y\in \mathbb{F}_p$. 

\begin{figure}[t]
    \centering
    \includegraphics[width=0.75\linewidth]{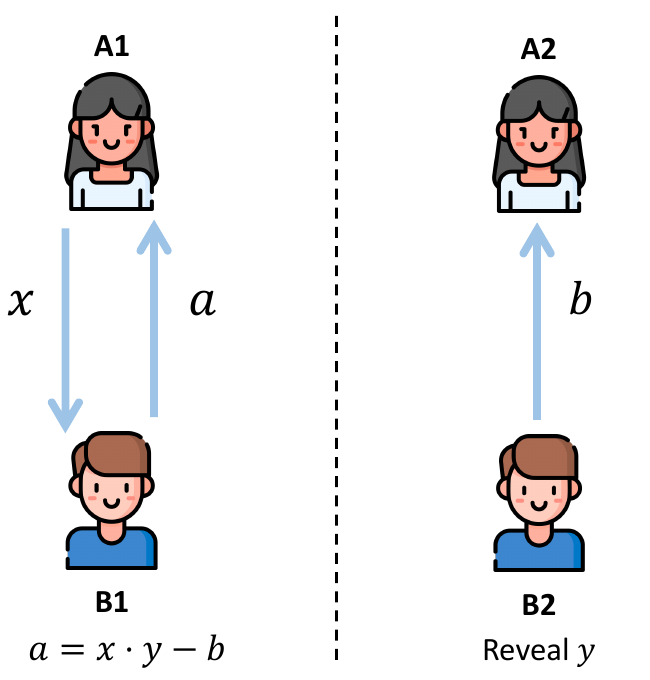}
     \caption{\textbf{Schematic of $\mathbb{F}_p$ RBC.} There are two agents of Alice (Bob), denoted by A1 and A2 (B1 and B2), which are spatially separated to satisfy the no-signaling principle. A1 sends $x\in \mathbb{F}_q$ to B1, and B1 responds with $a\in \mathbb{F}_q$, where $a=x\cdot y-b$. Here $y\in \mathbb{F}_p$ is the committed value and $b\in\mathbb{F}_q$ is a pre-shared uniformly random element held by B1 and B2. To reveal $y$, B2 sends the value $b$ to A2.}
    \label{fig:RBC}
\end{figure}

\begin{definition}[$\mathbb{F}_p$ RBC]
The $\mathbb{F}_p$ relativistic bit commitment proceeds as follows (see Fig.~\ref{fig:RBC}):

\begin{enumerate}
    \item [0.] \textbf{Preparation phase:} A1 and A2 jointly generate a uniformly random query $x\in \mathbb{F}_q$. B1 and B2 jointly generate a uniformly random encoding key $b\in \mathbb{F}_q$.
    \item[1.]  \textbf{Commit phase:} A1 sends the query $x\in \mathbb{F}_q$ to B1. B1 immediately commits $a$ where $a=x\cdot y-b$ and $y$ is the value the provers want to commit. 
    \item[2.] \textbf{Reveal phase:} B2 provides A2 with $b$ and $y$ to let A2 check whether $a+b=x\cdot y$ holds. If the equation holds, the commitment is accepted.
\end{enumerate}

\end{definition}

\begin{theorem}[Sum-binding of $\mathbb{F}_p$ RBC]\label{thm:sumbinding_RBC}
The $\mathbb{F}_p$ relativistic bit commitment is $\varepsilon_b$-sum-binding with 
\begin{equation}
    \varepsilon_b=\frac{1}{3}\left(\mathcal{C}(\Delta) -1\right),
\end{equation}
where $\mathcal{C}(\cdot)$ is defined in Eq.~\eqref{eqn:defofCdelta} and $\Delta = \frac{27p(p-1)^2}{q} - 2$. 
\end{theorem}
The proof of Theorem~\ref{thm:sumbinding_RBC} can be found in Appendix~\ref{apd:RBC}.

Moreover, the security parameter $\varepsilon_b$ can be chosen to guarantee the $\varepsilon_b$-sum-binding property of the $\mathbb{F}_p$-string relativistic bit commitment protocol. To achieve $\varepsilon_b$-sum-binding, we require $\frac{1}{3}\left(\mathcal{C}(\Delta) -1\right)= \varepsilon_b$, which determines the required value of $q$, and hence the number of communication bits $N=\log q$. 

To the best of our knowledge, the upper bound on the quantum value of ${\rm CHSH}_q(p)$ derived using our tight CMT method is tighter than the existing analytical bounds and, for $l\geq3$, the numerical SDP bounds as well, as summarized in Table~\ref{tab:l}. Therefore, applying Theorem~\ref{thm:sumbinding_RBC} leads to a substantially reduced communication cost $N$ in experimental implementations of RBC.

\subsubsection{\texorpdfstring{$m$-fold}{m-fold} parallel \texorpdfstring{$\mathbb{F}_p$}{Fp} relativistic bit commitment}\label{parallel_BC}
Certain cryptographic applications require multiple parallel bit commitments to commit to a long message to be revealed during later stages. However, the $\mathbb{F}_p$-string commitment scheme described earlier does not inherently support revelation of long messages within parallel commitments. Therefore, we require the $m$-fold parallel repetition of the $\mathbb{F}_p$-string commitment scheme ($\mathbb{F}_p^{m}$ RBC) as follows.

\begin{definition}[$\mathbb{F}_p^{m}$ RBC]\label{def_parallel_BC}

Let $\mathbb{F}_p$ and $\mathbb{F}_q$ be fields of order $p=r^t$ and $q=r^s$ respectively, with $t$ dividing $s$. The $\mathbb{F}_p^{m}$ relativistic bit commitment proceeds as follows:
\begin{itemize}
    \item[0.] \textbf{Preparation phase}: A1 and A2 pre-share $m$ random strings $\left(x_1,x_2,\cdots,x_m\right)\in \mathbb{F}^{m}_q$. B1 and B2 pre-share $m$ uniformly random strings $\left(b_1,b_2,\cdots,b_m\right)\in \mathbb{F}^{m}_q$.
    \item[1.] \textbf{Commit phase}: A1 sends $X=\left\{x_i\right\}_{i\in \left[m\right]}$ to B1 and B1 immediately replies with $A=\left\{a_i\right\}_{i\in \left[m\right]}  $, where $a_i=x_i\cdot y_i-b_i$. $y_i\in\mathbb{F}_p$ is what the provers want to commit and all the calculations are in the finite field $\mathbb{F}_q$.
    \item[2.] \textbf{Reveal phase:} Within the effective time interval $\tau_c=\frac{d}{c}$, where $d$ is the distance between B1 and B2 and $c$ is the speed of light, B2 reveals all the committed values to A2 along with $B=\left\{b_i\right\}_{i\in \left[m\right]}$ and $Y=\left\{y_i\right\}_{i\in \left[m\right]}$. A2 checks whether for each $i\in  \left[m\right]$, the relationship $x_i\cdot y_i=a_i+b_i$ holds. The commitment is accepted if all verifications succeed.
\end{itemize}

\end{definition}
  
Additionally, directly applying the sum-binding property in parallel commitments does not ensure composable security due to the lack of strong parallelism of sum-binding~\cite{kaniewski2013secure}. To address these challenges, we analyze the security of the $\mathbb{F}_p^{m}$ relativistic bit commitment, considering the parallel repetition as an integrated entity for sum-binding evaluation.

\begin{theorem}[Sum-binding of $\mathbb{F}_p^{m}$ RBC]\label{thm:sum_binding_parallel_RBC}
The $\mathbb{F}_p^{m}$ relativistic bit commitment is  $\varepsilon_b$-sum-binding with
\begin{equation}
    \varepsilon_b=\frac{1}{3}\left(\mathcal{C}(\Delta_m) -1\right),
\end{equation}
where $\mathcal{C}(\cdot)$ is defined in Eq.~\eqref{eqn:defofCdelta} and 
\begin{equation}
    \Delta_m = 27p^m \br{p^m-1} \br{\br{1+\frac{p-1}{q}}^m-1} - 2. 
\end{equation}
\end{theorem}
The proof of Theorem~\ref{thm:sum_binding_parallel_RBC} can be found in Appendix~\ref{apd:RBC_parallel}.

\subsection{Soundness analysis of relativistic zero-knowledge proofs}
\label{sec:rzko-application}

The coupled-game method applies to several relativistic zero-knowledge proofs (RZKPs) whose soundness can be formulated as a two-prover one-round nonlocal game with finite classical questions and answers. More precisely, for a protocol $\Pi$ and a fixed no-instance $z\notin L$, let $G_z$ denote the associated soundness game. Its quantum value is exactly the largest probability with which entangled cheating provers can make the verifier accept:
\begin{equation}
    s_{\rm q}(\Pi;z) := \sup_{\mathcal{S}_{\rm ent}} \Pr\!\left[ \Pi\text{ accepts }z \,\middle|\, \mathcal{S}_{\rm ent} \right] = \omega^*(G_z),
    \label{eq:rzkp-soundness-game-value}
\end{equation}
where the supremum is taken over all entangled prover strategies. For fixed protocol parameters, the one-round quantum soundness error of $\Pi$ is therefore
\begin{equation}
    s_{\rm q}(\Pi)  := \sup_{z\notin L}s_{\rm q}(\Pi;z) = \sup_{z\notin L}\omega^*(G_z).
    \label{eq:rzkp-overall-soundness}
\end{equation}
Thus, a uniform upper bound on $\omega^*(G_z)$ over all no-instances directly gives a quantum soundness bound for the protocol.

We apply this perspective to three representative RZKPs: the Hamiltonian-Cycle protocol of Chailloux and Leverrier~\cite{chailloux2017relativistic}, and the homomorphic-commitment protocols for Subset Sum and $3\mathrm{SAT}$ proposed by Cr\'epeau and Stuart~\cite{crepeau_2023}. In all three protocols, the second prover receives a uniformly random binary challenge. Winning $G_z$ corresponds to producing an accepting answer to the randomly selected challenge, whereas producing accepting answers to both possible challenges is exactly the winning event of the associated coupled game $G_{z,{\rm cp}}$. Consequently, an upper bound on $\omega^*(G_{z,{\rm cp}})$, combined with our tight CMT, yields an upper bound on $\omega^*(G_z)$ and hence on the protocol's quantum soundness.

Throughout the following applications, $q$ denotes the order of the finite field $\mathbb F_q$ used by the underlying relativistic commitment scheme of RZKPs, where $q$ is a prime power and all field arithmetic is performed over $\mathbb F_q$. The original soundness proofs of the three protocols share an additional protocol-specific property: jointly accepting answers to both binary challenges determine an element sampled uniformly from $\mathbb F_q$ by the verifier. Since this element is unavailable to the spatially separated second prover, the no-signaling principle gives a coupled-game upper bound of $1/q$. Although the same upper bound occurs in all three applications, it is established separately from the structure of each protocol and is not a generic property of RZKPs.

Our analysis does not modify these protocols or the overall structure of their original soundness proofs. We only replace the consecutive-measurement bound used in the coupled-game step with our tight CMT, while all other steps remain unchanged. Complete protocol descriptions and the remaining proof details can be found in Refs.~\cite{chailloux2017relativistic,crepeau_2023}.

More generally, a nonlocal game is called $S$-projective if, for every fixed pair of questions and every fixed answer of the first prover, at most $S$ answers of the second prover are accepting.

\begin{proposition}[Coupled-game bound for $S$-projective games]
\label{prop:s-projective-coupled}
Let $G$ be a uniformly distributed $S$-projective nonlocal game, and suppose that the second prover has $r\geq2$ possible questions. Then
\begin{equation}
    \omega^*(G_{\rm cp})  \geq  \frac{r^2}{(r-1)^2S}\,  \omega^*(G)  \left(  \max\left\{0,\omega^*(G)-\frac1r\right\}  \right)^2.  
    \label{eq:s-projective-coupled}
\end{equation}
In particular, consider a two-challenge $S$-projective nonlocal game where $r=2$ whose coupled game satisfies $\omega^*(G_{\rm cp})\leq\eta$, where $0<\eta\leq1/S$. Then

\begin{equation}
    \omega^*(G) \leq \Omega(S,\eta) := \frac{1}{6} \left[ 2+\mathcal{C}\!\left(54S\eta-2\right) \right],
    \label{eq:s-projective-analytic}
\end{equation}
where $\mathcal{C}(\cdot)$ is defined in Eq.~\eqref{eqn:defofCdelta}.
\end{proposition}

Proposition~\ref{prop:s-projective-coupled} follows from the multi-outcome extension of our tight CMT, stated as Proposition~\ref{thm:consec_mmt_general}. Its proof is given in Appendix~\ref{app:rzko}.

The first application is the Hamiltonian-Cycle protocol of Ref.~\cite{chailloux2017relativistic}. For an $n$-vertex graph, its soundness game $G_{\rm HAM}$ is $n!$-projective, and the protocol-specific no-signaling argument gives $\omega^*(G_{{\rm HAM},{\rm cp}})\leq\frac1q$. Substituting $S=n!$ and $\eta=1/q$ into Proposition~\ref{prop:s-projective-coupled} gives the following result.

\begin{proposition}[Quantum soundness of the Hamiltonian-Cycle RZKP]
\label{prop:rzko-hamiltonian}
For an $n$-vertex graph with no Hamiltonian cycle and an admissible field order $q\geq n!$, the acceptance probability of entangled cheating provers satisfies
\begin{equation}
    \omega^*(G_{\rm HAM}) \leq \frac{1}{6} \left[ 2+\mathcal{C}\!\left(\frac{54n!}{q}-2\right) \right].
    \label{eq:ham-exact-soundness}
\end{equation}
Consequently, for $k\geq1$, any admissible prime-power field order satisfying
\begin{equation}
    q\geq \frac{n!\,2^{2k-1}}{1+2^{1-k}}
    \label{eq:ham-exact-field-order}
\end{equation}
ensures quantum soundness at most $1/2+2^{-k}$.
\end{proposition}

The second application is the homomorphic-commitment RZKP for the Subset Sum problem proposed in Ref.~\cite{crepeau_2023}. For an instance containing $n$ elements, its soundness game $G_{\rm SS}$ is $2^n$-projective, and its protocol-specific no-signaling argument gives $\omega^*(G_{{\rm SS},{\rm cp}})\leq\frac1q$.
Substituting $S=2^n$ and $\eta=1/q$ into Proposition~\ref{prop:s-projective-coupled} gives the following result.

\begin{proposition}[Quantum soundness of the Subset Sum RZKP]
\label{prop:rzko-subset-sum}
For a Subset Sum no-instance containing $n$ elements and an admissible field order $q\geq2^n$, the acceptance probability of entangled cheating provers satisfies
\begin{equation}
    \omega^*(G_{\rm SS}) \leq \frac{1}{6} \left[ 2+\mathcal{C}\!\left(\frac{54\cdot2^n}{q}-2\right) \right].
    \label{eq:ss-exact-soundness}
\end{equation}
Consequently, for $k\geq1$, any admissible prime-power field order satisfying
\begin{equation}
    q\geq \frac{2^{n+2k-1}}{1+2^{1-k}}
    \label{eq:ss-exact-field-order}
\end{equation}
ensures quantum soundness at most $1/2+2^{-k}$.
\end{proposition}

The third application is the homomorphic-commitment RZKP for the $3\mathrm{SAT}$ problem proposed in Ref.~\cite{crepeau_2023}. For a formula containing $m$ clauses, its soundness game $G_{\rm 3SAT}$ is $3^m$-projective, and its protocol-specific no-signaling argument gives $\omega^*(G_{{\rm 3SAT},{\rm cp}})\leq\frac1q$. Substituting $S=3^m$ and $\eta=1/q$ into Proposition~\ref{prop:s-projective-coupled} gives the following result.

\begin{proposition}[Quantum soundness of the $3\mathrm{SAT}$ RZKP]
\label{prop:rzko-3sat}
For an unsatisfiable $3$-CNF formula containing $m$ clauses and an admissible field order $q\geq3^m$, the acceptance probability of entangled cheating provers satisfies
\begin{equation}
    \omega^*(G_{\rm 3SAT}) \leq \frac{1}{6} \left[ 2+\mathcal{C}\!\left(\frac{54\cdot3^m}{q}-2\right) \right].
    \label{eq:3sat-exact-soundness}
\end{equation}
Consequently, for $k\geq1$, any admissible prime-power field order satisfying
\begin{equation}
    q\geq \frac{3^m2^{2k-1}}{1+2^{1-k}}
    \label{eq:3sat-exact-field-order}
\end{equation}
ensures quantum soundness at most $1/2+2^{-k}$.
\end{proposition}

Compared with the original analyses of Refs.~\cite{chailloux2017relativistic,crepeau_2023}, this reduces the exponent in the required field order $q$ and thus the coefficient in the communication randomness cost $\lceil\log_2 q\rceil$ of these protocols. The exact analytic bound provides an additional finite-parameter improvement over the simplified bounds of Ref.~\cite{shi2024relativistic}, although it does not change the asymptotic $q=\mathcal{O}(S2^{2k})$ scaling.

\section{Robust CMT in fidelity}

In addition to the tight CMT, we introduce, for the first time, a fidelity-based robust CMT in which measurements are performed on two slightly different quantum states.

In MQC, we rarely assume adversaries operate on ideal states following honest strategies. A malicious party might easily deviate from a protocol by introducing a small error, using a noisy state, or performing operations that slightly corrupt the initial quantum system. Consequently, consecutive measurements in a cheating strategy may essentially be applied to two slightly different states rather than a single ideal one. To establish rigorous security proofs in these adversarial scenarios, we must bound the adversary's information gain even when the final states contain such errors. 
We employ the robust CMT in fidelity when our security analysis relies on purification techniques, as fidelity is exceptionally well-behaved under purifications due to Uhlmann's theorem \cite{Watrous_2018}. 
Conversely, we utilize the trace distance formulation (presented in Section~\ref{sec:TRCMT}) in contexts where the physical distinguishing probability between states is the primary operational metric of interest.

\begin{theorem}[Robust CMT in fidelity]
\label{thm:RCMT_F}
    Given two projections $P_0$ and $P_1$, and two quantum states $\sigma_0$ and $\sigma_1$ in some finite-dimensional Hilbert space $\H$, define 
    \begin{equation}\label{eqn:VEdelta}
    \begin{aligned}
        V & = \frac{1}{2}\br{\Tr(P_0 \sigma_0) +  \Tr(P_1 \sigma_1)}, \\
        E & = \frac{1}{2}\br{\Tr(P_1P_0 \sigma_0 P_0 P_1) + \Tr(P_0P_1 \sigma_1 P_1 P_0)},
    \end{aligned}  
    \end{equation}
    Let $F=\|\sqrt{\sigma_0}\sqrt{\sigma_1}\|_1$ denote the fidelity. 
    Then 
    \begin{equation}
        \label{eqn:rcmt_fidelity}
        \begin{aligned}
            E \geq  V \left(\max\left\{0,(2V-1)F - 2\sqrt{V(1-V)}\sqrt{1-F^2}\right\}\right)^2.
        \end{aligned} 
    \end{equation}
    Furthermore, for any $v,f\in[0,1]$, there exist states $\sigma_0$, $\sigma_1$ and projectors $P_0$, $P_1$ such that $F=f$, $V=v$ and Eq.~\eqref{eqn:rcmt_fidelity} holds with equality. 
\end{theorem}

When $F=1$, our robust CMT reduces to the tight CMT for $n=2$. The proof, deferred to Appendix~\ref{sec:FRCMT}, uses the Bures angle and several trigonometric arguments.

\section{Application of Robust CMT: oblivious transfer and reductions}

Oblivious transfer~\cite{Rabin_1981,Even_1985,Crepeau_1988} is a cornerstone of quantum cryptography which serves as the fundamental building block of secure multi-party computation~\cite{Yao_1986,Goldreich_1988,Killian_1988}. The goal is three-fold: Alice transfers $x_b$ out of two data bits $(x_0,x_1)$ upon Bob's request $b$; Alice remains oblivious to $b$; Bob remains oblivious to $x_{\overline{b}}$. Quantum oblivious transfer was later investigated with the aim of achieving perfect security~\cite{Crepeau_1994}. While fundamental principles prohibit perfect QOT~\cite{Lo_1997}, quantum resources still strictly enhance its security, prompting extensive recent efforts to establish rigorous no-go theorems that bound the achievable security parameters of imperfect QOT~\cite{Chailloux_2013,Chailloux_2016,Amiri_2021,Hu_2023,Stroh_2023}. Our robust CMT provides a natural toolkit for analyzing cheating strategies in QOT, where an adversary might perform consecutive measurements to illegally extract both bits. By analyzing the success probability of this sequential cheating strategy, our approach establishes a tighter no-go theorem for QOT security parameters in certain regimes. 

We first introduce the definition and the properties of oblivious transfer:

\begin{definition}[Oblivious transfer~{\cite[Definition 1]{Amiri_2021}}]
    \label{def:qot}
    An oblivious transfer protocol with $\delta$-completeness, $P_A^\star$-soundness against cheating Alice and $P_B^\star$-soundness against cheating Bob is a two-party protocol where Alice begins with $(x_0,x_1)\in \{0,1\}^2$ and Bob begins with $b\in \{0,1\}$, and Alice ends with ${\rm output}_A\in\{{\Accept},{\Abort}\}$, and Bob ends with ${\rm output}_B \in\{0,1\}\cup\{\Abort\}$. If Bob does not output $\Abort$, we say that Bob accepts and outputs $\hat{x}\in\{0,1\}$.
    
    \vspace{2mm}$\mathrm{OT}:$
    \vspace{-2mm}\begin{align}
        \begin{array}{ccc}
             {\rm Alice} \times {\rm Bob} & \rightarrow & {\rm Alice} \times {\rm Bob}  \\
             \{0,1\}^2 \times \{0,1\} & \rightarrow & \{{\Accept},{\Abort}\}  \times  \{0,1,\Abort\}\\
             (x_0,x_1)\times b  & \mapsto & {\rm output}_A \times  {\rm output}_B. 
        \end{array} 
    \end{align}
\end{definition}

\begin{definition}[Completeness]
    If Alice and Bob are both honest, then both parties accept, and $\hat{x}=x_{b}$ with probability at least $1-\delta$.
    \begin{align}
        \Pr[{\rm Both\ accept}] =1,\quad {\rm and} \quad \Pr[\hat{x}=x_{b}] \geq 1-\delta. 
    \end{align}
    The above equations should hold for any choice $(x_0,x_1)$. 
\end{definition}

\begin{definition}[Soundness against a cheating Bob]
Let Alice's $(x_0,x_1)$ be uniformly random. With probability at most $\frac{1}{2}+\epsilon_B$, a cheating Bob can guess $(\hat{x}_0,\hat{x}_1)$ for an honest Alice's $(x_0,x_1)$ correctly and Alice accepts, 
\begin{align}
    \Pr[((\hat{x}_0,\hat{x}_1)=(x_0,x_1)) \land ({\rm Alice\ accepts}) ] \leq  \frac{1}{2} + \epsilon_B . 
\end{align}
\end{definition}

\begin{definition}[Soundness against a cheating Alice]
Suppose that Bob's $b$ is uniformly random. With probability at most $\frac{1}{2}+\epsilon_A$, a cheating Alice can guess $\hat{b}$ for an honest Bob's $b$ correctly and Bob accepts, 
\begin{align}
    \Pr[(\hat{b}=b)\land({\rm Bob\ accepts})] \leq \frac{1}{2} + \epsilon_A. 
\end{align} 
\end{definition}

\begin{figure}[t]
    \centering
       \includegraphics[width=1\linewidth]{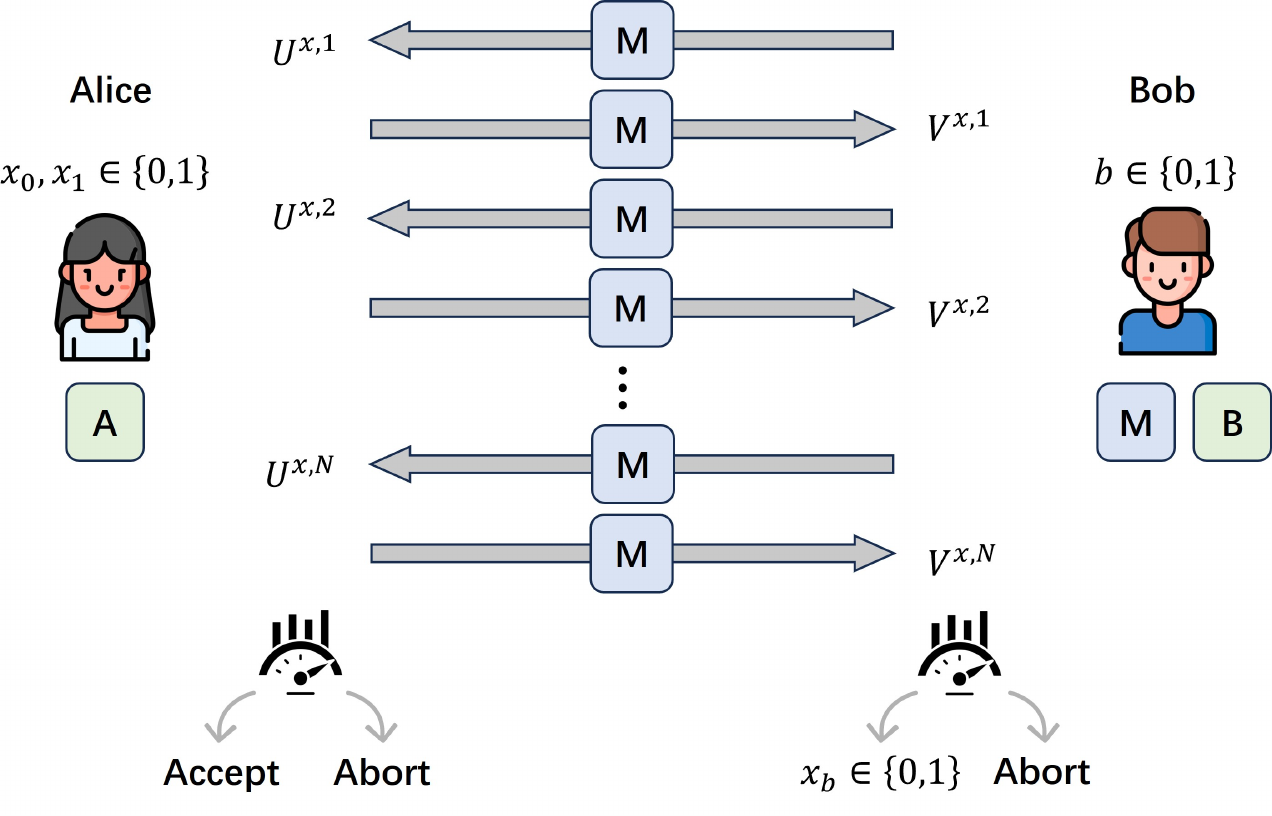}
    \caption{\textbf{A general protocol for QOT.} Alice and Bob exchange quantum messages with each other to transfer $x_b$ out of two data bits $(x_0,x_1)$ to Bob upon Bob's request $b$. Alice remains oblivious to $b$ and Bob remains oblivious to $x_{\overline{b}}$. }
    \label{fig:qot}
\end{figure}

\begin{algorithm}[H]
    \caption{QOT with $N$ rounds of communication}
    \label{prot:qot}
    \begin{algorithmic}[1]
        \Input Alice: $(x_0,x_1)\in\lbrace 0,1\rbrace ^2$; Bob: $b\in\lbrace 0,1\rbrace $.
        \Output $\textnormal{output}_A\in\lbrace \Accept,\Abort\rbrace $, $\textnormal{output}_{B}\in\lbrace 0,1\rbrace \cup\lbrace \Abort\rbrace$. 
        \State Alice: Prepare the state $\ket{0}$ in her system $A$.
        \State Bob: Prepare the pure state $\ket{\rho^b}_{BM}$ in his system $B$ and the message system $M$.  
        \For{$\ell = 1$ to $\ell = N$} 
            \State Bob: Send the system $M$ to Alice. 
            \State Alice: Perform a unitary $U^{x_0x_1,\ell}_{MA}$ to the composite system $MA$ according to her input $(x_0,x_1)$.
            \State Alice: Send $M$ back to Bob. 
            \State Bob: Perform a unitary $V^{b,\ell}_{BM}$ to the composite system $BM$ according to his input $b$. 
        \EndFor
        \State Alice: Perform a PVM ${P_{\mathrm{Acc}}^{x_0x_1}, P_{\mathrm{Abt}}^{x_0x_1}}$, outputting Accept if the outcome corresponds to $P_{\mathrm{Acc}}^{x_0x_1}$, and Abort otherwise.
        \State Bob: Perform a PVM ${Q_{\mathrm{Acc}}^b, Q_{\mathrm{Abt}}^b}$, outputting Abort if the outcome corresponds to $Q_{\mathrm{Abt}}^b$.
        \State Bob: Perform a PVM ${Q_0^{b}, Q_1^{b}}$, outputting $0$ if the outcome corresponds to $Q_0^{b}$, and $1$ otherwise.
     \end{algorithmic}
\end{algorithm}

Any oblivious transfer with $N$ rounds of communication can be described by a general scheme described in Protocol~\ref{prot:qot} and illustrated in Figure~\ref{fig:qot}. Alice and Bob keep their memories private and exchange their messages publicly. Each time either Alice or Bob receives the message, they apply a unitary that acts jointly on the memory and the message. In the last step, they measure their state to obtain their output.

In QOT, two parties implement OT through noiseless quantum communication channels to transmit quantum states between them. The tightest known no-go theorem for QOT before our work is $\epsilon_B+2\epsilon_A + 4\sqrt{\delta}\geq \frac{1}{2}$. Cases $\delta=0$ and $\delta>0$ were explored in~\cite{Chailloux_2016,Amiri_2021} and~\cite{Hu_2023}, respectively. Upper bounds are better studied for $\delta=0$ than for $\delta>0$. For $\delta=0$, a protocol achieves this bound when cheating Bob is required to output $x_b$ with certainty in~\cite{Chailloux_2013}. Without assumptions on adversary, the best protocol achieves $\max\{\epsilon_A,\epsilon_B\}=0.240$ at $\delta=0$~\cite{Chailloux_2013,Amiri_2021}.

It is natural for Bob to perform the following honest-but-curious cheating strategy in QOT: after finishing all rounds of communication honestly, Bob performs a measurement to learn $x_i$ followed by another measurement to try to learn $x_{\overline{i}}$. QOT requires that the information gained in the first measurement is large to satisfy completeness, while that in both measurements is small to enforce soundness against cheating Bob. However, this requirement violates the CMT, which establishes an impossibility result for QOT. Because a possible cheating strategy is sufficient to prove an impossibility result, it is sufficient to assume an honest-but-curious cheating strategy for both parties. Using our robust CMT Theorem \ref{thm:RCMT_F}, we obtain a new no-go theorem:
\begin{theorem}[No-go theorem for QOT]
    \label{thm:bound_RCMT}
    The security parameters $\epsilon_A$, $\epsilon_B$ and $\delta\leq \frac{1}{2}$ for QOT must satisfy    
    \begin{multline}
         \frac{1}{2} + \epsilon_B \geq (1-\delta)\cdot \\\left( \max
          \left\lbrace 0,(1-2\delta)(1-2\epsilon_A)-4\sqrt{\delta(1-\delta)\epsilon_A(1-\epsilon_A)} \right\rbrace\right)^2.
    \end{multline}
\end{theorem}
The proof of Theorem~\ref{thm:bound_RCMT} can be found in Appendix~\ref{apd:ot}. 

As shown in Fig.~\ref{fig:tradeoff}, the tradeoff bound in Theorem~\ref{thm:bound_RCMT} is more fault-tolerant, in the sense that it remains meaningful even for relatively large values of $\delta$. Furthermore, Theorem~\ref{thm:bound_RCMT} improves upon the previously tightest known no-go theorem $\epsilon_B + 2\epsilon_A + 4\sqrt{\delta} \geq \frac{1}{2}$~\cite[Theorem 9]{Hu_2023} when $\delta$ is not too small. At $\epsilon_A=0$, our QOT bound remains nontrivial for $\delta<0.1226$, whereas the previous bound becomes trivial for $\delta\geq1/64$. 

\begin{figure}[t]
    \centering    
    \includegraphics[width=1\linewidth]{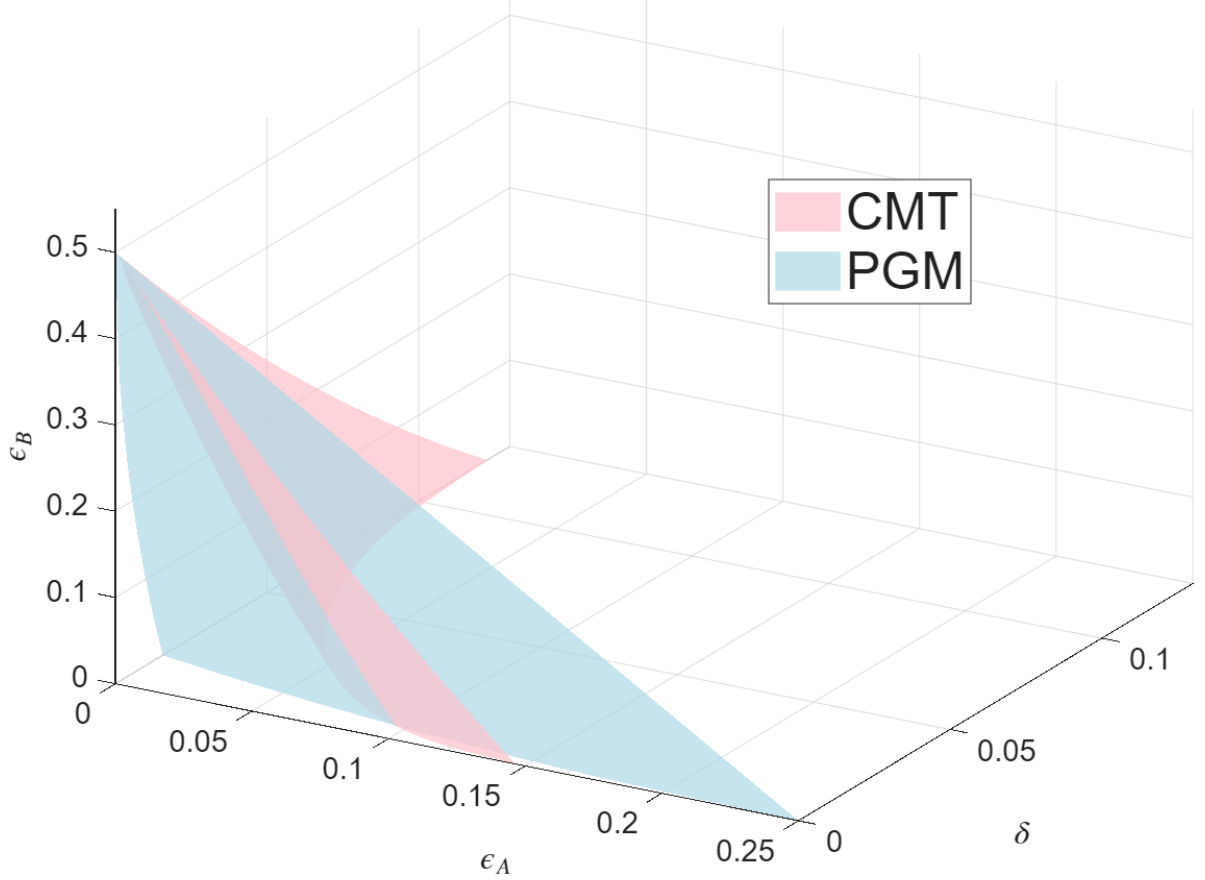}
    \caption{\textbf{Comparison of different no-go theorems for QOT.} The pink surface is the no-go theorem for QOT in Theorem~\ref{thm:bound_RCMT} derived from consecutive measurement (CMT) while the blue surface is the theorem in Refs.~\cite{Amiri_2021,Hu_2023} derived from pretty good measurement (PGM).} 
    \label{fig:tradeoff}
\end{figure}

Cryptographic reductions allow security constraints for one primitive to be directly mapped to another. Our QOT bound can be translated into bounds for various cryptographic primitives via reductions. As a concrete example, we apply our improved QOT bound to QHE, the Swiss army knife of quantum cryptography, to establish a tighter QHE bound. 

A QHE protocol is a secure two-party computation protocol in which Bob asks Alice to evaluate a channel in one round of communication. The protocol is specified by the set $\mathscr{F}$ of channels that Bob wants to delegate, Bob's key generation map $\KeyGen$, Bob's encryption map $\Enc_k:\text{Plain-state}\to\text{Cipher-state}$ with key $k$, Alice's evaluation map $\hat{\mathcal{F}}:\text{Cipher-state}\to \text{Cipher-state}$ for $\mathcal{F}\in\mathscr{F}$ and Bob's decryption map $\Dec_k:\text{Cipher-state}\to \text{Plain-state}$ with key $k$.

A QHE encryption protocol is characterized by the following three properties~\cite{Hu_2023}, i.e. correctness, data privacy, and circuit privacy. 

\begin{definition}[Correctness]\label{def:qhe_correctness}
    A QHE protocol is $\epsilon$-correct if, for every channel $\mathcal{F}\in \mathscr{F}$, every input of Bob and its purification $\ket{\psi'}$ and every key $k$, 
    \begin{equation}
        \frac{1}{2}\left\|( \Dec_k \hat{\mathcal{F}} \Enc_k\otimes \mathcal{I})[\proj{\psi'}] - (\mathcal{F}\otimes\mathcal{I})(\proj{\psi'})\right\|_1 \leq \epsilon. 
    \end{equation}
\end{definition}

\begin{definition}[Data privacy]\label{def:qhe_data_privacy}
    A QHE protocol is $\epsilon_d$-data private if for any two inputs of Bob $\rho,\rho'$, it holds that
    \begin{equation}
        \frac{1}{2}\|\mathbbm{E}_k(\Enc_k[\rho])-\mathbbm{E}_k(\Enc_k[\rho'])\|_1\leq \epsilon_d, 
    \end{equation}
\end{definition}

\begin{definition}[Circuit privacy]\label{def:qhe_circuit_privacy}
    A QHE protocol is $\epsilon_c$-circuit private if for any purification of Bob's message $\ket{\psi}$, there exists a purification of Bob's input $\ket{\psi'}$ and a post-processing channel $\mathcal{N}$ such that for any $\mathcal{F}\in \mathscr{F}$
    \begin{align}\label{eqn:circuit_privacy}
        \frac{1}{2}\left\|(\hat{\mathcal{F}}\otimes \mathcal{I})[\proj{\psi}] - \mathcal{N}\left((\mathcal{F}\otimes \mathcal{I})(\proj{\psi'})\right) \right\|_1 \leq \epsilon_c, 
    \end{align}
\end{definition}

The bound for QOT in Theorem~\ref{thm:bound_RCMT} can be transferred into a bound for QHE in Corollary~\ref{thm:bound_QHE} using~\cite[Theorem 21]{Hu_2023} as was done in~\cite[Corollary 22]{Hu_2023}: 

\begin{corollary}[No-go theorem for QHE]
\label{thm:bound_QHE}
    The security parameters $\epsilon_c$, $\epsilon_d$ and $\delta$ for QHE must satisfy
    \begin{multline}
        \frac{1}{2} + \epsilon_c \geq (1-\delta)\cdot \\
        \left(\max\left\lbrace0,(1-2\delta)(1-\epsilon_d)-2\sqrt{\delta(1-\delta)\epsilon_d(2-\epsilon_d)} \right\rbrace\right)^2.
    \end{multline}
\end{corollary}

\section{Robust CMT in trace distance}
\label{sec:TRCMT}

In certain scenarios of MQC, more than two binary projective measurements may arise (for example, in QPQ, which we discuss later). In such cases, Theorem~\ref{thm:RCMT_F} is no longer applicable. We require the following result, which establishes a robust CMT for the setting with $n$ possible binary projective measurements.

\begin{theorem}\label{thm:RCMT_TD}
    Let $n\geq 2$ be an integer, let $P_1,\dots,P_n$ be $n$ projectors and $\sigma_1,\dots,\sigma_n$ be $n$ quantum states in a finite-dimensional Hilbert space $\H$. Let 
\begin{equation}
\begin{aligned}
    V & = \frac{1}{n} \sum_{i=1}^n\Tr(P_i\sigma_i),  \\
    E & = \frac{1}{n(n-1)} \sum_{i\neq j} \Tr(P_j P_i \sigma_i P_i P_j) ,\\
    \Delta&=\frac{1}{n(n-1)}\sum_{i<j}\norm{\sigma_i-\sigma_j}_1.
\end{aligned}
\end{equation}
It holds that
\begin{equation}\label{eqn:rcmt_trace_distance}
    E\geq 4 V \left(\max\left\{0,V-\frac{1}{2}\right\}\right)^2-\Delta,
\end{equation}
 as well as 
\begin{equation}\label{eqn:rcmt_trace_tight}
    E\geq  \frac{n^2}{(n-1)^2}V \left(\max\left\{0,V-\frac{1}{n}\right\}\right)^2-\frac{4n}{n-1}\Delta.
\end{equation}
Each of these two bounds is stronger than the other in a different regime. 

\end{theorem}

To prove the first inequality in this theorem ([Part 1]), we first prove the case $n=2$. We then extend the argument to general $n$ using convexity. For the basic case $n=2$, we first establish a reduction rule: if $\tilde{\sigma}_0,\tilde{\sigma}_1$ produce the same $E$ and $V$ but a smaller $\Delta$ than $\sigma_0,\sigma_1$, and if Eq.~\eqref{eqn:rcmt_trace_distance} holds for $\tilde{\sigma}_0,\tilde{\sigma}_1$, then it also holds for $\sigma_0,\sigma_1$. Using this reduction rule, we first use Jordan's lemma to block diagonalize $P_0,P_1$ and pinch $\sigma_0,\sigma_1$ with respect to these blocks. After that, we prove the theorem on each block and use the convexity to combine those blocks. On each block, geometric arguments in Bloch representation are used to prove the theorem by constructing symmetric states with respect to the angle bisection of projectors from reduced states following the reduction rule.  

\section{Application of Robust CMT: Quantum Private Query}
Quantum private query is a cryptographic primitive for private information retrieval, first proposed in~\cite{Giovannetti_2008}. As a generalization of QOT, QPQ allows Bob to retrieve an entry $x_b$ from Alice's database $x\in[k]^n$ while satisfying two privacy constraints: Alice learns nothing about Bob's request $b$, and Bob learns almost nothing about the rest of the database. 

While fundamental no-go theorems determine that perfectly secure QPQ is impossible~\cite{Lo_1997}, quantum features such as the no-cloning theorem and measurement disturbance provide practical security advantages that have no classical analogue in private access to remote databases. Specifically, these properties enable cheat-sensitive protocols, where any attempt by Alice to illegally extract information about Bob's query induces a detectable disturbance~\cite{Giovannetti_2008,jakobi2011practical,olejnik2011secure}. This interplay between quantum capabilities and fundamental limits has made QPQ a prominent setting to study secure remote database access~\cite{gao2012flexible}.

\begin{figure}[t]
    \centering    
    \includegraphics[width=0.95\linewidth]{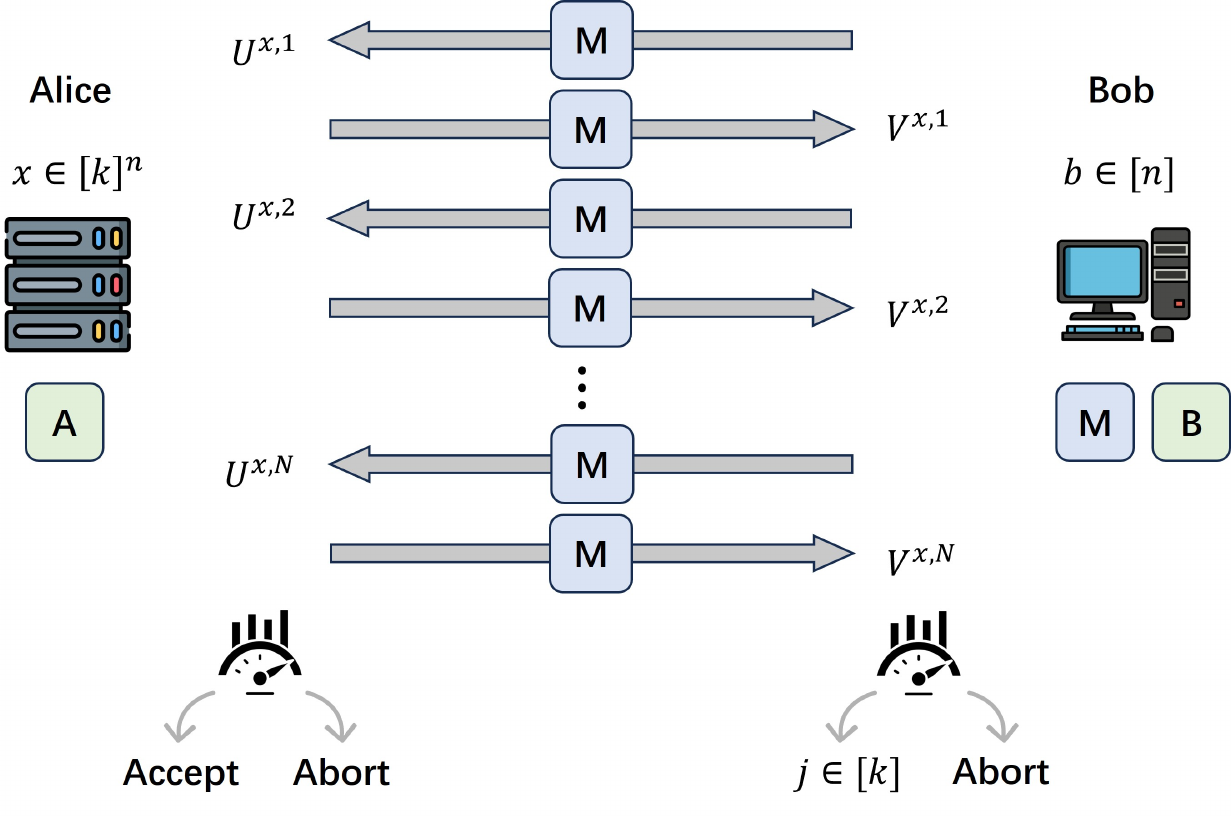}
    \caption{{\bf A general protocol for QPQ.} Bob communicates quantum messages with Alice to retrieve an entry $x_b$ from Alice's database $x\in[k]^n$. Alice remains oblivious to $b$ and Bob remains oblivious to other entries except $x_b.$}
    \label{fig:qpq}
\end{figure}

A general protocol for QPQ is presented in Protocol~\ref{prot:qpq}, and illustrated in Figure~\ref{fig:qpq}.

In an ideal QPQ protocol, Bob outputs $j=x_b$ with high probability. Furthermore, it should be secure for both Alice and Bob. 
We use the formal definition in \cite{Hanggi_2025}.

\begin{algorithm}[H]
    \caption{QPQ with $N$ rounds of communication}
    \label{prot:qpq}
    \begin{algorithmic}[1]
        \Input Alice: $x\in[k] ^n$; Bob: $b\in[n] $.
        \Output $\textnormal{output}_A\in\lbrace \Accept,\Abort\rbrace $, $\textnormal{output}_{B}\in[k] \cup\lbrace \Abort\rbrace$. 
        \State Alice: Prepare the state $\ket{0}$ in her system $A$.
        \State Bob: Prepare the pure state $\ket{\rho^b}_{BM}$ in his system $B$ and the message system $M$.  
        \For{$\ell = 1$ to $\ell = N$} 
            \State Bob: Send the system $M$ to Alice. 
            \State Alice: Perform a unitary $U^{x,\ell}_{MA}$ to the composite system $MA$ according to her input $x$.
            \State Alice: Send $M$ back to Bob. 
            \State Bob: Perform a unitary $V^{b,\ell}_{BM}$ to the composite system $BM$ according to his input $b$. 
        \EndFor
        \State Alice: Perform a PVM ${P_{\mathrm{Acc}}^{x}, P_{\mathrm{Abt}}^{x}}$, outputting Accept if the outcome corresponds to $P_{\mathrm{Acc}}^{x}$, and Abort otherwise.
        \State Bob: Perform a PVM ${Q_{\mathrm{Acc}}^b, Q_{\mathrm{Abt}}^b}$, outputting Abort if the outcome corresponds to $Q_{\mathrm{Abt}}^b$.
        \State Bob: Perform a PVM $\{Q_j^{b}\}_{j\in[k]}$, outputting $j$ if the outcome corresponds to $Q_j^{b}$.
     \end{algorithmic}
\end{algorithm}

\begin{definition}[Correctness]
    A QPQ protocol is $\delta$-correct if in any execution of the protocol where both players are honest
    \begin{align}
        \Pr[{\rm Both\ accept}\wedge j=x_{b}] \geq 1-\delta. 
    \end{align}
    The above equation should hold for any choice $x\in[k]^n$.
\end{definition}

By the principle of deferred measurement, we assume that all measurements are deferred until the end of the protocol. Let $\rho_A^b$ denote the quantum state held by Alice immediately before measurements where  Bob's query is $b$. 

\begin{definition}[Weak User Privacy]
 A QPQ protocol is weakly $(\epsilon_A,\delta_A)$-secure for Bob if for any strategy of a dishonest Alice under which Bob accepts with probability at least $1-\delta_A$, and any queries $b,b'\in[n]$,
 \[\frac{1}{2}\norm{\rho_A^b-\rho_A^{b'}}_1\leq2\epsilon_A.\]
\end{definition}

\begin{definition}[Weak Data Privacy]
A QPQ protocol is weakly $(\epsilon_B,\delta_B)$-secure for Alice if for any strategy of a dishonest Bob under which Alice accepts with probability at least $1-\delta_B$, 
\begin{multline}
    \Pr[\text{Bob guesses }x_b\text{ and }x_{b'}\text{ correctly for some }b\ne b'] \\\leq \frac{1}{k} + \epsilon_B. 
\end{multline} 
\end{definition}

\begin{figure}[t]
   \centering
   \begin{overpic}[width=0.85\columnwidth]{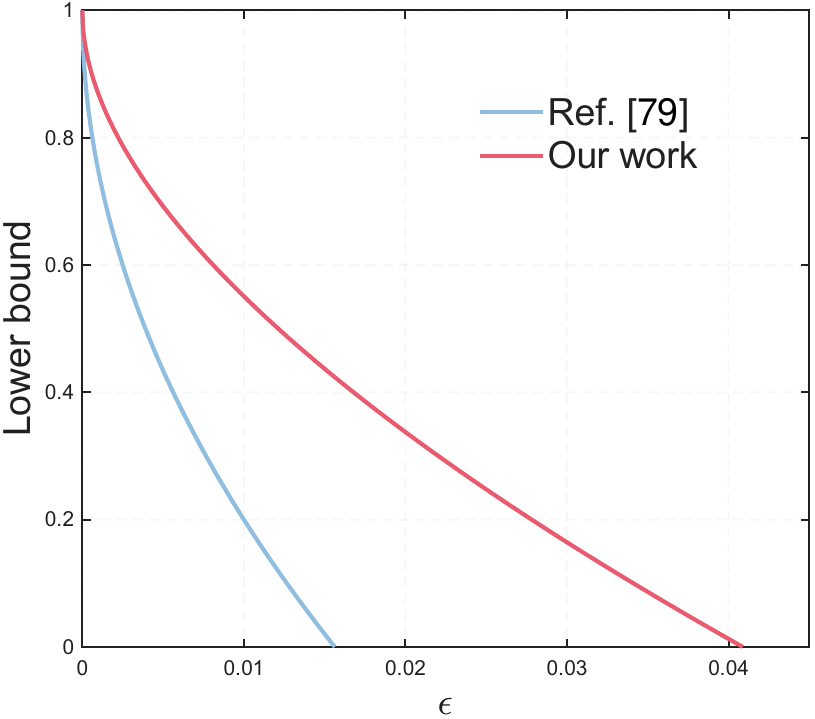}
    \put(67,65){\color{white}\rule{40pt}{25pt}}
    \put(67.5,73.5){\small Ref.~\cite{Hanggi_2025}}
    \put(67.5,67.5){\small Our work}
    \end{overpic}
   \caption{\textbf{Comparison of the bound in Theorem~\ref{thm:QPQ} (red) with that of~\cite[Theorem 1]{Hanggi_2025} (blue) when $\delta=\epsilon=\epsilon_A$.} The horizontal axis is $\epsilon$, and the vertical axis is the lower bound for the probability that dishonest Bob can retrieve two database entries.} 
   \label{fig:qpq_compare}
\end{figure}

Unfortunately, H\"{a}nggi and Winkler \cite{Hanggi_2025} proved the impossibility of unconditionally secure QPQ. They showed a generic post-processing attack: any protocol that is $\varepsilon$-correct and $(\epsilon,\delta)$-secure for the user allows a dishonest Bob to retrieve $m$ database entries with probability at least $1-2m^2\sqrt{\varepsilon}$, for all $2\le m\le n$. In particular, for $m=2$ this implies that Bob can obtain two entries with probability at least $1-8\sqrt{\epsilon}$. 
 
In contrast to the setting where correctness and user-privacy errors are tied to the same $\epsilon$, we allow the correctness parameter $\delta$ and the weak user-privacy parameter $\epsilon_A$ to be independent, and we derive a lower bound on the probability that Bob can retrieve two database entries.

\begin{theorem}[No-go theorem for QPQ]\label{thm:QPQ}
    For any QPQ protocol which is $\delta$-correct, weakly $(\epsilon_A,\delta)$-secure for Bob (where $\delta\leq \frac{1}{2}$), dishonest Bob can retrieve two database entries with probability at least
    \begin{equation}\label{eqn:QPQrobust}
        4(1-\delta)\left(\max\left\lbrace 0,\frac{1}{2}-\delta \right\rbrace\right)^2-4\sqrt{\epsilon_A}. 
    \end{equation}
    In particular, if $n$ is large and $\epsilon_A$ is sufficiently small, we can obtain an even tighter bound
    \begin{equation}\label{eqn:QPQtight}
         \frac{n^2}{(n-1)^2}(1-\delta)\left(\max\left\lbrace0,1-\frac{1}{n}-\delta \right\rbrace\right)^2-\frac{16n}{n-1}\sqrt{\epsilon_A}. 
    \end{equation}
\end{theorem}

The proof is deferred to Appendix~\ref{sec:QPQ}. To compare our result with \cite[Theorem 1]{Hanggi_2025}, we set $\delta=\epsilon=\epsilon_A$. As shown in Figure~\ref{fig:qpq_compare}, our bound is strictly tighter than that in \cite{Hanggi_2025}.

A notable feature of our robust CMTs is the different bounding behavior between the fidelity-based and trace-distance-based bounds. This analytical divergence arises because fidelity admits a mathematically simpler pure-state representation via Uhlmann's theorem, whereas trace distance lacks such a direct pure-state equivalence for arbitrary mixed states.

\section{Discussion}

In this work, we establish a tight consecutive measurement theorem and introduce two robust variants formulated in terms of fidelity and trace distance. To the best of our knowledge, these are the first robust CMTs that accommodate measurements performed on different, but nearby, quantum states.

We apply these results to several primitives in mistrustful quantum cryptography, using CMTs in two complementary directions. In coupled-game analyses, an upper bound on the consecutive-measurement success probability $E$, derived from the no-signaling principle, can be combined with our tight CMT to upper-bound the single-measurement success probability $V$. This yields tighter bounds for nonlocal $\mathrm{CHSH}_q(p)$ games and their parallel repetitions, thereby reducing the communication resources required by the corresponding RBC protocols. The same approach also improves the quantum soundness analyses of RZKP protocols. Conversely, in QOT and QPQ, protocol correctness provides a lower bound on $V$, which our robust CMTs convert into a lower bound on the success probability $E$ of an explicit sequential cheating strategy, leading to stronger impossibility results. To the best of our knowledge, this is the first use of CMTs to derive no-go theorems for mistrustful quantum cryptography.  These applications illustrate how CMTs can be used in opposite directions, although relating a cryptographic security experiment to the CMT variables remains
protocol dependent.

The applicability of our CMTs requires a protocol-specific reduction from the relevant security experiment to single- and consecutive-measurement success probabilities. Our theorems apply to ordered projective measurements on a common state, or on nearby states satisfying the closeness assumptions of the robust CMTs.  

Neither application assumes that sequential strategies are generally optimal. These are sufficient patterns rather than a characterization of mistrustful quantum cryptography, and therefore do not directly cover settings such as quantum coin flipping and the semidefinite-programming methods used to analyze it~\cite{Spekkens_2002,Kitaev_2003,Ambainis_2004,Mochon_2007,Aharonov_2016,Bozzio_2020}, measurement-based blind quantum computing~\cite{Broadbent_2009,Barz_2012,Fitzsimons_2017}, splitting-channel BB84-state RBC~\cite{Lunghi2013Experimental}, or cloning-channel quantum money counterfeiting~\cite{Molina_2013}. Moreover, the coupled-game route is restricted to projective nonlocal games with uniform inputs and finite classical alphabets, and yields a useful bound only when an independent nontrivial upper bound on $\omega^\ast(G_{\mathrm{cp}})$ is available.

Currently, we do not know which other MQC protocols can be analyzed using CMTs, and this remains an interesting open question. Their applicability depends on both the structure of the protocol and its specific security requirement. For each protocol, one must construct a suitable sequential cheating strategy, identify the measurements corresponding to the relevant cheating objectives, and relate their single and consecutive success probabilities to the cryptographic security parameters. Developing systematic criteria for identifying CMT-compatible security experiments, as well as new CMT formulations adapted to other protocol structures, is an important direction for future research.

Finally, some cryptographic tasks, such as $m$-out-of-$n$ oblivious transfer~\cite{Naor_1999,Naor_2005,Chailloux_2013}, device-independent randomness certification~\cite{Bowles_2020,Senno_2023,Padovan_2024} and three-challenge relativistic zero-knowledge proofs~\cite{chailloux2021relativistic}, naturally involve more than two consecutive measurements. Extending the present results to longer and adaptive measurement sequences is therefore another promising direction. More broadly, determining which security experiments in other mistrustful primitives can be converted into suitable consecutive-measurement scenarios will help clarify both the reach and the boundaries of the CMT toolkit.

\section*{Acknowledgments}

C.-X.W., MQ and YH contributed equally to this work.

YH and MT are supported by the National Research Foundation, Singapore and A*STAR under its Quantum Engineering Programme (NRF2021-QEP2-01-P06). MQ, YH and MT are also supported by the National Research Foundation, Singapore through the National Quantum Office, hosted in A*STAR, under its Centre for Quantum Technologies Funding Initiative (S24Q2d0009). C.-X.W. appreciates the hospitality of the Centre for Quantum Technologies and the financial support from the China Scholarship Council (No. 202406190220).

The images for Alice and Bob in Fig.~\ref{fig:RBC} and~\ref{fig:qot} are \href{https://www.flaticon.com/free-icons/boy}{boy} and \href{https://www.flaticon.com/free-icons/girl}{girl} icons created by Freepik - Flaticon. 

\appendix

\section{Proof of tight CMT}
\label{apd:TCMT}

\begin{reptheorem}{thm:consec_mmt}
Let $n\geq 2$ be an integer, let $P_1,\ldots,P_n$ be $n$ projectors in a finite-dimensional Hilbert space $\H$, let $\sigma$ be any quantum state in $\H$, and let
\begin{equation}
\begin{aligned}
    V & = \frac{1}{n} \sum_{i=1}^n \Tr(P_i\sigma),  \\
    E & = \frac{1}{n(n-1)} \sum_{i\neq j} \Tr(P_j P_i \sigma P_i P_j) .
\end{aligned}
\end{equation}
It holds that
\begin{align}\tag{\ref{eqn:CMT_inequality}}
    E\geq  \frac{n^2}{(n-1)^2 }V \left(\max\left\{0,V-\frac{1}{n}\right\}\right)^2.
\end{align}
Moreover, for any $n\geq 2$ and $v\in[0,1]$, there exists a state $\sigma$ and projectors $P_1,\ldots,P_n$ in some finite-dimensional Hilbert space $\H$, such that $V=v$ and Eq.~\eqref{eqn:CMT_inequality} holds with equality. 
\end{reptheorem}

To prove Theorem~\ref{thm:consec_mmt}, we require the following lemma, which expresses $E$ and $V$ as inner products of pure quantum states. 

\begin{lemma}[{\cite[Lemma 3]{chailloux2017relativistic}}]
\label{lem:VEvec}
Given $n$ projectors $P_1,\ldots,P_n$ on some Hilbert space $\H$. Let $\sigma$ be any quantum state on $\H$, and $V$ and $E$ be defined as in Theorem~\ref{thm:consec_mmt}. 
Consider a purification $\ket{\phi}$ of $\sigma$ in some space $\H\otimes\E$. For $i\in[n]$, define
\begin{equation}
    \label{eqn:phiis}
    \ket{\phi_i}:=\frac{(P_i\otimes\id_\E)\ket{\phi}}{\norm{(P_i\otimes\id_\E)\ket{\phi}}}.
\end{equation}
Then
\begin{equation}
\begin{aligned}
    V & =\frac{1}{n}\sum_{i=1}^n\abs{\braket{\phi}{\phi_i}}^2, \\
    E & \geq\frac{1}{n(n-1)}\sum_{i,j\neq i}\abs{\braket{\phi}{\phi_i}}^2\abs{\braket{\phi_i}{\phi_j}}^2.
\end{aligned}  
\end{equation}
\end{lemma}

\begin{proof}[Proof of Theorem \ref{thm:consec_mmt}]
    We will first prove the lower bound. 
    
    By Lemma~\ref{lem:VEvec}, we have
    \begin{equation}
    \begin{aligned}
        E &  \geq V\sum_{i\neq j}\frac{\abs{\braket{\phi}{\phi_i}}^2+\abs{\braket{\phi}{\phi_j}}^2}{2n(n-1)V}\cdot\abs{\braket{\phi_i}{\phi_j}}^2 \\
        & \overset{(a)}{\geq} V\br{\sum_{i\neq j}\frac{\abs{\braket{\phi}{\phi_i}}^2+\abs{\braket{\phi}{\phi_j}}^2}{2n(n-1)V}\cdot\abs{\braket{\phi_i}{\phi_j}}}^2\\
        & \overset{(b)}{\geq}\frac{V}{(n-1)^2}\br{\sum_{i\neq j}\frac{\abs{\braket{\phi_j}{\phi}\!\braket{\phi}{\phi_i}\!\braket{\phi_i}{\phi_j}}}{nV}}^2\\
        & \overset{(c)}{=}\frac{V}{(n-1)^2}\br{\sum_{i,j=1}^n\frac{\abs{\braket{\phi_j}{\phi}\!\braket{\phi}{\phi_i}\!\braket{\phi_i}{\phi_j}}}{nV}-1}^2,
    \end{aligned}
    \end{equation}
    Here, $(a)$ follows from Jensen's inequality, applied to the convex function $f(x)=x^2$, and the identity 
    \begin{equation}
        \sum_{i\neq j}\br{ \abs{\braket{\phi}{\phi_i}}^2 + \abs{\braket{\phi}{\phi_j}}^2}=2n(n-1)V,  
    \end{equation}
    given by Lemma \ref{lem:VEvec}. Moreover, $(b)$ follows from the arithmetic-geometric mean inequality, and $(c)$ uses the fact that $\braket{\phi_i}{\phi_i}=1$ for any $i\in[n]$. Define 
    \begin{equation}
        M:=\sum_{i=1}^n\ketbra{\phi_i}{\phi_i}=M^\dagger,
    \end{equation}
    We observe that 
    \begin{equation}
    \begin{aligned}
        \Tr\br{M\ketbra{\phi}{\phi}} & = nV,  \\ 
        \Tr\br{M^2\ketbra{\phi}{\phi}} & = \sum_{i,j}\braket{\phi_j}{\phi}\!\braket{\phi}{\phi_i}\!\braket{\phi_i}{\phi_j}  \\
        & \leq \sum_{i,j}\abs{\braket{\phi_j}{\phi}\!\braket{\phi}{\phi_i}\!\braket{\phi_i}{\phi_j}}.
    \end{aligned}
    \end{equation}
    Therefore,
    \begin{equation}
        \begin{aligned}
        E & \geq\frac{V}{(n-1)^2}\br{\max\left\{0,\frac{\Tr\br{M^2\ketbra{\phi}{\phi}}}{\Tr\br{M\ketbra{\phi}{\phi}}}-1\right\}}^2\\
        & \overset{(d)}{\geq}\frac{V}{(n-1)^2}\br{\max\left\{0,\Tr\br{M\ketbra{\phi}{\phi}}-1\right\}}^2\\
        & \overset{(e)}{=}\frac{n^2}{(n-1)^2}V\br{\max\left\{0,V-\frac{1}{n}\right\}}^2,
    \end{aligned} 
    \end{equation}
    where $(d)$ follows from the Cauchy-Schwarz inequality, 
    \begin{equation}
        \Tr\br{M^2\ketbra{\phi}{\phi}} =\bra{\phi} M^\dagger M\ket{\phi}\braket{\phi}{\phi} \geq \abs{\bra{\phi}M\ket{\phi}}^2, 
    \end{equation}
    and $(e)$ follows from the fact that $\Tr\br{M\ketbra{\phi}{\phi}}=nV$. 

    We now prove the tightness of the bound. 
    
    Consider the case where $v\in(\frac{1}{n},1]$. Let $\epsilon = \frac{nv-1}{n-1}$, then $\epsilon\in(0,1]$. We explain how to choose $\lbrace\ket{\phi_i}\rbrace_{i\in [n]}$ such that $\braket{\phi_i}{\phi_j}=\epsilon$ for $i\ne j$. For any $\epsilon\in(0,1]$, we take 
    \begin{equation}
        \ket{\phi_i} =\sum_j  \frac{1+t \delta_{ij}}{\sqrt{n+2t+t^2}} \ket{j},
    \end{equation}
    for all $i\in[n]$. To ensure $\braket{\phi_i}{\phi_j}=\epsilon$, we take $t\in[0,+\infty)$ satisfying 
    \begin{equation}
        \epsilon = 1 - \frac{t^2}{n+2t+t^2}.
    \end{equation}
    Let
    \begin{equation}
        \ket{\phi} =  \frac{1}{\sqrt{n+n(n-1)\epsilon}} \sum_{i}\ket{\phi_i}. 
    \end{equation}
    Therefore,
    \begin{equation}
        \braket{\phi}{\phi_i} =\sqrt{\frac{1+ (n-1)\epsilon}{n}} = \sqrt{v}. 
    \end{equation}
    Now we set $\sigma = \ketbra{\phi}{\phi}$ and $P_i = \ketbra{\phi_i}{\phi_i}$. 
    Then 
    \begin{equation}
        V = \frac{1}{n} \sum_i \Tr\br{P_i\sigma } = \frac{1}{n} \sum_i \abs{\braket{\phi}{\phi_i}}^2 = v,
    \end{equation}
    and
    \begin{equation}
    \begin{aligned}
        E & = \frac{1}{n(n-1)} \sum_{i\neq j} \Tr(P_iP_jP_i\sigma) \\
        & = \frac{1}{n(n-1)} \sum_{i\neq j } \abs{\braket{\phi}{\phi_i}}^2 \abs{\braket{\phi_i}{\phi_j}}^2 \\
        & =  \frac{n^2}{(n-1)^2}v \br{v-\frac{1}{n}}^2. 
    \end{aligned}
    \end{equation}
    
    Now consider the case where $v\in[0,\frac{1}{n}]$. We take a complete orthonormal basis $\lbrace\ket{i}\rbrace_{i\in[n+1]}$ on $\H$. We set 
    \begin{equation}
        \sigma = nv \ketbra{1}{1} + (1-nv) \ketbra{n+1}{n+1}, \quad P_i = \ketbra{i}{i}. 
    \end{equation}
    Then 
    \begin{equation}
        V = \frac{1}{n} \sum_i \Tr(P_i\sigma ) = v, 
    \end{equation}
    and 
    \begin{equation}
        E = \frac{1}{n(n-1)} \sum_{i\neq j} \Tr( P_i P_j P_i \sigma ) = 0.
    \end{equation}
    
    Combining both cases, we conclude that for any $v\in[0,1]$, there exists a state $\sigma$ and $n$ projectors $P_1,\ldots, P_n$ in a finite-dimensional Hilbert space $\H$ such that $V=v$ and 
    \begin{equation}
        E = \frac{n^2}{(n-1)^2} v \left(\max \left\{0, v-\frac{1}{n}\right\}\right)^2, 
    \end{equation}
    which concludes our proof. 
\end{proof}

Our tight CMT can be generalized to the case where each measurement has at most $S$ desired outcomes: 

\begin{proposition}
\label{thm:consec_mmt_general}
Given integers $n,S\in\N$ and $n$ ensembles of projectors $\lbrace P_i^s\rbrace_{s=1}^{S}$ for $i\in[n]$ in a finite-dimensional Hilbert space $\H$. Let $\sigma$ be any quantum state on $\H$, and let 
\begin{equation}
\begin{aligned}
    V & = \frac{1}{n} \sum_{i=1}^n\sum_{s=1}^S \Tr(P_i^s\sigma),  \\
    E & = \frac{1}{n(n-1)} \sum_{i\neq j} \sum_{s,t=1}^S \Tr(P_j^t P_i^s \sigma P_i^s P_j^t) .
\end{aligned}
\end{equation}
It holds that
\begin{align}
    E\geq  \frac{n^2}{(n-1)^2 S} V \left(\max\left\{0,V-\frac{1}{n}\right\}\right)^2.
\end{align}
\end{proposition}

The proof follows the same routine as in Theorem~\ref{thm:consec_mmt} and we do not elaborate it here.

\section{Coupled-game method and applications}

In this section, we detail the coupled-game method and bound the quantum values of the ${\rm CHSH}_q(p)$ and ${\rm CHSH}_q(p)^{\otimes m}$ games, subsequently applying these results to analyze the security of the $\mathbb{F}_p$ and $\mathbb{F}^{\otimes m}_p$ relativistic bit commitment protocols. In addition, we also detail the application of the coupled-game method in the soundness analysis of relativistic ZKPs.

\subsection{Nonlocal game and its coupled game}

\subsubsection{Formal definitions}\label{subsec:formaldef}

An important scenario in quantum information involves two spatially separated parties attempting to cooperatively satisfy a specific winning condition without communicating. This is a framework formally known as a nonlocal game.

\begin{definition}[Nonlocal game]
    A nonlocal game $G$ is defined by the tuple $(\mathcal{I}_A, \mathcal{I}_B, \mathcal{O}_A, \mathcal{O}_B, V, \pi)$, where $\mathcal{I}_A$ and $\mathcal{I}_B$ are the input sets; $\mathcal{O}_A$ and $\mathcal{O}_B$ are output sets; $V: \mathcal{I}_A \times \mathcal{I}_B \times \mathcal{O}_A \times \mathcal{O}_B \to \{0, 1\}$ is the verification function, where $V(x, y, a, b) = 1$ indicates a win and 0 a loss for inputs $x, y$ and outputs $a, b$; and $\pi: \mathcal{I}_A \times \mathcal{I}_B \to [0, 1]$ is the input distribution, satisfying $\sum_{(x,y) \in \mathcal{I}_A \times \mathcal{I}_B} \pi(x, y) = 1$. 
\end{definition}

In many cryptographic scenarios, games possess a deterministic structure where one player's input and output uniquely determine the other player's correct response. We formally capture this strict correlation as follows:

\begin{definition}[projective]
     A game $G=(\mathcal{I}_A, \mathcal{I}_B, \mathcal{O}_A, \mathcal{O}_B, V, \pi)$ is projective if for any $(x, y, a)\in\mathcal{I}_A\times\mathcal{I}_B\times\mathcal{O}_A$, there is a unique $b\in\mathcal{O}_B$ such that $V(x,y,a,b) = 1$.
\end{definition}

To analyze the physical capabilities of the players in these games, we must specify the strategy, which is characterized by shared entanglement and local measurements. 

\begin{definition}[Strategy]
    Given finite-dimensional Hilbert spaces $\H_A$ and $\H_B$, a strategy $ \mathscr{S} $ for a nonlocal game $ G $ is a tuple $ (\rho, \mathcal{Q}, \mathcal{P}) $, where $ \rho $ is a quantum state in $\H_A\otimes \H_B$; $ \mathcal{Q} = \{\mathcal{Q}_x\}_{x \in \mathcal{I}_A} $, where each $ \mathcal{Q}_x = \{Q_x^a\}_{a \in \mathcal{O}_A} $ is a POVM on $\H_A$ for all $x$; similarly, $ \mathcal{P} = \{\mathcal{P}_y\}_{y \in \mathcal{I}_B} $, where each $ \mathcal{P}_y = \{P_y^b\}_{b \in \mathcal{O}_B} $ is a POVM on $\H_B$ for all $y$. The winning probability of the strategy is 
    \begin{equation}
        \omega(G,\mathscr{S}) = \sum_{x,y,a,b} \pi(x,y) V(x,y,a,b) \Tr\br{\br{Q_x^a\otimes P_y^b} \rho}.
    \end{equation}
\end{definition}

The ultimate metric of interest for our security analysis is the maximum success probability achievable by players restricted only by the laws of quantum mechanics, or the quantum value of the game. 

\begin{definition}[Quantum value]
    The quantum value $\omega^*(G)$ for a nonlocal game $G$ is defined as the maximal winning probability over all possible strategies: 
    \begin{align}
        \omega^*(G) = \max_{\mathscr{S}\in \mathcal{S}} \omega(G,\mathscr{S}), 
    \end{align}
    where $\mathcal{S}$ denotes the set of all valid strategies.
\end{definition}

Calculating or bounding this quantum value directly is generally intractable. A powerful method to bypass this difficulty involves analyzing a related scenario where one player receives two consecutive inputs, known as the coupled game~\cite{chailloux2017relativistic}.

\begin{definition}[Coupled game]
    For any nonlocal game $G=(\mathcal{I}_A, \mathcal{I}_B, \mathcal{O}_A, \mathcal{O}_B, V, \pi)$  with uniform input distribution, its coupled game is defined as a nonlocal game $G_{\text{cp}}=(\mathcal{I}_A, \mathcal{I}_B\times\mathcal{I}_B, \mathcal{O}_A, \mathcal{O}_B\times\mathcal{O}_B, V_{\text{cp}}, \pi_{\text{cp}})$, where
    $V_{\text{cp}}(x,(y,y'),a,(b,b'))=V(x,y,a,b)\cdot V(x,y',a,b')$ and $\pi_{\text{cp}}(x,(y,y'))=\frac{1-\delta_{y,y'}}{|\mathcal{I}_A||\mathcal{I}_B|(|\mathcal{I}_B|-1)}$ for any $x\in\mathcal{I}_A$, $y,y'\in\mathcal{I}_B$, $a\in\mathcal{O}_A$ and $b,b'\in\mathcal{O}_B$. 
\end{definition}

To connect the quantum value of the original game to its coupled counterpart, we construct a specific sequential-measurement strategy for the coupled game based directly on the measurements used in the original game.

\begin{definition}[Induced strategy]
    Let $\mathscr{S}=(\rho,\mathcal{Q},\mathcal{P})$ be a strategy for $G$ where $\mathcal{Q}_x$ and $\mathcal{P}_y$ are projective measurements for all $x\in\mathcal{I}_A$ and $y\in\mathcal{I}_B$. Then we define $\mathscr{S}_{\text{cp}}= (\rho,\mathcal{Q},\mathcal{P}_{\text{cp}})$ where $\mathcal{P}_{\text{cp}} = \set{\mathcal{P}_{\text{cp,}y,y'}}_{y\neq y'\in \mathcal{I}_B}$ and $\mathcal{P}_{\text{cp,}y,y'} = \set{ P_y^b P_{y'}^{b'}P_y^b}_{b,b'\in\mathcal{O}_B}$ to be the induced strategy of $\mathscr{S}$ for $G_{\text{cp}}$. 
\end{definition}

\subsubsection{Lower bound of \texorpdfstring{$\omega^{*}(G_{\text{cp}})$}{omega(Gcp)}}\label{apd:LBwG}

With definitions established above, we can now apply our tight CMT to quantify the precise relationship between the quantum value of a projective game and its coupled game.

\begin{repproposition}{thm:coupledgamevalue}
Let $ G $ be a projective nonlocal game with uniform input distribution, and let $ n = |\mathcal{I}_B| $. Then the quantum values of $ G $ and its coupled game $ G_{\text{cp}} $ satisfy:
\begin{equation}
    \tag{\ref{eqn:omega_relation}}
    \omega^*(G_{\text{cp}}) \geq \frac{n^2}{(n-1)^2} \omega^*(G)\br{\max\left\{0,\omega^*(G)-\frac{1}{n}\right\}}^2.
\end{equation}   
\end{repproposition}
\begin{proof}[Proof of Proposition \ref{thm:coupledgamevalue}]

Consider that Alice and Bob use the strategy $\mathscr{S}=(\rho,\mathcal{Q},\mathcal{P})$ for $G$, where $\rho$ is a shared quantum state, and $ \mathcal{Q} = \{\mathcal{Q}_x\}_{x \in \mathcal{I}_A} $ and $ \mathcal{P} = \{\mathcal{P}_y\}_{y \in \mathcal{I}_B} $ are the projective measurement collections used by Alice and Bob, respectively. Let $\mathscr{S}_{\text{cp}}$ denote the induced strategy of $\mathscr{S}$ for $G_{\text{cp}}$.
We fix Alice's input-output pair as $(x,a)$. Then Bob's reduced state conditioned on $(x,a)$ is 
\begin{equation}
    \rho^B_{x,a} = \frac{\Tr_A(Q_x^a \rho Q_x^a)}{\Tr(Q_x^a \rho )}. 
\end{equation}
Conditioned on Alice's input-output pair $(x,a)$, the winning probability of $G$ is
\begin{equation}
    \begin{aligned}
        \Pr[\text{win}\: G|(x,a)]  & = \frac{1}{|\mathcal{I}_B|} \sum_{y,b} V(x,y,a,b) \Tr\br{P_y^{b}\rho^B_{x,a} } \\
        & = \frac{1}{n} \sum_{y=1}^n \Tr\br{P_y^{b_y} \rho^B_{x,a}} \quad =: V_{x,a},\\
    \end{aligned}  
\end{equation}
where $b_y$ is the unique output of Bob such that $V(x,y,a,b_y)=1$ since $ G $ is a projective game. 

Similarly, conditioned on Alice's input-output pair $(x,a)$, the winning probability of $G_{\text{cp}}$ is 
\begin{equation}
    \begin{aligned}
        & \Pr[\text{win}\: G_{\text{cp}}|(x,a)] \\
        & = \frac{1}{|\mathcal{I}_B|(|\mathcal{I}_B|-1)}
        \sum_{y,y'\neq y,b,b'} \\
        & \quad\quad  V(x,y,a,b) V(x,y',a,b')\Tr\br{\br{P_y^b  P_{y'}^{b'}  P_y^b }\rho^B_{x,a}} \\
        & = \frac{1}{n(n-1)} \sum_{y,y'\neq y} \Tr((P_y^{b_y}  P_{y'}^{b'_{y'}}  P_y^{b_y} )\rho^B_{x,a}) \quad =: E_{x,a},
    \end{aligned}
\end{equation}
where $b_y$ and $b'_{y'}$ are the unique outputs of Bob such that $V(x,y,a,b_y)=1$ and $V(x,y',a,b'_{y'})=1$, respectively.
We can now apply Theorem~\ref{thm:consec_mmt} to $V_{x,a}$ and $E_{x,a}$,
\begin{equation}
    \label{eqn:ExaVxa}
    E_{x,a} \geq \frac{n^2}{(n-1)^2} V_{x,a} \br{\max\left\{ 0, V_{x,a} - \frac{1}{n}\right\}}^2. 
\end{equation}
Note that 
\begin{equation}
    \begin{aligned}
        \omega(G,\mathscr{S}) & =  \sum_{x,a} \Pr[(x,a)] \Pr[\text{win}\: G|(x,a)] \\
        & = \sum_{x,a}  \Pr[(x,a)]  V_{x,a}, \\
        \omega(G_{\text{cp}},\mathscr{S}_{\text{cp}}) &  =  \sum_{x,a} \Pr[(x,a)] \Pr[\text{win}\: G_{\text{cp}}|(x,a)] \\
        & = \sum_{x,a}  \Pr[(x,a)]  E_{x,a}. 
    \end{aligned}
\end{equation}
We then obtain
\begin{equation}
    \begin{aligned}
    & \omega(G_{\text{cp}},\mathscr{S}_{\text{cp}})  = \sum_{x,a} \Pr[(x,a)] E_{x,a} \\
    & \overset{(a)}{\geq} \sum_{x,a} \Pr[(x,a)] \frac{n^2}{(n-1)^2} V_{x,a}\br{\max\left\{0,V_{x,a}-\frac{1}{n}\right\}}^2 \\
    & \overset{(b)}{\geq}  \frac{n^2}{(n-1)^2} \br{\sum_{x,a} \Pr[(x,a)]V_{x,a}} \\
    & \quad \quad \br{\max\left\{0,\sum_{x,a} \Pr[(x,a)]V_{x,a}-\frac{1}{n}\right\}}^2 \\
    & = \frac{n^2}{(n-1)^2} \omega(G,\mathscr{S})\br{\max\left\{0,\omega(G,\mathscr{S})-\frac{1}{n}\right\}}^2
    \end{aligned}
\end{equation}
where (a) follows from Eq.~\eqref{eqn:ExaVxa} and (b) follows from the convexity of $f(V) = V\br{\max\left\{0,nV-1\right\}}^2$, see Appendix~\ref{apd:convexity}. 
We conclude our proof by choosing $\mathscr{S}$ to be an optimal strategy achieving $\omega^*(G)$, which results in 
\begin{equation}
    \omega^*(G_{\text{cp}}) \geq \frac{n^2}{(n-1)^2} \omega^*(G)\br{\max\left\{0,\omega^*(G)-\frac{1}{n}\right\}}^2.
\end{equation} 
\end{proof}

\subsubsection{Convexity of \texorpdfstring{$f(V)=V(nV-1)^2$}{f(V)}}
\label{apd:convexity}
\begin{lemma}
    \label{lem:convexity}
    The function $f(V) = V\br{\max\{0,nV-1\}}^2$ is convex for any positive integer $n$. 
\end{lemma}
\begin{proof}
    For $V\geq \frac{1}{n}$, $f(V)$ is convex because $f''(V) =6n^2 V -4n>0$. Therefore, for $\frac{1}{n}\leq V_1\leq V_2$ and $p_1+p_2=1$,
    \begin{align}
        p_1 f(V_1) + p_2 f(V_2)\geq f(p_1V_1+p_2 V_2). 
    \end{align}
    $f(V)$ is convex when $V< \frac{1}{n}$ because $f(V) = 0$. It holds that for $V_1\leq V_2 \leq \frac{1}{n}$ and $p_1+p_2=1$,
    \begin{align}
        p_1 f(V_1) + p_2 f(V_2)\geq f(p_1V_1+p_2 V_2). 
    \end{align}
    For $V_1\leq \frac{1}{n}\leq V_2$ and $p_1+p_2=1$, 
    \begin{equation}
    \begin{aligned}
        p_1f(V_1) + p_2 f(V_2) & \overset{(a)}{=} p_1 f\br{\frac{1}{n}} + p_2 f(V_2) \\
        & \overset{(b)}{\geq} f\br{p_1\frac{1}{n}+p_2 V_2 } \\
        & \overset{(c)}{\geq} f(p_1 V_1 + p_2 V_2), 
    \end{aligned}
    \end{equation}
    where $(a)$ uses $f(V)=0$ for $V\leq \frac{1}{n}$, $(b)$ uses the convexity of $f(V)$ when $V\geq \frac{1}{n}$ and $(c)$ uses that  $f(V)$ is monotonically non-decreasing. In conclusion, it holds for any $V_1$ and $V_2$ that 
    \begin{align}
        p_1 f(V_1) + p_2 f(V_2)\geq f(p_1V_1+p_2 V_2), 
    \end{align}
    which means $f(V)$ is convex on $\mathbb{R}$. 
\end{proof}

\subsubsection{Upper bound of \texorpdfstring{$\omega^{*}({\rm CHSH}_q(p))$}{CHSHqp}}
\label{apd:chsh_upper}

\begin{repproposition}{thm:chsh_upper}[Upper bound of $\omega^{*}$(${\rm CHSH}_q(p)$)]
Let $p=r^t$ and $q=r^s$ be powers of the same prime $r$, with $t$ dividing $s$. Then the upper bound on the quantum value of the ${\rm CHSH}_q(p)$ game is given by:
\begin{equation}
\tag{\ref{eqn:chshqp_analytical}}
\begin{aligned}
    \omega^*({\rm CHSH}_q(p)) \leq \frac{1}{3p} \br{2 + \mathcal{C}(\Delta) },
\end{aligned}
\end{equation}
where $\Delta = \frac{27p(p-1)^2}{q} - 2$ and
\begin{equation}\tag{\ref{eqn:defofCdelta}}
   \mathcal{C}(\Delta) =\begin{cases}
    2\cosh\!\left(\dfrac{1}{3}\operatorname{arccosh}   \dfrac{\Delta}{2}\right), & \Delta\in[2,\infty), \\
    2\cos\!\left(\dfrac{1}{3}\arccos  \dfrac{\Delta}{2}\right), & \Delta\in (-2,2). 
\end{cases} 
\end{equation}
\end{repproposition}
\begin{proof}
Let $G$ denote ${\rm CHSH}_q(p)$ and $G_{\text{cp}}$ denote ${\rm CHSH}_q(p)_\text{cp}$ coupled game. First, fix Alice's input/output pair $(x, a)$ and randomly select two distinct inputs $y, y^{\prime}$ for Bob. Bob then outputs $(b, b^{\prime})$ corresponding to these inputs.  A win in $G$ requires that $a+b=x\cdot y$ and $a+b^{\prime}=x\cdot y^{\prime}$, which implies $x = \frac{b - b^{\prime}}{y - y^{\prime}}$. Thus, to win $G_\text{cp}$, Bob must correctly guess Alice's input $x$, an event with probability $\frac{1}{q}$ due to the no-signaling principle. Considering $\omega^*(G_\text{cp})=\frac{1}{q}$ and Proposition~\ref{thm:coupledgamevalue}, we have

\begin{equation}\label{eqn:wGeq}
\omega^*(G) \br{\max\left\{0,\omega^*(G)-\frac{1}{p}\right\}}^2\leq \frac{(p-1)^2}{qp^2},
\end{equation}
where $|\mathcal{I}_B| = p$ and $S = 1$. We obtain the upper bound on the quantum value of the ${\rm CHSH}_q(p)$ game
\begin{equation}
\tag{\ref{eqn:chshqp_analytical}}
\begin{aligned}
    \omega^*({\rm CHSH}_q(p)) \leq \frac{1}{3p} \br{2 + \mathcal{C}(\Delta) },
\end{aligned}
\end{equation}
where $\mathcal{C}(\cdot)$ is defined in Eq.~\eqref{eqn:defofCdelta} and $\Delta = \frac{27p(p-1)^2}{q} - 2$. Note that the right-hand side of Eq.~\eqref{eqn:chshqp_analytical} is the largest real root of the cubic equation obtained by saturating Eq.~\eqref{eqn:wGeq} in the regime $x\geq1/p$:
\begin{equation}
    x\br{x-\frac{1}{p}}^2= \frac{(p-1)^2}{qp^2}.
\end{equation}
\end{proof}

\subsubsection{Upper bound of \texorpdfstring{${\rm CHSH}_q(p)^{\otimes m}$}{CHSHqpm} game}
\label{apd:chsh_m_upper}

\begin{repproposition}{thm:n_upper}[Upper bound of $\omega^{*}$(${\rm CHSH}_{q}(p)^{\otimes m}$)]
Let $p=r^t$ and $q=r^s$ be powers of the same prime $r$, with $t$ dividing $s$. Then the quantum value of $ {\rm CHSH}_{q}(p)^{\otimes m} $ is upper bounded by:

\begin{equation}\tag{\ref{eqn:chshqpm_analytical}}
    \begin{aligned}        
        \omega^*({\rm CHSH}_q(p)^{\otimes m}) \leq  \frac{1}{3p^m} \br{2 +  \mathcal{C}(\Delta_m) }, \\
    \end{aligned}
\end{equation}
where $\mathcal{C}(\cdot)$ is defined in Eq.~\eqref{eqn:defofCdelta} and 
\begin{equation}
    \Delta_m = 27p^m \br{p^m-1} \br{\br{1+\frac{p-1}{q}}^m-1} - 2. 
\end{equation}
If $m\gg \frac{q}{p-1}$, the quantum value of $ {\rm CHSH}_{q}(p)^{\otimes m} $ is asymptotically upper bounded by:
\begin{equation}
    \omega^{*}({\rm CHSH}_{q}(p)^{\otimes m})\leq \frac{1}{p^m} + \frac{1}{p^{\frac{m}{3}}}\br{1+\frac{p-1}{q}}^{\frac{m}{3}}.
\end{equation}   
\end{repproposition}
\begin{proof}
    Let $G$ denote ${\rm CHSH}_q(p)^{\otimes m}$ game and $G_{\text{cp}}$ denote ${\rm CHSH}_q(p)^{\otimes m}_\text{cp}$ coupled game. Fix Alice's input/output pair $(X, A)$ and choose two distinct input strings for Bob, $Y$ and $Y^{\prime}$, where $Y = (y_1, \dots, y_m)$ and $Y^{\prime} = (y^{\prime}_1, \dots, y^{\prime}_m)$ with $Y \ne Y^{\prime}$. Bob outputs $(B, B^{\prime})$, where $B = (b_1, \dots, b_m)$ and $B^{\prime} = (b^{\prime}_1, \dots, b^{\prime}_m)$, corresponding to the inputs $Y$ and $Y^{\prime}$, respectively. The coupled game is won if and only if 
    \begin{equation}\label{eqn:mfoldwinning}
        a_i+b_i=x_i\cdot y_i \quad\text{and}\quad a_i+b_i^{\prime}=x_i\cdot y_i^{\prime}
    \end{equation}
    for all $i\in [m]$. Let $ \tau $ denote the set of indices where $ Y $ and $ Y' $ differ, i.e., $ \tau = \{ i \in [m] \mid y_i \neq y'_i \} $. For each $ i \in \tau $, the winning condition in Eq.~\eqref{eqn:mfoldwinning} requires $ x_i = \frac{b_i - b'_i}{y_i - y'_i} $. Thus, to win the game, Bob must correctly guess all $ x_i $ for $ i \in \tau $, which occurs with probability $ \frac{1}{q^{|\tau|}} $.  
    For a fixed $ Y $, the probability that $ Y' $ differs from $ Y $ in exactly $ |\tau| $ positions is $ \binom{m}{|\tau|} \frac{(p - 1)^{|\tau|}}{p^m - 1} $, where $ \binom{a}{b} = \frac{a!}{b!(a - b)!} $. Therefore,
    \begin{equation}
        \omega^*(G_{\text{cp}}) = \sum_{|\tau|=1}^m \binom{m}{|\tau|}\frac{(p-1)^{|\tau|}}{(p^m - 1) q^{|\tau|}} = \frac{\left(1 + \frac{p-1}{q}\right)^m - 1}{p^m-1}.
    \end{equation}
    Thus we upper bound the quantum value of ${\rm CHSH}_q(p)^{\otimes m}$ by setting $n=p^m$ in Proposition~\ref{thm:coupledgamevalue}, yielding
    \begin{equation}
    \label{eqn:chshqpm}
    \begin{aligned}
        \br{\frac{p^m}{p^m-1}}^2\omega^*(G) \br{\omega^*(G) - \frac{1}{p^m}}^2 
        \leq \frac{\left(1 + \frac{p-1}{q}\right)^m - 1}{p^m-1}.
    \end{aligned}
    \end{equation} 
    The cubic inequality Eq.~\eqref{eqn:chshqpm} can be solved analytically, which results in 
    \begin{equation}\tag{\ref{eqn:chshqpm_analytical}}
    \begin{aligned}        
        \omega^*({\rm CHSH}_q(p)^{\otimes m}) \leq  \frac{1}{3p^m} \br{2 +  \mathcal{C}(\Delta_m) }, 
        \end{aligned}
    \end{equation}
    where $\mathcal{C}(\cdot)$ is defined in Eq.~\eqref{eqn:defofCdelta} and 
    \begin{equation}
        \Delta_m = 27p^m \br{p^m-1} \br{\br{1+\frac{p-1}{q}}^m-1} - 2. 
    \end{equation}    
    However, the asymptotic behavior of the analytical solution is difficult to analyze. So we derive an asymptotic solution as well.
    
    Consider the asymptotic regime where $ m \to \infty $ while $ \frac{p - 1}{q} $ remains finite. Let us denote $ \omega^*({\rm CHSH}_q(p)^{\otimes m}) = \frac{\alpha}{p^m} $, where $ \alpha \geq 1 $. Then we get
    \begin{align}
    \label{eqn:chshqpm_ancilla}
        \alpha \br{\alpha - 1}^2 \leq p^{m}\br{p^m-1}\br{ \br{1+\frac{p-1}{q}}^m -1 } .
    \end{align}
    As $p^m \to +\infty$ and $(1+\frac{p-1}{q})^{m} \to +\infty$, we can approximate $p^m-1$ and $\br{1+\frac{p-1}{q}}^m-1$ by $p^m$ and $\br{1+\frac{p-1}{q}}^m$, respectively, in Eq.~\eqref{eqn:chshqpm} without significant loss of accuracy. 
    If equality holds in Eq.~\eqref{eqn:chshqpm_ancilla}, then $\alpha\to+\infty$ as $m\to\infty$, since $p^{2m} \br{1+\frac{p-1}{q}}^m\to +\infty$. In this regime, the term $\alpha(\alpha-1)^2$ can be well approximated by $(\alpha-1)^3$, which results in
    \begin{align}
        \label{eqn:chshqpm_asymptotic}
        \omega^*(G) \leq \frac{1}{p^m} + \frac{1}{p^{\frac{m}{3}}}\br{1+\frac{p-1}{q}}^{\frac{m}{3}}. 
    \end{align}
    The above asymptotic expression closely matches the asymptotic behavior of the analytical solution.
\end{proof}

${\rm CHSH}_q(p)$ is the $m=1$ case of ${\rm CHSH}_q^{\otimes m}(p)$. For the special case where $m=1$, $\omega^*({\rm CHSH}_q(p))$ can be bounded by 
\begin{equation}
\begin{aligned}
    \omega^*({\rm CHSH}_q(p)) \leq \frac{1}{3p} \br{2 +  \mathcal{C}(\Delta)},
\end{aligned}
\end{equation}
where $\mathcal{C}(\cdot)$ is defined in Eq.~\eqref{eqn:defofCdelta} and $\Delta = \frac{27p(p-1)^2}{q} - 2$.

\begin{figure}[htbp!]
    \centering
    \includegraphics[width=0.92\linewidth]{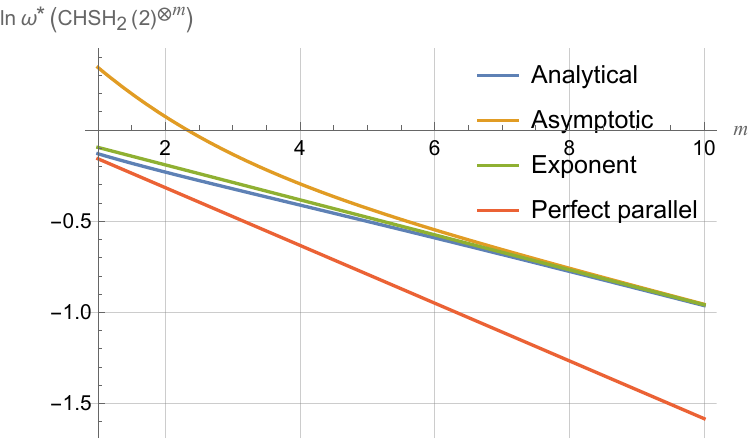}
    \caption{\textbf{Upper bounds on $\ln \omega^*({\rm CHSH}_2^{\otimes m}(2))$.} Blue line, the analytical solution (Eq.~\eqref{eqn:chshqpm_analytical}) to Eq.~\eqref{eqn:chshqpm}; orange line, the asymptotic solution (Eq.~\eqref{eqn:chshqpm_asymptotic}) to Eq.~\eqref{eqn:chshqpm}; green line, the exponent $0.909^m$; red line, the perfect parallel bound $0.853^m$~\cite{cleve2008perfect}. }
    \label{fig:chshqpm_asymptotic} 
\end{figure}

Fig.~\ref{fig:chshqpm_asymptotic} provides an intuitive illustration of the existing bounds on $\ln \omega^*({\rm CHSH}_2^{\otimes m}(2))$. 

\subsection{\texorpdfstring{$\mathbb{F}_p$ RBC}{Fp RBC}}\label{apd:RBC}
\begin{reptheorem}{thm:sumbinding_RBC}[Sum-binding of $\mathbb{F}_p$ RBC]
The $\mathbb{F}_p$ relativistic bit commitment is $\varepsilon_b$-sum-binding with 
\begin{equation}
    \varepsilon_b=\frac{1}{3}\left(\mathcal{C}(\Delta) -1\right),
\end{equation}
where $\mathcal{C}(\cdot)$ is defined in Eq.~\eqref{eqn:defofCdelta} and $\Delta = \frac{27p(p-1)^2}{q} - 2$. 
\end{reptheorem}

\begin{proof}
Consider a cheating strategy $(\text{Com}^{*},\text{Open}^{*})$ where provers B1 and B2 share entanglement. Upon receiving $x\in \mathbb{F}_q$, B1 performs a quantum measurement on the shared entangled state to produce $a\in \mathbb{F}_q$ and sends $a$ to the verifier. For a random $y\in \mathbb{F}_p$ that B2 wants to reveal, B2 performs a measurement on the shared entangled state and outputs $b\in \mathbb{F}_q$. We will have
\begin{multline}
    \frac{1}{p}\sum_{y=0}^{p-1} \Pr[\text{B1 and B2 successfully reveal \textit{y}} \, | \, (\text{Com}^*, \text{Open}^*)]\\
    =\Pr[x\cdot y=a+b].
\end{multline}

This cheating strategy in bit commitment is actually equivalent to a strategy for the ${\rm CHSH}_q(p)$ game, thus we have
\begin{equation}
    \Pr[x\cdot y=a+b]\le \omega^*({\rm CHSH}_q(p))\le \frac{1}{3p}(2+\mathcal{C}(\Delta)),
\end{equation}
where the upper bound on the quantum value of ${\rm CHSH}_q(p)$ game is given by Proposition~\ref{thm:chsh_upper}. This indicates that  
\begin{multline}
    \sum_{y=0}^{p-1} \Pr[\text{B1 and B2 successfully reveal \textit{y}} \, | \, (\text{Com}^*, \text{Open}^*)]\\
    \leq 1+\frac{1}{3}\left(\mathcal{C}(\Delta) -1\right),
\end{multline}
where we can obtain $\varepsilon_b=\frac{1}{3}\left(\mathcal{C}(\Delta)-1\right)$.
\end{proof}

\subsection{\texorpdfstring{$m$-fold parallel $\mathbb{F}_p$-string RBC}{m-fold parallel Fp-string RBC}}\label{apd:RBC_parallel}

\begin{reptheorem}{thm:sum_binding_parallel_RBC}[Sum-binding of $\mathbb{F}_p^{m}$ RBC]
The $\mathbb{F}_p^{m}$ relativistic bit commitment is  $\varepsilon_b$-sum-binding with
\begin{equation}
    \varepsilon_b=\frac{1}{3}\left(\mathcal{C}(\Delta_m) -1\right),
 \end{equation}
where $\mathcal{C}(\cdot)$ is defined in Eq.~\eqref{eqn:defofCdelta} and 
\begin{equation}
    \Delta_m = 27p^m \br{p^m-1} \br{\br{1+\frac{p-1}{q}}^m-1} - 2. 
\end{equation}
\end{reptheorem}

\begin{proof}
Consider a quantum cheating strategy $(\text{Com}^{*},\text{Open}^{*})$ where B1 and B2 share quantum entanglement. After receiving $X=\{x_i\}_{i\in [m]}\in \mathbb{F}^{\otimes m}_q$, B1 performs a quantum measurement on the shared quantum system to produce an output $A=\{a_i\}_{i\in [m]}\in \mathbb{F}^{\otimes m}_q$ and sends $A$ to the verifier. For the string $Y=\{y_i\}_{i\in \left[m\right]}\in \mathbb{F}^{\otimes m}_p$ that B2 wants to reveal, B2 performs a measurement on the shared quantum system and outputs $B  =\{b_i\}_{i\in \left[m\right]}\in \mathbb{F}^{\otimes m}_q$. We will have
\begin{multline}
     \frac{1}{p^{m}}\sum_{Y\in \mathbb{F}_p^{m}} \Pr[\text{ Successfully reveal}\,\, Y \, | \, (\text{Com}^*, \text{Open}^*)] \\
     =\Pr[x_i\cdot y_i=a_i+b_i, \forall i\in \left[m\right]].
\end{multline}
This cheating strategy is equivalent to winning the $m$-fold parallel ${\rm CHSH}_q(p)^{\otimes m}$ game, thus we have
\begin{multline}
    \Pr[x_i\cdot y_i=a_i+b_i, \forall i\in\left[m\right]]\le \omega^*({\rm CHSH}_q(p)^{\otimes m})\\
    \leq\frac{1}{3p^m} \br{2 + \mathcal{C}(\Delta_m) },
\end{multline}
where the upper bound on winning probability of the ${\rm CHSH}_q(p)^{\otimes m}$ game is given by Proposition~\ref{thm:n_upper}. 
\begin{equation}
\begin{aligned}
   & \sum_{Y\in \mathbb{F}_p^{m}} \Pr[\text{Successfully reveal}\,\, Y  |   (\text{Com}^*, \text{Open}^*)]\\                   
   &\leq p^{m}\cdot \omega^*({\rm CHSH}_q(p)^{\otimes m})= 1+\frac{1}{3}\left(\mathcal{C}(\Delta_m) -1\right).
\end{aligned}
\end{equation}

\end{proof}

\subsection{Relativistic zero-knowledge proofs}
\label{app:rzko}

In this appendix, we first establish a general coupled-game bound for $S$-projective games and then specialize it to the three relativistic zero-knowledge protocols considered in the main text. Importantly, $S$-projectivity alone does not provide an upper bound on the coupled-game value. Such an upper bound must be established separately from the algebraic and no-signaling properties of each protocol.

\subsubsection{Coupled-game bound for $S$-projective games}

Recall that a nonlocal game is $S$-projective if, for every fixed question pair and first-prover answer, at most $S$ answers of the second prover are accepting. Let $G$ be such a game with a uniform input distribution, and suppose that the second prover has $r$ possible questions.

Consider an arbitrary entangled strategy for $G$. Conditioned on a question-answer pair $(x,a)$ of the first prover having nonzero probability, let $\sigma^{x,a}$ denote the normalized conditional state held by the second prover. For every second-prover question $y$, let $W_{x,a,y}$ be the set of accepting answers and let $\{P_y^b:b\in W_{x,a,y}\}$ denote their corresponding mutually orthogonal projectors. By $S$-projectivity,
\begin{equation}
    |W_{x,a,y}|\leq S.
\end{equation}
Define the conditional single-measurement and consecutive-measurement success probabilities by
\begin{align}
    V_{x,a}  & =  \frac{1}{r}  \sum_y \sum_{b\in W_{x,a,y}} \Tr\!\left(P_y^b\sigma^{x,a}\right), \label{eq:app-s-projective-V}\\
    E_{x,a}  &=  \frac{1}{r(r-1)}  \sum_{y\neq y'}  \sum_{\substack{b\in W_{x,a,y}\\  b'\in W_{x,a,y'}}} \Tr\!\left(  P_{y'}^{b'}P_y^b\sigma^{x,a}P_y^bP_{y'}^{b'} \right).
    \label{eq:app-s-projective-E}
\end{align}
Applying the multi-outcome tight CMT of Proposition~\ref{thm:consec_mmt_general} gives
\begin{equation}
    E_{x,a} \geq \frac{r^2}{(r-1)^2S} V_{x,a} \left( \max\left\{0,V_{x,a}-\frac{1}{r}\right\} \right)^2.
    \label{eq:conditional-s-projective}
\end{equation}

Let
\begin{equation}
    f_r(v):= v\left(\max\left\{0,v-\frac{1}{r}\right\}\right)^2.
\end{equation}
The function $f_r$ is convex on $[0,1]$. Indeed, it vanishes on $[0,1/r]$, whereas for $v>1/r$,
\begin{equation}
    f_r''(v)=6v-\frac{4}{r}>0,
\end{equation}
and its first derivative is continuous at $v=1/r$.

Averaging Eq.~\eqref{eq:conditional-s-projective} over $(x,a)$ and applying Jensen's inequality therefore yields
\begin{equation}
    E \geq \frac{r^2}{(r-1)^2S} V \left( \max\left\{0,V-\frac{1}{r}\right\} \right)^2,
    \label{eq:app-averaged-s-projective}
\end{equation}
where $V$ is the winning probability of the original strategy. The sequential measurements appearing in $E$ define a valid strategy for the coupled game $G_{\rm cp}$, and hence
\begin{equation}
    \omega^*(G_{\rm cp})\geq E.
\end{equation}
Taking a sequence of strategies whose winning probabilities approach $\omega^*(G)$ proves
\begin{equation}
    \omega^*(G_{\rm cp}) \geq \frac{r^2}{(r-1)^2S} \omega^*(G) \left( \max\left\{0,\omega^*(G)-\frac{1}{r}\right\} \right)^2.
\end{equation}

For a binary-challenge game, $r=2$. Suppose, in addition, that a protocol-specific argument establishes
\begin{equation}
    \omega^*(G_{\rm cp})\leq\eta.
    \label{eq:app-generic-eta}
\end{equation}
Equation~\eqref{eq:s-projective-coupled} then gives the cubic inequality
\begin{equation}
    4\omega^*(G) \left( \max\left\{0,\omega^*(G)-\frac{1}{2}\right\} \right)^2 \leq S\eta.
    \label{eq:app-binary-cubic}
\end{equation}
Analytically inverting this inequality, in the same manner as in Proposition~\ref{thm:chsh_upper}, yields
\begin{equation}
 \omega^*(G) \leq \Omega(S,\eta) := \frac{1}{6} \left[ 2+\mathcal{C}\!\left(54S\eta-2\right) \right],
\end{equation}
where $\mathcal{C}$ is defined in Eq.~\eqref{eqn:defofCdelta}. This result is conditional on the independent upper bound in Eq.~\eqref{eq:app-generic-eta}; such a bound does not follow from $S$-projectivity alone.

More generally, for a target soundness error
\begin{equation}
    \omega^*(G)\leq\frac{1}{2}+\epsilon,
\end{equation}
Eq.~\eqref{eq:app-binary-cubic} shows that it is sufficient to have
\begin{equation}
    S\eta \leq 4\left(\frac{1}{2}+\epsilon\right)\epsilon^2.
    \label{eq:app-generic-target}
\end{equation}
We next determine $S$ and $\eta$ separately for each of the three RZKP soundness games.

\subsubsection{Hamiltonian-Cycle protocol}

We first consider the relativistic zero-knowledge proof for the Hamiltonian-Cycle problem introduced by Chailloux and Leverrier~\cite{chailloux2017relativistic}. Let $H$ be an $n$-vertex graph that contains no Hamiltonian cycle. The quantum soundness error is the maximum acceptance probability of entangled cheating provers in the corresponding two-prover one-round game, which we denote by $G_{\rm HAM}$.

The verifier sends a uniformly random binary challenge to the second prover. For fixed questions and a fixed answer of the first prover, an accepting answer to the second challenge is determined by a permutation of the $n$ vertices. Consequently, the soundness game is $n!$-projective:
\begin{equation}
    S_{\rm HAM}=n!.
    \label{eq:app-hc-projectivity}
\end{equation}

In the coupled game, the second prover must provide accepting answers to both possible challenges while remaining consistent with the same first-prover answer. For a no-instance, two such accepting answers determine the uniformly random field element used in the first prover's commitment. Since the provers cannot communicate after receiving their questions, the no-signaling principle implies
\begin{equation}
    \omega^*\!\left(G_{{\rm HAM},{\rm cp}}\right) \leq\frac{1}{q},
    \label{eq:app-hc-coupled-upper}
\end{equation}
where $q$ is the order of the finite field used by the protocol.

Substituting $S=n!$ and $\eta=1/q$ into Eq.~\eqref{eq:s-projective-analytic} gives
\begin{equation}
    \omega^*\!\left(G_{\rm HAM}(H)\right) \leq \frac{1}{6} \left[ 2+\mathcal{C}\!\left(\frac{54n!}{q}-2\right) \right].
    \label{eq:app-hc-exact}
\end{equation}
Thus, the quantum soundness error is at most $1/2+2^{-k}$ whenever
\begin{equation}
    q  \geq \frac{n!\,2^{2k-1}}{1+2^{1-k}},
    \label{eq:app-hc-field-order}
\end{equation}
with $q$ chosen as an admissible prime-power field order. This proves Proposition~\ref{prop:rzko-hamiltonian}.

\subsubsection{Homomorphic-commitment protocols}

We next consider the relativistic zero-knowledge protocols of Cr\'epeau and Stuart~\cite{crepeau_2023}. Their commitment primitive has the algebraic form
\begin{equation}
    w=a\cdot b+c,
    \label{eq:app-homomorphic-commitment}
\end{equation}
where $a\in\mathbb{F}_q$ is uniformly random and held by the verifier, whereas $b$ and $c$ encode the prover's committed data. In each protocol, the verifier chooses a uniformly random binary challenge.

If the second prover can answer both challenges consistently, the two answers can be combined with Eq.~\eqref{eq:app-homomorphic-commitment} to recover the uniformly random element $a$. The no-signaling principle therefore gives, separately for both protocols,
\begin{equation}
    \omega^*(G_{\rm cp})\leq\frac{1}{q}.
    \label{eq:app-homomorphic-coupled-upper}
\end{equation}
The remaining parameter $S$ depends on the underlying NP relation.

\paragraph{Subset Sum.}

Consider a Subset Sum instance with $n$ elements that is a no-instance. For fixed questions and a fixed first-prover answer, an accepting second-prover answer is determined once a binary selection vector in $\{0,1\}^n$ is fixed. Hence the soundness game $G_{\rm SS}$ is $2^n$-projective:
\begin{equation}
    S_{\rm SS}=2^n.
    \label{eq:app-ss-projectivity}
\end{equation}
Combining this value with Eq.~\eqref{eq:app-homomorphic-coupled-upper} and applying
Eq.~\eqref{eq:s-projective-analytic} yields 
\begin{equation}
    \omega^*(G_{\rm SS}) \leq \frac{1}{6} \left[ 2+\mathcal{C}\!\left(\frac{54\cdot2^n}{q}-2\right) \right].
    \label{eq:app-ss-exact}
\end{equation}
Consequently, the quantum soundness error is at most $1/2+2^{-k}$ whenever
\begin{equation}
    q \geq \frac{2^{n+2k-1}}{1+2^{1-k}}.
    \label{eq:app-ss-field-order}
\end{equation}
This proves Proposition~\ref{prop:rzko-subset-sum}.

\paragraph{$3\mathrm{SAT}$.}

Consider a $3\mathrm{SAT}$ formula with $m$ clauses that is unsatisfiable. For each clause, an accepting response can select one of at most three literals. Thus, after fixing the questions and the first-prover answer, there are at most $3^m$ accepting answers for the second prover. The corresponding soundness game $G_{3{\rm SAT}}$ is therefore $3^m$-projective:
\begin{equation}
    S_{3{\rm SAT}}=3^m.
    \label{eq:app-3sat-projectivity}
\end{equation}
Together with Eq.~\eqref{eq:app-homomorphic-coupled-upper}, the analytic CMT bound gives
\begin{equation}
    \omega^*(G_{3{\rm SAT}})  \leq  \frac{1}{6}  \left[ 2+\mathcal{C}\!\left(\frac{54\cdot3^m}{q}-2\right) \right].
    \label{eq:app-3sat-exact}
\end{equation}
It follows that the quantum soundness error is at most $1/2+2^{-k}$ whenever
\begin{equation}
    q  \geq \frac{3^m2^{2k-1}}{1+2^{1-k}}.
    \label{eq:app-3sat-field-order}
\end{equation}
This proves Proposition~\ref{prop:rzko-3sat}.

\section{Robust CMT in fidelity}\label{sec:FRCMT}
\begin{reptheorem}{thm:RCMT_F}
    Given two projections $P_0$ and $P_1$, and two quantum states $\sigma_0$ and $\sigma_1$ in some finite-dimensional Hilbert space $\H$, define 
    \begin{equation}\tag{\ref{eqn:VEdelta}}
    \begin{aligned}
        V & = \frac{1}{2}\br{\Tr(P_0 \sigma_0) +  \Tr(P_1 \sigma_1)}, \\
        E & = \frac{1}{2}\br{\Tr(P_1P_0 \sigma_0 P_0 P_1) + \Tr(P_0P_1 \sigma_1 P_1 P_0)},
    \end{aligned}  
    \end{equation}
    Let $F=\|\sqrt{\sigma_0}\sqrt{\sigma_1}\|_1$  denote the fidelity. 
    Then 
    \begin{equation}
        \tag{\ref{eqn:rcmt_fidelity}}
        \begin{aligned}
            E \geq  V \left(\max\left\{0,(2V-1)F - 2\sqrt{V(1-V)}\sqrt{1-F^2}\right\}\right)^2.
        \end{aligned} 
    \end{equation}
    Furthermore, for any $v,f\in[0,1]$, there exist states $\sigma_0$, $\sigma_1$ and projectors $P_0$, $P_1$ such that $F=f$, $V=v$ and Eq.~\eqref{eqn:rcmt_fidelity} holds with equality. 
\end{reptheorem}

\begin{proof}[Proof of Theorem~\ref{thm:RCMT_F}]
    We will first prove the lower bound. 
    
    If $2V-1<\sqrt{1-F^2}$, $\mathrm{RHS}=0$, the conclusion trivially holds. From now on we assume that $2V-1\geq\sqrt{1-F^2}$. 

    According to Uhlmann's theorem~\cite[Theorem 3.22]{Watrous_2018}, we can find purifications $\psi_0=\ketbra{\psi_0}{\psi_0}$ and $\psi_1=\ketbra{\psi_1}{\psi_1}$ of states $\sigma_0$ and $\sigma_1$ in some space $\H\otimes\E$, respectively, such that
    \begin{equation}
        F(\psi_0,\psi_1) = F(\sigma_0,\sigma_1). 
    \end{equation}
    We have
    \begin{equation}
        V(\psi_0,\psi_1,P_0\otimes\id_\E,P_1\otimes \id_\E) = V(\sigma_0,\sigma_1,P_0,P_1) ,
    \end{equation}
    and 
    \begin{equation}
        E(\psi_0,\psi_1,P_0\otimes\id_\E,P_1\otimes \id_\E) = E(\sigma_0,\sigma_1,P_0,P_1). 
    \end{equation}
    Define 
    \begin{equation}
        \ket{\phi_0}:=\frac{(P_0\otimes\id_\E)\ket{\psi_0}}{\norm{(P_0\otimes\id_\E)\ket{\psi_0}}},\quad
        \ket{\phi_1}:=\frac{(P_1\otimes\id_\E)\ket{\psi_1}}{\norm{(P_1\otimes\id_\E)\ket{\psi_1}}}.
    \end{equation}
    Then by a similar argument of Lemma \ref{lem:VEvec}, we have
    \begin{align}
        V(\sigma_0,\sigma_1,P_0,P_1) = \frac{1}{2} (|\braket{\phi_0}{\psi_0}|^2 + |\braket{\phi_1}{\psi_1}|^2),
    \end{align}
    and 
    \begin{align}
        E(\sigma_0,\sigma_1,P_0,P_1) \geq \frac{1}{2} |\braket{\phi_0}{\phi_1}|^2 (|\braket{\phi_0}{\psi_0}|^2 + |\braket{\phi_1}{\psi_1}|^2). 
    \end{align}
    We can prove the theorem by proving the following inequality instead 
    \begin{align}
        |\braket{\phi_0}{\phi_1}| \geq (2V-1) F - \sqrt{1-(2V-1)^2} \sqrt{1-F^2}. 
    \end{align}
    We set $|\braket{\phi_0}{\phi_1}| = \cos\theta$, $|\braket{\phi_0}{\psi_0}| = \cos\varphi_0$, $|\braket{\phi_1}{\psi_1}| = \cos\varphi_1$ and $|\braket{\psi_0}{\psi_1}| = \cos\tau$, where $\theta$, $\varphi_0$, $\varphi_1$ and $\tau$ are all in $[0,\frac{\pi}{2}]$. Then $2V -1 = \cos^2\varphi_0+\cos^2 \varphi_1 -1=\cos(\varphi_0 + \varphi_1)\cos (\varphi_0-\varphi_1)$ and $F = \cos\tau$. Substituting into the above inequality, it suffices to show 
    \begin{equation}
    \begin{aligned}
        \cos\theta  & \geq \cos(\varphi_0 + \varphi_1) \cos (\varphi_0-\varphi_1) \cos\tau \\
        & \quad - \sqrt{1-\cos^2(\varphi_0 + \varphi_1) \cos^2 (\varphi_0-\varphi_1)} \sin\tau,
    \end{aligned}
    \end{equation}
    Let $\cos \varphi = \cos (\varphi_0+\varphi_1) \cos (\varphi_0-\varphi_1)$ where $\varphi\in [0,\frac{\pi}{2}]$. It suffices to prove 
    \begin{align}
        \cos\theta \geq \cos\varphi\cos\tau - \sin\varphi \sin\tau  = \cos (\varphi+\tau). 
    \end{align}
    Using the monotonicity of cosine function in $[0,\pi]$, it suffices to prove
    \begin{align}
        \theta \leq \varphi + \tau.
    \end{align}
    Recall that $\varphi, \varphi_0, \varphi_1 \in [0,\frac{\pi}{2}]$ as is secured by $2V-1\geq \sqrt{1-F^2}$. Therefore, 
    \begin{align}
        \cos \varphi \leq \cos(\varphi_0+\varphi_1), 
    \end{align}
    which means 
    \begin{align}
        \varphi \geq \varphi_0+\varphi_1.
    \end{align}
    It suffices for us to prove 
    \begin{align}
        \theta \leq \varphi_0 + \varphi_1 + \tau.
    \end{align}
    Indeed, this is true, since $d(\psi,\phi) = \arccos |\braket{\psi}{\phi}|$ defines a metric called the Bures angle (or the angle in~\cite[Section 9.2.2]{Nielsen_2010}) and satisfies triangle inequality, 
    \begin{equation}
    \begin{aligned}
        \theta & = d(\phi_0,\phi_1) \leq d(\phi_0,\psi_1) + d(\psi_1,\phi_1)  \\ 
        & \leq d(\phi_0,\psi_0) + d(\psi_0,\psi_1) + d(\psi_1,\phi_1) \\
        & = \varphi_0 + \varphi_1 + \tau. 
    \end{aligned}
    \end{equation}

    We will now prove the tightness. 

    Let $\alpha = \arccos \sqrt{v}$ and $\beta = \frac{1}{2}\arccos f$ where $\alpha\in [0,\frac{\pi}{2}]$ and $\beta\in [0,\frac{\pi}{4}]$. We first consider the case where $2v-1 > \sqrt{1-f^2}$, which corresponds to $\alpha + \beta < \frac{\pi}{4}$. We define
    \begin{equation}
            \begin{aligned}
        \ket{\psi_0} & = (\cos \beta , \sin \beta), \\
        \ket{\psi_1} & = (\cos \beta , -\sin \beta), \\
        \ket{\phi_0} & = (\cos (\alpha + \beta) , \sin(\alpha + \beta)), \\
        \ket{\phi_1} & = (\cos (\alpha + \beta) , -\sin(\alpha + \beta)), 
    \end{aligned}
    \end{equation}
    We set $\sigma_0 = \ketbra{\psi_0}{\psi_0}$, $\sigma_1 = \ketbra{\psi_1}{\psi_1}$, $P_0 = \ketbra{\phi_0}{\phi_0}$ and $P_1 = \ketbra{\phi_1}{\phi_1}$. Then 
    \begin{equation}
       \begin{aligned}
        F & = |\braket{\psi_0}{\psi_1}| = \cos 2\beta = f, \\
        V & = \frac{1}{2} (\Tr(P_0\sigma_0) + \Tr(P_1\sigma_1)) \\
        & = \frac{1}{2} (|\braket{\phi_0}{\psi_0}|^2
        + |\braket{\phi_1}{\psi_1}|^2) = \cos^2\alpha = v, 
        \end{aligned} 
    \end{equation}
    and
    \begin{equation}
        \begin{aligned}
        E & = \frac{1}{2} (\Tr(P_0 P_1 P_0 \sigma_0) + \Tr(P_1 P_0 P_1 \sigma_1) )\\ &= \frac{1}{2} |\braket{\phi_0}{\phi_1}|^2 ( |\braket{\phi_0}{\psi_0}|^2 +  |\braket{\phi_1}{\psi_1}|^2) \\
        & = \cos^2 \alpha \cos^2 (2\alpha + 2\beta) \\ & =  v (\cos 2\alpha \cos 2\beta -\sin2\alpha \sin 2\beta)^2 \\&= v \br{(2v-1)f - \sqrt{1-(2v-1)^2} \sqrt{1-f^2}}^2. 
        \end{aligned}
    \end{equation}
    We now consider the case where $2v-1 \leq \sqrt{1-f^2}$, which corresponds to $\alpha + \beta \geq \frac{\pi}{4}$. We define 
    \begin{equation}
        \begin{aligned}
        \ket{\psi_0} & = (\cos\alpha, t , \sqrt{\sin^2\alpha - t^2}), \\
        \ket{\psi_1} & = ( t, \cos\alpha,\sqrt{\sin^2\alpha - t^2} ), \\
        \ket{\phi_0} & = (1,0,0), \\
        \ket{\phi_1} & = (0,1,0), 
        \end{aligned}  
    \end{equation}
    where $t = \cos\alpha -\sqrt{2} \sin \beta$. The above states are eligible if $|t| \leq \sin \alpha$, which exactly corresponds to $\alpha + \beta \geq \frac{\pi}{4}$. We set $\sigma_0 = \ketbra{\psi_0}{\psi_0}$, $\sigma_1 = \ketbra{\psi_1}{\psi_1}$, $P_0 = \ketbra{\phi_0}{\phi_0}$ and $P_1 = \ketbra{\phi_1}{\phi_1}$. Then 
    \begin{equation}
        \begin{aligned}
        F & = |\braket{\psi_0}{\psi_1}| = 2t\cos\alpha +\sin^2\alpha - t^2 \\
        & = 1-2\sin^2 \beta = \cos 2\beta = f, \\
        V & = \frac{1}{2} (\Tr(P_0\sigma_0) + \Tr(P_1\sigma_1)) \\
        & = \frac{1}{2} (|\braket{\phi_0}{\psi_0}|^2
        + |\braket{\phi_1}{\psi_1}|^2) = \cos^2\alpha = v, 
        \end{aligned}   
    \end{equation}    
    and 
    \begin{equation}
    \begin{aligned}
        E & = \frac{1}{2} (\Tr(P_0 P_1 P_0 \sigma_0) + \Tr(P_1 P_0 P_1 \sigma_1) ) \\
        & = \frac{1}{2} |\braket{\phi_0}{\phi_1}|^2 ( |\braket{\phi_0}{\psi_0}|^2 +  |\braket{\phi_1}{\psi_1}|^2) = 0.
    \end{aligned}   
    \end{equation}
    Combining both cases, we conclude that for any $v,f\in[0,1]$, there exist states $\sigma_0$, $\sigma_1$ and projectors $P_0$, $P_1$ in a finite-dimensional Hilbert space $\H$ such that $F = f$, $V = v$ and 
    \begin{equation}
        E = v \br{\max\left\{ 0, (2v-1) f - 2\sqrt{v(1-v)}\sqrt{1-f^2}\right\}}^2, 
    \end{equation}
    which is exactly our statement. 
\end{proof}

\section{Lower bound for oblivious transfer} 
\label{apd:ot}

\begin{reptheorem}{thm:bound_RCMT}[No-go theorem for QOT from robust CMT]
    The security parameters $\epsilon_A$, $\epsilon_B$ and $\delta\leq \frac{1}{2}$ for quantum oblivious transfer satisfy  
    \begin{multline}
         \frac{1}{2} + \epsilon_B \geq (1-\delta)\cdot\\
         \left(\max\left\lbrace 0,(1-2\delta)(1-2\epsilon_A)-4\sqrt{\delta(1-\delta)\epsilon_A(1-\epsilon_A)} \right\rbrace \right)^2. 
    \end{multline}
\end{reptheorem}

We will construct cheating strategies for Alice and Bob in the proof. Bob's cheating strategy involves performing two consecutive measurements in the last step to try to learn Alice's input. Alice can again input a well-chosen superposition at the beginning and measure at the end to try to learn Bob's output.

\begin{proof}[Proof of Theorem~\ref{thm:bound_RCMT}]
    Consider an oblivious transfer protocol with $\delta$-completeness as in Protocol~\ref{prot:qot}. After $N$ iterations, Alice and Bob's joint state is denoted by 
    \begin{multline}
        \ket{\psi^{x_0x_1,b}}_{BMA}=\\
        V^{b,N}_{BM} U^{x_0x_1,N}_{MA}\dots V^{b,1}_{BM} U^{x_0x_1,1}_{MA}\circ\br{\ket{0}_A\otimes \ket{\rho^b}_{BM}}.
    \end{multline}
    Bob's reduced state is denoted by
    \begin{equation}
        \sigma^{x_0x_1,b}_{BM}=\Tr_A\psi^{x_0x_1,b}_{BMA},
    \end{equation}
    where $\psi^{x_0x_1,b}_{BMA}=\proj{\psi^{x_0x_1,b}_{BMA}}$. 
    The $\delta$-completeness requires that for any $(x_0,x_1)\in\{0,1\}^2$ and $b\in\{0,1\}$,      
    \begin{equation}
    \begin{aligned}
        & \Pr[\text{Bob accepts}]=\Tr\br{Q_{\mathrm{Acc}}^b\sigma^{x_0x_1,b}_{BM}}=1, \\
        & \Pr[\hat{x}=x_b]=\Tr\br{Q_{x_b}^b\sigma^{x_0x_1,b}_{BM}}\geq1-\delta.
    \end{aligned}
    \end{equation}

    We first discuss the soundness against Alice. Let Bob input $b$ uniformly at random. Alice can cheat by preparing a superposition of all possible $(x_0,x_1)$ in an auxiliary system $A'$, and making a measurement on the composite system $AA'$ in the end to try to learn Bob's input. More precisely, instead of following Protocol~\ref{prot:qot}, Alice prepares the pure state $\frac{1}{2}\sum_{x_0,x_1} \ket{x_0,x_1}_{A'}\ket{0}_A$ at Step 1. In Step 5 of each round of communication, Alice applies the unitary 
    \begin{equation}
        \sum_{x_0,x_1\in\lbrace0,1\rbrace}\proj{x_0,x_1}_{A'}\otimes U^{x_0x_1,\ell}_{MA}.
    \end{equation}
    Before Step 9, the joint state of both is the superposition $\ket{\Phi^b}_{A'BMA}=\frac{1}{2} \sum_{x_0,x_1} \ket{x_0,x_1}_{A'} \ket{\psi^{x_0x_1,b}}_{BMA}$. A cheating Alice will not be caught by an honest Bob because $Q_{\mathrm{Acc}}^b\ket{\psi^{x_0x_1,b}}_{BMA} = \ket{\psi^{x_0x_1,b}}_{BMA}$ for all $(x_0,x_1)$ and $b$, hence 
    \begin{equation}
        Q_{\mathrm{Acc}}^b \ket{\Phi^b}_{A'BMA} =  \ket{\Phi^b}_{A'BMA} . 
    \end{equation}
    Alice's reduced density matrix $\rho_{AA'}^{b}$ is 
    \begin{equation}
        \rho_{AA'}^{b} = \Tr_{BM} \Phi^{b}_{A'BMA}. 
    \end{equation}
    Therefore, a cheating Alice can distinguish $\rho_{AA'}^{0}$ and $\rho_{AA'}^{1}$ and thus guess an honest Bob's $b$, and both parties accept with probability at least 
    \begin{equation}
        \Pr[\hat{b}=b]\geq \frac{1}{2}\left(1+\frac{1}{2}\left\|\rho_{AA'}^{0}-\rho_{AA'}^{1}\right\|_1\right). 
    \end{equation}
    Hence, 
    \begin{equation}
        \frac{1}{2} +\epsilon_A \geq \frac{1}{2}\left(1+\frac{1}{2}\left\|\rho_{AA'}^{0}-\rho_{AA'}^{1}\right\|_1\right). 
    \end{equation}
    By Fuchs-van de Graaf inequality, 
    \begin{equation}
        F\left(\rho_{AA'}^{0},\rho_{AA'}^{1}\right)\geq 1 - 2\epsilon_A.
    \end{equation}
    By Uhlmann's theorem, there exists a unitary $W_{BM}$ such that 
    \begin{equation}
        F(\tilde{\Phi}^{0}_{A'BMA}, \tilde{\Phi}^{1}_{A'BMA}) \geq 1-2\epsilon_A, 
    \end{equation}
    where $\tilde{\Phi}^{0}_{A'BMA} = \Phi^{0}_{A'BMA}$ and $\tilde{\Phi}^{1}_{A'BMA} = W_{BM}(\Phi^{1}_{A'BMA})W_{BM}^\dagger$. 
    
    Second we discuss the soundness against Bob. Bob can cheat by honestly following the protocol, making two consecutive measurements on his state in the end to try to learn Alice's input. More precisely, instead of following Protocol~\ref{prot:qot}, Bob inputs $b$ uniformly at random. Before Step 11, if Bob's input is $1$, then he applies $W_{BM}$. 
    After that, Bob's reduced state is 
    \[
    \begin{cases}
        \tilde{\sigma}^{x_0x_1,0}_{BM}=\Tr_A\psi^{x_0x_1,0}_{BMA}&\text{if $b=0$},\\ \tilde{\sigma}^{x_0x_1,1}_{BM}=\Tr_AW_{BM}\psi^{x_0x_1,1}_{BMA}W_{BM}^\dagger&\text{if $b=1$}.
    \end{cases}
    \]
     Then Bob measures $\set{\tilde{Q}^b_0,\tilde{Q}^b_1}$ and subsequently $\set{\tilde{Q}^{\bar{b}}_0,\tilde{Q}^{\bar{b}}_1}$, where $\tilde{Q}^0_{\hat{x}}={Q}^0_{\hat{x}}$ and $\tilde{Q}^1_{\hat{x}}=W_{BM}{Q}^1_{\hat{x}}W_{BM}^\dagger$, respectively. The average probability that an honest Bob can guess an honest Alice's $x_b$ is 
     \begin{align}
     V :=~&\frac{1}{8}\sum_{b,x_0,x_1\in\lbrace0,1\rbrace} \Tr\br{Q_{x_b}^b\sigma^{x_0x_1,b}_{BM}}\\
     =~&\frac{1}{2}\sum_{b,x_0,x_1\in\lbrace0,1\rbrace}\Tr\br{\br{\proj{x_0,x_1}_{A'}\otimes\tilde{Q}^b_{x_b}}\tilde{\Phi}^{b}_{A'BMA}}.
     \end{align}
    Therefore, the average probability that a cheating Bob can guess an honest Alice's $(x_0,x_1)$ and both parties accept is at least 
    \begin{align*}
        E:=~& \frac{1}{8} \sum_{b,x_0,x_1\in\lbrace0,1\rbrace} \Tr\br{\tilde{Q}^b_{x_b}\tilde{Q}^{\bar{b}}_{x_{\bar{b}}}\tilde{Q}^b_{x_b}\sigma^{x_0x_1,b}_{BM}}\\
        =~& \frac{1}{2}\sum_{b,x_0,x_1\in\lbrace0,1\rbrace} \\
        &\Tr\br{\br{\proj{x_0,x_1}_{A'}\otimes\tilde{Q}^b_{x_b}\tilde{Q}^{\bar{b}}_{x_{\bar{b}}}\tilde{Q}^b_{x_b}}\tilde{\Phi}^{b}_{A'BMA} } . 
    \end{align*}
    We now define two PVMs $\set{R^0_0,R^0_1}$ and $\set{R^1_0,R^1_1}$ in the composite system $A'BMA$ such that  
    \begin{equation}\begin{aligned}
        R^0_0 & = \sum_{x_0,x_1\in\lbrace0,1\rbrace} \proj{x_0,x_1}_{A'}\otimes\tilde{Q}^0_{x_0}, \\
        R^0_1 & = \id_{A'BMA} - R^0_0, \\
        R^1_0 & = \sum_{x_0,x_1\in\lbrace0,1\rbrace} \proj{x_0,x_1}_{A'}\otimes\tilde{Q}^1_{x_1}, \\
        R^1_1 & = \id_{A'BMA} - R^1_0. 
    \end{aligned}\end{equation}
    Then 
    \begin{equation}\begin{aligned}
        & V   = \frac{1}{2}\sum_{b\in\lbrace0,1\rbrace}  \Tr( R^b_0\tilde{\Phi}^{b}_{A'BMA}), \\  
        & E  \geq \frac{1}{2} \sum_{b\in\lbrace0,1\rbrace}\Tr( R^b_0R^{\bar{b}}_0R^b_0\tilde{\Phi}^{b}_{A'BMA}).
    \end{aligned}\end{equation}
    Recall that 
    \begin{equation}\begin{aligned}
        & V   \geq 1-\delta, \\
        & E  \leq \frac{1}{2} + \epsilon_B, \\
        & F(\tilde{\Phi}^{0}_{A'BMA}, \tilde{\Phi}^{1}_{A'BMA}) \geq 1-2\epsilon_A, 
    \end{aligned}\end{equation}
    Because the bounding function is monotonically increasing with respect to $V$ and $F$ in our regime of interest, substituting our limits yields the valid lower bound. Then we obtain by Theorem~\ref{thm:RCMT_F} 
    \begin{multline}
        \frac{1}{2} + \epsilon_B \geq(1-\delta)\cdot \\
        \left(\max\left\lbrace0,(1-2\delta)(1-2\epsilon_A)-4\sqrt{\delta(1-\delta)\epsilon_A(1-\epsilon_A)} \right\rbrace\right)^2. 
    \end{multline}
\end{proof}

\section{Robust CMT in trace distance}
\label{apd:RCMT_TD}
We have also proved a trace distance version of robust CMT, i.e. Theorem~\ref{thm:RCMT_TD}. We prove two inequalities in Theorem~\ref{thm:RCMT_TD} in two parts, i.e., Eq.~\eqref{eqn:rcmt_trace_distance} in Part 1 and Eq.~\eqref{eqn:rcmt_trace_tight} in Part 2. 

\subsection{Part 1}
\begin{reptheorem}{thm:RCMT_TD}[Part 1]
    Given $n$ projectors $P_1,\dots,P_n$ and $n$ quantum states $\sigma_1,\dots,\sigma_n$ in a finite-dimensional Hilbert space $\H$. Let 
\begin{equation}
\begin{aligned}
    V & = \frac{1}{n} \sum_{i=1}^n\Tr(P_i\sigma_i),  \\
    E & = \frac{1}{n(n-1)} \sum_{i\neq j} \Tr(P_j P_i \sigma_i P_i P_j) ,\\
    \Delta&=\frac{1}{n(n-1)}\sum_{i<j}\norm{\sigma_i-\sigma_j}_1
\end{aligned}
\end{equation}
It holds that
\begin{align}\tag{\ref{eqn:rcmt_trace_distance}}
    E\geq 4 V \left(\max\left\{0,V-\frac{1}{2}\right\}\right)^2-\Delta.
\end{align}
\end{reptheorem}

We will first prove Theorem \ref{thm:RCMT_TD} for the case $n=2$, stated as a separate lemma below.
\begin{lemma}\label{lem:RCMT_TD2}
Given two projections $P_0$ and $P_1$, and two quantum states $\sigma_0$ and $\sigma_1$ in some finite-dimensional Hilbert space $\H$, define 
\begin{equation}
    \begin{aligned}
        \Delta(\sigma_0,\sigma_1) &=\frac{1}{2}\norm{\sigma_0-\sigma_1}_1,\\
        V(P_0,P_1,\sigma_0,\sigma_1) & = \frac{1}{2}\br{\Tr(P_0 \sigma_0) +  \Tr(P_1 \sigma_1) }, \\
        E(P_0,P_1,\sigma_0,\sigma_1) & = \frac{1}{2}\br{\Tr(P_0P_1P_0 \sigma_0) + \Tr(P_1P_0P_1 \sigma_1)}.
    \end{aligned} 
\end{equation}
Then we have
\begin{equation}
    E \geq 4V\left(\max\left\{0,V-\frac{1}{2}\right\}\right)^2 - \Delta.
\end{equation}
\end{lemma}
\begin{remark}
Suppose there exist quantum states $\tilde{\sigma}_0$ and $\tilde{\sigma}_1$ such that 
\begin{equation}
     \begin{aligned}
        \Delta(\tilde{\sigma}_0,\tilde{\sigma}_1) & \leq \Delta(\sigma_0,\sigma_1), \\
        V(P_0,P_1,\tilde{\sigma}_0,\tilde{\sigma}_1) & = V(P_0,P_1,\sigma_0,\sigma_1), \\
        E(P_0,P_1,\tilde{\sigma}_0,\tilde{\sigma}_1) & = E(P_0,P_1,\sigma_0,\sigma_1). 
    \end{aligned}   
\end{equation}
If Lemma~\ref{lem:RCMT_TD2} holds for $\tilde{\sigma}_0$ and $\tilde{\sigma}_1$, then Lemma~\ref{lem:RCMT_TD2} naturally holds for $\sigma_0$ and $\sigma_1$. 
\end{remark}

We will use Jordan's lemma~\cite{Jordan_1875} to prove Lemma \ref{lem:RCMT_TD2}. 

\begin{lemma}[Jordan's lemma]
    Let $P_0$ and $P_1$ be projectors on $\H$. There exists an orthogonal decomposition of $\H$ into mutually orthogonal subspaces
    \begin{equation}\label{eqn:Hdecomposition}
        \H=\bigoplus_{j=1}^n\H_j
    \end{equation}
    for some positive integer $n$, such that each $\H_j$ is either one-dimensional or two-dimensional, and both $P_0$ and $P_1$ leave each $\H_j$ invariant. In addition, whenever $\dim \H_j = 2$, both $P_0|_{\H_j}$ and $P_1|_{\H_j}$ are projectors of rank exactly one.
\end{lemma}

Using Jordan's lemma, we may assume that $P_0$ and $P_1$ are block-diagonalized with respect to the decomposition; we may further assume without loss of generality that $\sigma_0$ and $\sigma_1$ are also block-diagonal with respect to the decomposition, since applying the pinching map
\begin{equation}
    \mathcal{E}(\rho)=\sum_{j=1}^n\Pi_j\rho\Pi_j,
\end{equation}
where $\Pi_j$ is the projector onto $\H_j$, does not change $V$ and $E$, and does not increase $\Delta$. 

In fact, we may further assume that $\dim \H_j = 2$ for all $j\in[n]$, as shown in the following lemma:
\begin{lemma}\label{lem:all2Dim}
    There exists an extension Hilbert space $\widetilde{\H}$ containing $\H$:
    \begin{equation}
    \widetilde{\H}=\bigoplus_{j=1}^n\widetilde{\H}_j
\end{equation}
where each $\widetilde{\H}_j$ is a two-dimensional subspace satisfying $\H_j \subseteq \widetilde{\H}_j$ for all $j \in [n]$, together with extended projectors $\widetilde{P}_0$, $\widetilde{P}_1$ and extended quantum states $\widetilde{\sigma}_0$, $\widetilde{\sigma}_1$ in $\widetilde{\H}$, such that 
\begin{enumerate}
    \item each $\widetilde{\H}_j$ is invariant under $\widetilde{P}_0$, $\widetilde{P}_1$, $\widetilde{\sigma}_0$ and $\widetilde{\sigma}_1$,
    \item the restrictions $\widetilde{P}_0|_{\widetilde{\H}_j}$ and $\widetilde{P}_1|_{\widetilde{\H}_j}$ are projectors of rank one for every $j \in [n]$,
    \item and the following equalities hold:
    \begin{equation}
        \begin{aligned}
        V(P_0,P_1,\sigma_0,\sigma_1) & = V(\widetilde{P}_0,\widetilde{P}_1,\widetilde{\sigma}_0,\widetilde{\sigma}_1), \\
        E(P_0,P_1,\sigma_0,\sigma_1) & = E(\widetilde{P}_0,\widetilde{P}_1,\widetilde{\sigma}_0,\widetilde{\sigma}_1),\\
        \Delta(\sigma_0,\sigma_1) &=\Delta(\widetilde{\sigma}_0,\widetilde{\sigma}_1).
    \end{aligned}
    \end{equation}
    \end{enumerate}
\end{lemma}

\begin{proof}[Proof of Lemma \ref{lem:all2Dim}]
    For each $j\in[n]$, if $\dim\H_j=2$, then let $\widetilde{\H}_j=\H_j,$ $\widetilde{P}_i|_{\widetilde{\H}_j} = P_i|_{\H_j}$ and $\widetilde{\sigma}_i|_{\widetilde{\H}_j} = \sigma_i|_{\H_j}$ for $i\in\{0,1\}$. Otherwise $\dim\H_j=1$, let $\widetilde{\H}_j = \H_j \oplus \mathbb{C}$, $\widetilde{\sigma}_i|_{\widetilde{\H}_j} = \sigma_i|_{\H_j}\oplus0$, and define $\widetilde{P}_i$ by:
    \begin{itemize}
        \item If $P_i|_{\H_j} = 0$, define $\widetilde{P}_i|_{\widetilde{\H}_j} = 0 \oplus 1$,
        \item If $P_i|_{\H_j} = 1$, define $\widetilde{P}_i|_{\widetilde{\H}_j} = 1 \oplus 0$.
    \end{itemize}
    It is easy to verify that all the items are satisfied.
\end{proof}

From now on, we consider the case where $P_0$, $P_1$, $\sigma_0$ and $\sigma_1$ are all block-diagonal with respect to the orthogonal decomposition of $\H$ in Eq.~\eqref{eqn:Hdecomposition}, each $\H_j$ is of dimension 2, and $\restoj{P_0},\restoj{P_1}$ are both projectors of rank 1. 
The following lemma indicates that we can assume that $\Tr\br{\restoj{\sigma_0}}=\Tr\br{\restoj{\sigma_1}}$ for all $j\in[n]$ without loss of generality:

\begin{lemma}\label{lem:sameBlockTrace}
    There exist quantum states $\widetilde{\sigma}_0$ and $\widetilde{\sigma}_1$ which are block-diagonal with respect to the orthogonal decomposition of $\H$ in Eq.~\eqref{eqn:Hdecomposition}, such that $\Tr\br{\restoj{\widetilde{\sigma}_0}}=\Tr\br{\restoj{\widetilde{\sigma}_1}}$ for all $j\in[n]$, and that
\begin{equation}
    \begin{aligned}
        V(P_0,P_1,\widetilde{\sigma}_0,\widetilde{\sigma}_1) & = V(P_0,P_1,\sigma_0,\sigma_1), \\
        E(P_0,P_1,\widetilde{\sigma}_0,\widetilde{\sigma}_1) & = E(P_0,P_1,\sigma_0,\sigma_1),\\
        \Delta(\widetilde{\sigma}_0,\widetilde{\sigma}_1) &\leq\Delta(\sigma_0,\sigma_1).
    \end{aligned}   
\end{equation}

\end{lemma}

\begin{proof}[Proof of Lemma \ref{lem:sameBlockTrace}]
We use the Bloch-sphere representation of $\restoj{P_i}$ and $\restoj{\sigma_i}$ for $i\in\{0,1\}$:
\begin{equation}
    \restoj{P_i}=\frac{\id+\mathbf{r}_i^j\cdot\vec{\sigma}}{2},\quad\restoj{\sigma_i}=p_i^j\cdot\frac{\id+\mathbf{s}_i^j\cdot\vec{\sigma}}{2},
\end{equation}
where $\mathbf{r}_i^j$ and $\mathbf{s}_i^j$ are 3-dimensional real vectors with Euclidean norms $\|\mathbf{r}_i^j\|=1,$ $\|\mathbf{s}_i^j\|\leq1,$ $\vec{\sigma}=(X,Y,Z)$ is the vector of Pauli matrices, and $p_i^j=\Tr\br{\restoj{\sigma_i}}$.

Now we can explicitly write down $V$, $E$ and $\Delta$ in the Bloch-sphere representation, i.e., 
\begin{align*}
    V(P_0,P_1,\sigma_0,\sigma_1) &= \frac{1}{4} \br{2+\sum_j\br{ p_0^j \mathbf{r}_0^j\cdot \mathbf{s}_0^j + p_1^j \mathbf{r}_1^j \cdot \mathbf{s}_1^j}},\\
    E(P_0,P_1,\sigma_0,\sigma_1) &=  \frac{1}{8}\sum_j(1+\mathbf{r}_0^j\cdot \mathbf{r}_1^j)\cdot\\
    &\qquad\quad(p_0^j+p_1^j+ p_0^j \mathbf{r}_0^j\cdot \mathbf{s}_0^j + p_1^j \mathbf{r}_1^j \cdot \mathbf{s}_1^j ),\\
    \Delta(\sigma_0,\sigma_1) &= \frac{1}{2}\sum_j \max \lbrace |p_0^j-p_1^j|, \|p_0^j\mathbf{s}_0^j-p_1^j\mathbf{s}_1^j\|\rbrace.
\end{align*}
We consider the symmetric vector $\widehat{\mathbf{s}}_i^j$ of $\mathbf{s}_i^j$ with respect to $\mathbf{r}_0^j+\mathbf{r}_1^j$, i.e.,
\begin{equation}\label{eqn:hatSdef}
    \widehat{\mathbf{s}}_i^j=\frac{2\mathbf{s}_i^j\cdot\br{\mathbf{r}_0^j+\mathbf{r}_1^j}}{\|\mathbf{r}_0^j+\mathbf{r}_1^j\|^2}\br{\mathbf{r}_0^j+\mathbf{r}_1^j}-\mathbf{s}_i^j.
\end{equation}
Then we have
\begin{equation}\label{eqn:rsSym}
    \mathbf{r}_1^j\cdot\widehat{\mathbf{s}}_0^j=\mathbf{r}_0^j\cdot \mathbf{s}_0^j,\quad \mathbf{r}_0^j\cdot\widehat{\mathbf{s}}_1^j=\mathbf{r}_1^j\cdot \mathbf{s}_1^j.
\end{equation}
We also have 
\begin{equation}\label{eqn:equalNorm}
    \|p_0^j\mathbf{s}_0^j-p_1^j\mathbf{s}_1^j\|=\|p_0^j\widehat{\mathbf{s}}_0^j-p_1^j\widehat{\mathbf{s}}_1^j\|.
\end{equation}
Since the transformation $\mathbf{s} \to \widehat{\mathbf{s}}$ defined in Eq.~\eqref{eqn:hatSdef} is a Householder reflection (which is an orthogonal isometry), it preserves Euclidean distances. Eq.~\eqref{eqn:rsSym} and~\eqref{eqn:equalNorm} are intuitively illustrated by Fig.~\ref{fig:illustration}.

\begin{figure}[t]
    \centering
    \includegraphics[width=0.95\linewidth]{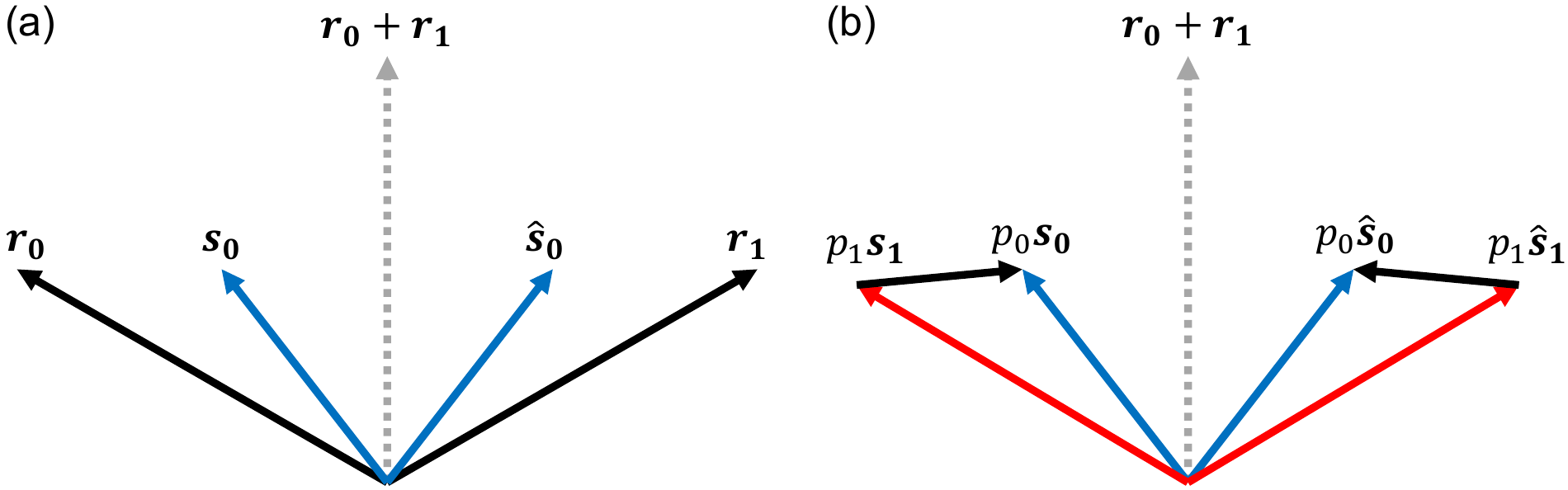}
    \caption{\textbf{Illustration for Eq.~\eqref{eqn:rsSym} and \eqref{eqn:equalNorm}.}  For simplicity, superscripts are omitted. (a) Eq.~\eqref{eqn:rsSym}. Vectors $\mathbf{r}_0$ and $\mathbf{r}_1$ are symmetric with respect to $\mathbf{r}_0 + \mathbf{r}_1$, as are $\mathbf{s}_0$ and $\widehat{\mathbf{s}}_0$, implying $\mathbf{r}_0 \cdot \mathbf{s}_0 = \mathbf{r}_1 \cdot \widehat{\mathbf{s}}_0$. (b) Eq.~\eqref{eqn:equalNorm}. Vectors $p_0\mathbf{s}_0$ and $p_0\widehat{\mathbf{s}}_0$ are symmetric with respect to $\mathbf{r}_0 + \mathbf{r}_1$, as are $p_1\mathbf{s}_1$ and $p_1\widehat{\mathbf{s}}_1$, implying $\norm{p_0\mathbf{s}_0-p_1\mathbf{s}_1}=\norm{p_0\widehat{\mathbf{s}}_0-p_1\widehat{\mathbf{s}}_1}$.}
    \label{fig:illustration} 
\end{figure}

Define
\begin{equation}
    \widetilde{\mathbf{s}}_0^j=\frac{p_0^j\mathbf{s}_0^j+p_1^j\widehat{\mathbf{s}}_1^j}{p_0^j+p_1^j},\quad\widetilde{\mathbf{s}}_1^j=\frac{p_0^j\widehat{\mathbf{s}}_0^j+p_1^j\mathbf{s}_1^j}{p_0^j+p_1^j},\quad\widetilde{p}^j=\frac{p_0^j+p_1^j}{2},
\end{equation}
where $\widetilde{\mathbf{s}}_0^j$ and $\widetilde{\mathbf{s}}_1^j$ are eligible Bloch vectors because \[\|\widetilde{\mathbf{s}}_0^j\|\leq \frac{p_0^j}{p_0^j+p_1^j}\|\mathbf{s}_0^j\|+ \frac{p_1^j}{p_0^j+p_1^j}\|\widehat{\mathbf{s}}_1^j\| \leq 1\quad\mbox{and} \]\[\|\widetilde{\mathbf{s}}_1^j\|\leq \frac{p_0^j}{p_0^j+p_1^j}\|\widehat{\mathbf{s}}_0^j\|+\frac{p_1^j}{p_0^j+p_1^j}\|\mathbf{s}_1^j\| \leq 1.\] Then for all $j\in[n]$ we have
\begin{equation}
 \begin{aligned}
    &\widetilde{p}^j(\mathbf{r}_0^j\cdot \widetilde{\mathbf{s}}_0^j + \mathbf{r}_1^j \cdot \widetilde{\mathbf{s}}_1^j )\\
    =~&\frac{1}{2}\br{ \mathbf{r}_0^j\cdot (p_0^j\mathbf{s}_0^j+p_1^j\widehat{\mathbf{s}}_1^j) + \mathbf{r}_1^j \cdot (p_1^j\mathbf{s}_1^j+p_0^j\widehat{\mathbf{s}}_0^j)}\\
    \overset{(a)}{=}~&\frac{1}{2}\br{ p_0^j \mathbf{r}_0^j\cdot \mathbf{s}_0^j+p_1^j\mathbf{r}_1^j\cdot \mathbf{s}_1^j+ p_1^j \mathbf{r}_1^j \cdot \mathbf{s}_1^j+p_0^j\mathbf{r}_0^j\cdot \mathbf{s}_0^j}\\
    =~& p_0^j \mathbf{r}_0^j\cdot \mathbf{s}_0^j + p_1^j \mathbf{r}_1^j \cdot \mathbf{s}_1^j,\label{eqn:forVEeq}
\end{aligned}   
\end{equation}
where $(a)$ follows from Eq.~\eqref{eqn:rsSym}.
For $i\in\{0,1\},$ define
\begin{equation}
\restoj{\widetilde{\sigma}_i}=\widetilde{p}^j\cdot\frac{\id+\widetilde{\mathbf{s}}_i^j\cdot\vec{\sigma}}{2}.
\end{equation}
Then 
\begin{equation}
    \begin{aligned}
    &V(P_0,P_1,\widetilde{\sigma}_0,\widetilde{\sigma}_1)\\
    =~&\frac{1}{4} \br{2+\sum_j\widetilde{p}^j\br{  \mathbf{r}_0^j\cdot \widetilde{\mathbf{s}}_0^j +  \mathbf{r}_1^j \cdot \widetilde{\mathbf{s}}_1^j}}\\
    \overset{(b)}{=}~&\frac{1}{4} \br{2+\sum_j\br{ p_0^j \mathbf{r}_0^j\cdot \mathbf{s}_0^j + p_1^j \mathbf{r}_1^j \cdot \mathbf{s}_1^j}}\\
    =~&V(P_0,P_1,\sigma_0,\sigma_1),
\end{aligned}
\end{equation}
and
\begin{equation}
  \begin{aligned}
    &E(P_0,P_1,\widetilde{\sigma}_0,\widetilde{\sigma}_1)\\
    =~&  \frac{1}{8}\sum_j(1+\mathbf{r}_0^j\cdot \mathbf{r}_1^j)\widetilde{p}^j(2+ \mathbf{r}_0^j\cdot \widetilde{\mathbf{s}}_0^j + \mathbf{r}_1^j \cdot \widetilde{\mathbf{s}}_1^j )\\
    \overset{(c)}{=}~&\frac{1}{8}\sum_j(1+\mathbf{r}_0^j\cdot \mathbf{r}_1^j)(p_0^j+p_1^j+ p_0^j \mathbf{r}_0^j\cdot \mathbf{s}_0^j + p_1^j \mathbf{r}_1^j \cdot \mathbf{s}_1^j )\\
    =~&E(P_0,P_1,\sigma_0,\sigma_1),
\end{aligned}  
\end{equation}
where $(b)$ and $(c)$ follow from Eq.~\eqref{eqn:forVEeq}, and
\begin{equation}
    \begin{aligned}
    \Delta(\widetilde{\sigma}_0,\widetilde{\sigma}_1) & =\frac{1}{2}\sum_j  \widetilde{p}^j\|\widetilde{\mathbf{s}}_0^j-\widetilde{\mathbf{s}}_1^j\|\\
    & =\frac{1}{2}\sum_j\frac{1}{2}\norm{p_0^j\mathbf{s}_0^j+p_1^j\widehat{\mathbf{s}}_1^j-(p_0^j\widehat{\mathbf{s}}_0^j+p_1^j\mathbf{s}_1^j)}\\
    & \leq\frac{1}{2}\sum_j\frac{1}{2}\br{\norm{p_0^j\mathbf{s}_0^j-p_1^j\mathbf{s}_1^j}+\norm{p_0^j\widehat{\mathbf{s}}_0^j-p_1^j\widehat{\mathbf{s}}_1^j}}\\
    & \overset{(d)}{=}\frac{1}{2}\sum_j\norm{p_0^j\mathbf{s}_0^j-p_1^j\mathbf{s}_1^j}\\
    & \leq \Delta(\sigma_0,\sigma_1),
\end{aligned}
\end{equation}
where $(d)$ follows from Eq.~\eqref{eqn:equalNorm}.
\end{proof}

With this lemma, we prove Lemma~\ref{lem:RCMT_TD2} by analyzing each subspace $\H_j$ individually and then combining the results via a convexity argument. We first establish the result for each $\H_j$ in the following theorem, where the lower bound obtained is in fact tighter.

\begin{proposition}\label{thm:qubit}
    Given two projections $P_0$ and $P_1$ each of rank 1, and two quantum states $\sigma_0$ and $\sigma_1$ in some 2-dimensional Hilbert space $\H$, define 
\begin{equation}
    \begin{aligned}
        V & = \frac{1}{2}\br{\Tr(P_0 \sigma_0) +  \Tr(P_1 \sigma_1) }, \\
        E & = \frac{1}{2}\br{\Tr(P_0P_1P_0 \sigma_0) + \Tr(P_1P_0P_1 \sigma_1)},\\
        \Delta &=\frac{1}{2}\norm{\sigma_0-\sigma_1}_1.
    \end{aligned}
\end{equation}
Then 
    \begin{multline}\label{eqn:idealLowerBound}
        E \geq V \left(\max\left\{0,(2V-1)\sqrt{1-\Delta^2}\right.\right.\\\left.\left.-\sqrt{1-(2V-1)^2}\Delta\right\}\right)^2.
    \end{multline}
\end{proposition}
\begin{proof}
    If $2V-1<\Delta$, $\mathrm{RHS}=0$, the conclusion trivially holds. From now on we assume that $2V-1\geq\Delta$. We use the Bloch-sphere representation:
\begin{equation}
     \begin{aligned}
        P_i&=\frac{\id+\mathbf{r}_i\cdot\vec{\sigma}}{2},\quad \sigma_i=\frac{\id+\mathbf{s}_i\cdot\vec{\sigma}}{2}\\
    \end{aligned}   
\end{equation}
where $\mathbf{r}_i$ and $\mathbf{s}_i$ are 3-dimensional real vectors with Euclidean norms $\norm{\mathbf{r}_i}=1,$ $\norm{\mathbf{s}_i}\leq1,$  for $i\in\{0,1\}$, and $\vec{\sigma}=(X,Y,Z)$ is the vector of Pauli matrices. Then
\begin{equation}
    \begin{aligned}
        \label{eqn:VED_qubit}
        V & = \frac{1}{4}\br{2+\mathbf{r}_0\cdot \mathbf{s}_0 +  \mathbf{r}_1\cdot \mathbf{s}_1 }, \\
        E & = \frac{(1+\mathbf{r}_0\cdot \mathbf{r}_1)}{2}\cdot V,\\
        \Delta &=\frac{1}{2}\norm{\mathbf{s}_0-\mathbf{s}_1},
    \end{aligned}
\end{equation}
    where $\mathbf{r}_0\cdot \mathbf{s}_0$ denotes the inner product of $\mathbf{r}_0$ and $\mathbf{s}_0$. Define 
    \begin{equation}
        m  = \frac{\mathbf{s}_0+\mathbf{s}_1}{2},\quad d=\frac{\mathbf{s}_0-\mathbf{s}_1}{2}, 
    \end{equation}
    then we have $\norm{d}=\Delta$. By the parallelogram law, $\norm{m}^2+\norm{d}^2=\frac{\norm{\mathbf{s}_0}^2+\norm{\mathbf{s}_1}^2}{2}$, we have $\norm{m}\leq\sqrt{1-\Delta^2}$. Therefore,
    \begin{equation}
            \begin{aligned}
        2V-1=~&\frac{1}{2}\br{\mathbf{r}_0\cdot \mathbf{s}_0 +  \mathbf{r}_1\cdot \mathbf{s}_1}\\
        =~&\frac{1}{2}\br{(\mathbf{r}_0+\mathbf{r}_1)m+(\mathbf{r}_0-\mathbf{r}_1)d}\\
        \leq~&\frac{1}{2}\br{\norm{\mathbf{r}_0+\mathbf{r}_1}\norm{m}+\norm{\mathbf{r}_0-\mathbf{r}_1}\norm{d}}\\
        \leq~&\sqrt{\frac{1+\mathbf{r}_0\cdot \mathbf{r}_1}{2}}\sqrt{1-\Delta^2}+\sqrt{\frac{1-\mathbf{r}_0\cdot \mathbf{r}_1}{2}}\Delta
    \end{aligned}
    \end{equation}

    Let $2V-1=\cos\alpha$, $\Delta=\sin\beta$ and $\sqrt{\frac{1+\mathbf{r}_0\cdot \mathbf{r}_1}{2}}=\cos\theta$ for some $\alpha,\beta,\theta\in[0,\pi/2]$. Then $\sqrt{1-(2V-1)^2}=\sin\alpha$, $\sqrt{1-\Delta^2}=\cos\beta$ and $\sqrt{\frac{1-\mathbf{r}_0\cdot \mathbf{r}_1}{2}}=\sin\theta.$ Thus
\begin{equation}
    \cos\alpha\leq\cos\br{\theta-\beta}.
\end{equation}

Since $\alpha\in[0,\pi/2]$, $\theta-\beta\in[-\pi/2,\pi/2]$, we have $\alpha\geq\abs{\theta-\beta}\geq\theta-\beta.$ Since $\alpha+\beta\in[0,\pi]$, we have 
\begin{equation}
    \cos\theta\geq\cos\br{\alpha+\beta}.
\end{equation}

Now we re-express Eq.~\eqref{eqn:VED_qubit} with $\theta$, $\alpha$ and $\beta$ that
    \begin{align*}
        E=~&\frac{(1+\mathbf{r}_0\cdot \mathbf{r}_1)}{2}\cdot V\\
        =~&V\cos^2\theta \\
        \geq~&V\br{\max\{0,\cos\br{\alpha+\beta}\}}^2\\
        =~&V \left(\max\left\{0,(2V-1)\sqrt{1-\Delta^2}\right.\right.\\
        &-\left.\left.\sqrt{1-(2V-1)^2}\Delta\right\}\right)^2. 
    \end{align*}
\end{proof}

The following lemma shows that the lower bound in Proposition \ref{thm:qubit} is tighter than that in Theorem \ref{thm:RCMT_TD}:

\begin{lemma}\label{lem:looseLB}
For any $\Delta,V\in[0,1]$, 
\begin{equation}
\begin{aligned}
    & V\br{\max\left\{0,\br{2V-1}\sqrt{1-\Delta^2}-\sqrt{1-(2V-1)^2}\Delta\right\}}^2 \\
    & \quad\quad\quad \geq V \br{\max\left\{0,2V-1\right\}}^2-\Delta.
\end{aligned}
\end{equation}
\end{lemma}
\begin{proof}
    If $2V-1<\Delta$, $\mathrm{RHS}\leq\max\left\{0,2V-1\right\}-\Delta\leq0=\mathrm{LHS}$ since $V\leq1$ and $2V-1\leq1$. Now we consider the case where $2V-1\geq\Delta$.
 
    Let $2V-1=\cos\alpha$ and $\Delta=\sin\beta$ for some $\alpha\in[0,\pi]$ and $\beta\in[0,\pi/2].$ Then $\sqrt{1-(2V-1)^2}=\sin\alpha$ and $\sqrt{1-\Delta^2}=\cos\beta.$ Thus
    \begin{align*}
        &~\mathrm{LHS}-\mathrm{RHS} \\
        =~ & \frac{1+\cos\alpha}{2}\cdot\cos^2(\alpha+\beta)-\frac{1+\cos\alpha}{2}\cdot\cos^2\alpha+\sin\beta\\
         = ~&\frac{1+\cos\alpha}{2}\cdot\frac{\cos\br{2\alpha+2\beta}-\cos2\alpha}{2}+\sin\beta\\
         =~ &-\frac{1+\cos\alpha}{2}\cdot\sin\br{2\alpha+\beta}\sin\beta+\sin\beta\\
         =~&\sin\beta\br{1-\frac{1+\cos\alpha}{2}\cdot\sin\br{2\alpha+\beta}}\quad\geq0. 
    \end{align*}
\end{proof}

We are now ready to prove Lemma \ref{lem:RCMT_TD2}.
\begin{proof}[Proof of Lemma \ref{lem:RCMT_TD2}]
For each $i\in\{0,1\}$ and $j\in[n]$, denote 
\begin{equation}
   P_i^j=\restoj{P_i},\quad p_i^j=\Tr\restoj{\sigma_i},\quad\sigma_i^j=\frac{\restoj{\sigma_i}}{p_i^j}.
\end{equation}
By Lemma \ref{lem:sameBlockTrace}, we may assume $p_0^j=p_1^j=:p^j$ without loss of generality.
Define
\begin{equation}
    \begin{aligned}
        V_j & = \frac{1}{2}\br{\Tr(P_0^j \sigma_0^j) +  \Tr(P_1^j \sigma_1^j) }, \\
        E_j & = \frac{1}{2}\br{\Tr(P_0^jP_1^jP_0^j \sigma_0^j) + \Tr(P_1^jP_0^jP_1^j \sigma_1^j)},\\
        \Delta_j &=\frac{1}{2}\|\sigma_0^j-\sigma_1^j\|_1.
    \end{aligned}
\end{equation}
Then
    \begin{align}
        V=\sum_{j=1}^np^jV_j,\quad E=\sum_{j=1}^np^jE_j,\quad \Delta=\sum_{j=1}^np^j\Delta_j.
    \end{align}
    By Proposition \ref{thm:qubit}, we have
    \begin{multline}
        E_j \geq  V_j \left(\max\left\{0,(2V_j-1)\sqrt{1-\Delta_j^2}\right.\right.\\
        -  \left.\left.\sqrt{1-(2V_j-1)^2}\Delta_j \right\}\right)^2.
    \end{multline}
    By Lemma \ref{lem:looseLB}, 
    \begin{align}
        E_j \geq  V_j\br{\max\left\{0,2V_j-1\right\}}^2-\Delta_j.
    \end{align}
    Finally, 
    \begin{equation}
        \begin{aligned}
        E & =\sum_{j=1}^np^jE_j\\
        & \geq\sum_{j=1}^np^jV_j \br{\max\left\{0,2V_j-1\right\}}^2-\sum_{j=1}^np^j\Delta_j\\
        & = \sum_{j=1}^np^jV_j\br{\max\left\{0,2V_j-1\right\}}^2-\Delta\\
        & \overset{(a)}{\geq} \br{\sum_{j=1}^np^jV_j}\br{\max\left\{0,2\br{\sum_{j=1}^np^jV_j}-1\right\}}^2-\Delta\\
        & = V \br{\max\left\{0,2V-1\right\}}^2-\Delta,
    \end{aligned}
    \end{equation}
    where $(a)$ follows from the convexity of $f(V)=V\br{\max\left\{0,2V-1\right\}}^2$ (see Lemma~\ref{lem:convexity}).
\end{proof}

\begin{proof}[Proof of Theorem \ref{thm:RCMT_TD}]
    Denote 
    \begin{equation}
\begin{aligned}
    \Delta_{i,j} & =\frac{1}{2}\norm{\sigma_i-\sigma_j}_1, \\
    V_{i,j} & = \frac{1}{2} \br{\Tr(P_i\sigma_i)+\Tr(P_j\sigma_j)},  \\
    E_{i,j} & = \frac{1}{2} \br{\Tr(P_iP_jP_i\sigma_i)+\Tr(P_jP_iP_j\sigma_j)}.
\end{aligned}
\end{equation}
Then
\begin{equation}
\begin{aligned}
    V & = \frac{1}{n(n-1)} \sum_{i\ne j}V_{i,j},  \\
    E & = \frac{1}{n(n-1)} \sum_{i\neq j} E_{i,j} ,\\
    \Delta&=\frac{1}{n(n-1)}\sum_{i\ne j}\Delta_{i,j}.
\end{aligned}
\end{equation}
Denote $f(V)=4V \left(\max\left\{0,V-\frac{1}{2}\right\}\right)^2.$ Then
\begin{equation}
\begin{aligned}
    E & =\frac{1}{n(n-1)} \sum_{i\neq j} E_{i,j} \overset{(*)}{\geq} \frac{1}{n(n-1)} \sum_{i\neq j}\br{f(V_{i,j})-\Delta_{i,j}}\\
    & \overset{(**)}{\geq} f\br{\frac{1}{n(n-1)} \sum_{i\neq j}V_{i,j}}-\Delta =f(V)-\Delta,
\end{aligned}
\end{equation}
where $(*)$ follows from Lemma \ref{lem:RCMT_TD2} and $(**)$ follows from the convexity of $f(V).$
\end{proof}

\subsection{Part 2}
We can also prove a robust CMT in trace distance by perturbation in Theorem~\ref{thm:consec_mmt}, which results in Eq.~\eqref{eqn:rcmt_trace_tight}:
\begin{reptheorem}{thm:RCMT_TD}[Part 2]
    Given $n$ projectors $P_1,\dots,P_n$ and $n$ quantum states $\sigma_1,\dots,\sigma_n$ in a finite-dimensional Hilbert space $\H$. Let 
\begin{equation}
\begin{aligned}
    V & = \frac{1}{n} \sum_{i=1}^n\Tr(P_i\sigma_i),  \\
    E & = \frac{1}{n(n-1)} \sum_{i\neq j} \Tr(P_j P_i \sigma_i P_i P_j) ,\\
    \Delta&=\frac{1}{n(n-1)}\sum_{i<j}\norm{\sigma_i-\sigma_j}_1
\end{aligned}
\end{equation}
It holds that
\begin{align}\label{eqn:CMT_inequality_SI}
    E\geq  \frac{n^2}{(n-1)^2}V \left(\max\left\{0,V-\frac{1}{n}\right\}\right)^2-\frac{4n}{n-1}\Delta.
\end{align}
\end{reptheorem}
\begin{proof}
    Let $\sigma$ be the average state of $\sigma_1,\dots,\sigma_n$, i.e.,
    \[\sigma=\frac{1}{n}\sum_{i=1}^n\sigma_i.\]
    Then
    \[
        V=\frac{1}{n} \sum_{i=1}^n\Tr(P_i\sigma)+\frac{1}{n} \sum_{i=1}^n\Tr(P_i(\sigma_i-\sigma))
        =:V'+\delta V
    \]
    and 
    \begin{multline}
        E=\frac{1}{n(n-1)} \sum_{i\neq j} \Tr(P_j P_i \sigma P_i P_j)\\+\frac{1}{n(n-1)} \sum_{i\neq j} \Tr(P_j P_i (\sigma_i-\sigma) P_i P_j)
        =:E'+\delta E.
    \end{multline}
    We can bound $\delta V$ and $\delta E$ with $\Delta$ as follows:
    \begin{align*}
        \abs{\delta V}\leq~&\frac{1}{n} \sum_{i=1}^n\abs{\Tr(P_i(\sigma_i-\sigma))}\\
        \leq~&\frac{1}{2n} \sum_{i=1}^n\norm{\sigma_i-\sigma}_1\\
        =~&\frac{1}{2n} \sum_{i=1}^n\norm{\frac{1}{n}\sum_{j=1}^n(\sigma_i-\sigma_j)}_1\\
        \leq~&\frac{1}{2n^2} \sum_{i\ne j}\norm{\sigma_i-\sigma_j}_1\quad=\frac{(n-1)}{n}\Delta,
    \end{align*}
    and similarly,
    \begin{align*}
        \abs{\delta E}\leq~&\frac{1}{n} \sum_{i=1}^n\abs{\Tr(P_iP_jP_i(\sigma_i-\sigma))}\\
        \leq~&\frac{1}{2n} \sum_{i=1}^n\norm{\sigma_i-\sigma}_1\quad\leq\frac{(n-1)}{n}\Delta,
    \end{align*}
    Denote \[f(V)=\frac{n^2}{(n-1)^2}V \left(\max\left\{0,V-\frac{1}{n}\right\}\right)^2.\]
    Then 
    \begin{align*}
        E& = E'+\delta E\geq f(V')-\Delta\\
        & = f(V-\delta V)-\Delta = f(V)-f'(\xi)\delta V-\Delta\\
        & =f(V)-\max\left\lbrace 0,\frac{(n\xi-1)(3n\xi-1)}{(n-1)^2} \right\rbrace\delta V-\Delta\\
        & \geq f(V)-\frac{4n}{n-1}\Delta
    \end{align*}
    where in the second line we have used Lagrange's Mean Value Theorem for some $\xi\in[\min\set{V,V'},\max\set{V,V'}]$.
\end{proof}

\section{Quantum Private Query}\label{sec:QPQ}
In this section, we will prove Theorem~\ref{thm:QPQ}.
\begin{reptheorem}{thm:QPQ}
    For any QPQ protocol which is $\delta$-correct, weakly $(\epsilon_A,\delta)$-secure for Bob (where $\delta\leq \frac{1}{2}$), dishonest Bob can retrieve two database entries with probability at least
    \begin{equation}\tag{\ref{eqn:QPQrobust}}
        4(1-\delta)\left(\max\left\lbrace 0,\frac{1}{2}-\delta \right\rbrace\right)^2-4\sqrt{\epsilon_A}. 
    \end{equation}
    In particular, if $n$ is large and $\epsilon_A$ is sufficiently small, we can obtain an even tighter bound
    \begin{equation}\tag{\ref{eqn:QPQtight}}
         \frac{n^2}{(n-1)^2}(1-\delta)\left(\max\left\lbrace0,1-\frac{1}{n}-\delta \right\rbrace\right)^2-\frac{16n}{n-1}\sqrt{\epsilon_A}. 
    \end{equation}
\end{reptheorem}

We will construct cheating strategies for Alice and Bob in the proof. Bob's cheating strategy involves performing two consecutive measurements in the last step to try to learn Alice's input. Alice can again input a well-chosen superposition at the beginning and measure at the end to try to learn Bob's output.

\begin{proof}[Proof of Theorem~\ref{thm:QPQ}]
    Consider a QPQ protocol with $\delta$-completeness as in Protocol~\ref{prot:qpq}. After $N$ iterations, Alice and Bob's joint state is denoted by 
    \[
    \ket{\psi^{x,b}}_{BMA}=V^{b,N}_{BM} U^{x,N}_{MA}\dots V^{b,1}_{BM} U^{x,1}_{MA}\circ\br{\ket{0}_A\otimes \ket{\rho^b}_{BM}}.\]
    Denote $\psi^{x,b}=\proj{\psi^{x,b}}_{BMA}$.
    The $\delta$-completeness requires that for any $x\in[k]^n$ and $b\in[n]$, %both of these should hold: 
    \begin{multline}
        \Pr[\text{Both accept}\wedge (j=x_b)]=\Tr\br{\br{P_{\mathrm{Acc}}^x\otimes Q_{x_b}^b}\psi^{x,b}}\\ 
        \geq 1-\delta.
    \end{multline}

    We first discuss the soundness against Alice. Let Bob input $b$ uniformly at random. Alice can cheat by preparing a superposition of all possible $x$ in an auxiliary system $A'$, and making a measurement on the composite system $AA'$ in the end to try to learn Bob's input. More precisely, instead of following Protocol~\ref{prot:qpq}, Alice prepares the pure state $\frac{1}{k^{n/2}}\sum_{x} \ket{x}_{A'}\ket{0}_A$ at Step 1. In Step 5 of each round of communication, Alice applies the unitary 
    \begin{equation}
        \sum_{x\in[k]^n}\proj{x}_{A'}\otimes U^{x,\ell}_{MA}.
    \end{equation}
    Before Step 9, the joint state of both is the superposition $\ket{\Phi^b}_{A'BMA}=\frac{1}{k^{n/2}} \sum_{x} \ket{x}_{A'} \ket{\psi^{x,b}}_{BMA}$. 
    We denote $\Phi^{b}=\proj{\Phi^{b}}_{A'BMA}$. Then Alice's reduced density matrix $\rho_{AA'}^{b}$ is 
    \begin{equation}
        \rho_{AA'}^{b} = \Tr_{BM} \Phi^{b}. 
    \end{equation}
    
    By Fuchs-van de Graaf inequality, for any $b\in[n]$ and $b>1$, 
    \begin{equation}
        F\left(\rho_{AA'}^{b},\rho_{AA'}^{1}\right)\geq 1 - \frac{1}{2}\norm{\rho_{AA'}^{b}-\rho_{AA'}^{1}}_1\geq1-2\epsilon_A.
    \end{equation}
    By Uhlmann's theorem, there exists a unitary $W_{BM}^{b}$ such that 
    \begin{equation*}
        F(\tilde{\Phi}^{b}, \tilde{\Phi}^{1})=F\left(\rho_{AA'}^{b},\rho_{AA'}^{1}\right) \geq 1-2\epsilon_A, 
    \end{equation*}
    where $\tilde{\Phi}^{b} = W_{BM}^{b}\Phi^{b}W_{BM}^{b,\dagger}$ for $b\in[n]$ and $W_{BM}^{1}=\id$. 

    Again by Fuchs-van de Graaf inequality, we have
    \begin{equation*}
        \frac{1}{2}\norm{\tilde{\Phi}^{b}-\tilde{\Phi}^{1}}_1\leq \sqrt{1-F(\tilde{\Phi}^{b}, \tilde{\Phi}^{1})^2}\leq 2\sqrt{\epsilon_A}. 
    \end{equation*}

    By triangle inequality, for any $b,b'\in[n]$ and $b\ne b'$
    \begin{equation*}
        \frac{1}{2}\norm{\tilde{\Phi}^{b}-\tilde{\Phi}^{b'}}_1\leq 4\sqrt{\epsilon_A}. 
    \end{equation*}
    
    Second we discuss the soundness against Bob. Bob can cheat by honestly following the protocol, making two consecutive measurements on his state in the end to try to learn Alice's input. More precisely, instead of following Protocol~\ref{prot:qpq}, Bob inputs $b$ uniformly at random. Before Step 11, if Bob's input is $b$, then he applies $W_{BM}^b$. 
    Then Bob measures $\set{\tilde{Q}^b_j}_{j\in[k]}$ and subsequently $\set{\tilde{Q}^{b'}_j}_{j\in[k]}$, where $b'\ne b$ and $\tilde{Q}^b_{\hat{x}}=W_{BM}^b{Q}^b_{\hat{x}}W_{BM}^{b,\dagger}$. The average probability that an honest Bob can guess an honest Alice's $x_b$ is 
    \begin{align*}
        V :=~&\frac{1}{nk^{n}}\sum_{b=1}^n\sum_{x\in[k]^n} \Tr\br{\br{P_{\mathrm{Acc}}^x\otimes Q_{x_b}^b}\psi^{x,b}}\\
        =~&\frac{1}{n}\sum_{b=1}^n\sum_{x\in[k]^n}\Tr\br{\br{P_{\mathrm{Acc}}^x\otimes\proj{x}_{A'}\otimes\tilde{Q}^b_{x_b}}\tilde{\Phi}^{b}}.
    \end{align*}

    Therefore, the average probability that a cheating Bob can guess two database entries and both parties accept is at least 
    \begin{align*}
        E:=~& \frac{1}{n(n-1)k^n} \sum_{b\ne b'}\sum_{x\in[k]^n} \Tr\br{P_{\mathrm{Acc}}^x\otimes \tilde{Q}^b_{x_b}\tilde{Q}^{b'}_{x_{b'}}\tilde{Q}^b_{x_b}\psi^{x,b}}\\
        =~& \frac{1}{n(n-1)}\sum_{b\ne b'}\sum_{x\in[k]^n} \\
        &\qquad\qquad\Tr\br{\br{P_{\mathrm{Acc}}^x\otimes \proj{x}_{A'}\otimes\tilde{Q}^b_{x_b}\tilde{Q}^{b'}_{x_{b'}}\tilde{Q}^b_{x_b}}\tilde{\Phi}^{b}} . 
    \end{align*}
    Denote  
    \begin{equation}\begin{aligned}
        R^b & = \sum_{x\in[k]^n} P_{\mathrm{Acc}}^x\otimes \proj{x}_{A'}\otimes\tilde{Q}^b_{x_b}. 
    \end{aligned}\end{equation}
    Then 
    \begin{equation}\begin{aligned}
        & V   = \frac{1}{n}\sum_{b=1}^n  \Tr( R^b\tilde{\Phi}^{b}), \\  
        & E  \geq \frac{1}{n(n-1)} \sum_{b\ne b'}\Tr( R^bR^{b'}R^b\tilde{\Phi}^{b}).
    \end{aligned}\end{equation}
    Recall that 
    \begin{equation}
         V   \geq 1-\delta \quad\mbox{and}\quad
        \frac{1}{2}\norm{\tilde{\Phi}^{b}-\tilde{\Phi}^{b'}}_1\leq4\sqrt{\epsilon_A}, \end{equation}
    we obtain by Theorem~\ref{thm:RCMT_TD} 
    \begin{equation}
        E \geq 4(1-\delta)\left(\max\left\lbrace0,\frac{1}{2}-\delta\right\rbrace \right)^2-4\sqrt{\epsilon_A},
    \end{equation}
    as well as 
    \begin{equation}
        E\geq\frac{n^2(1-\delta)}{(n-1)^2}\left(\max\left\lbrace0,1-\frac{1}{n}-\delta \right\rbrace\right)^2-\frac{16n}{n-1}\sqrt{\epsilon_A}. 
    \end{equation}
\end{proof}

\bibliography{ref}

\end{document}